\documentclass[12pt]{article}
\usepackage{epsfig} \usepackage{graphicx}

\def\r{\vec r}

\def\ri {{\rm i}} \def\re {{\rm e}} \def\rd {{\rm d}}
\newcommand{\be}{\begin{equation}}
\newcommand{\ee}{\end{equation}}
\newcommand{\bea}{\begin{eqnarray}}
\newcommand{\eea}{\end{eqnarray}}
\newcommand{\nn}{\nonumber}
\begin{document}

\newcommand{\lra}{\leftrightarrow}    \newcommand{\Bq}{{\overline B}}                 \newcommand{\xq}{{\overline x}} 
\newcommand{\Kqz}{{\overline K}{}^0}  \newcommand{\lbt}{{\overline\lambda}{}^\prime}  \newcommand{\cq}{{\overline c}} 
\newcommand{\Dqz}{{\overline D}{}^0}  \newcommand{\jpsi}{J/\psi}                      \newcommand{\bq}{{\overline b}}
\newcommand{\Bqz}{{\overline B}{}^0}  \newcommand{\Mqz}{{\overline M}{}^0}            \newcommand{\reps}{{\rm Re\,\epsilon}} 
\newcommand{\Bqs}{{\overline B}{}_s}  \newcommand{\rdel}{{\rm Re\,\delta}} 
\newcommand{\Aq}{{\overline A}}       \newcommand{\idel}{{\rm Im\,\delta}}            \newcommand{\aq}{{\overline a}}

\setcounter{table}{1}

\title{T Violation and CPT Tests in Neutral-Meson Systems}
\author{K.~R.~Schubert
\\[8mm]
Institut f\"ur Kern- und Teilchenphysik, Technische Universit\"at Dresden, Germany\\
and\\
Institut f\"ur Kernphysik, Johannes Gutenberg-Universit\"at Mainz, Germany\\[6mm]
published in {\em Progress in Particle and Nuclear Physics}, December 29, 2014}
\maketitle
\begin{abstract} 
The review covers transitions $M^0\leftrightarrow\Mqz$ in all neutral-meson systems that can show
these weak-interaction transitions, $M^0=K^0,D^0,B^0$ and $B_s$. The four systems are ideal
laboratories for studying the discrete symmetries T and CPT. The properties of time reversal T
are discussed in classical and quantum mechanics. T violation in $K^0\Kqz$ transitions has been
observed 1970 independent of assuming CPT symmetry by using the Bell-Steinberger unitarity relation.
Improvements of this observation are reviewed including the latest results from 2013. They show with
high significance that CP violation in $K^0\Kqz$ transitions is only T violation without any 
CPT violation. Transitions in the other three systems do not show CP violation so far and, therefore,
neither T nor CPT violation. The large observed CP violation in $B^0\to J/\psi K^0$ decays is
presented to be compatible with only T and no CPT violation, the same for the significant but very small
CP violation in decays $K^0\to\pi\pi,~I=2$.
\end{abstract}

\tableofcontents

\section{Introduction}

Symmetries have always played an important role in the formulation of basic-physics
laws. The theorem of E.~Noether \cite{1918-Noether} relates the symmetry under time 
transformations $t\to t+t_0$ to energy conservation, under space transformations
${\vec x}\to {\vec x}+{\vec x_0}$ to momentum conservation, and rotations ${\vec x}\to
{\bf R}\,{\vec x}$ to the conservation of angular momentum. In relativistic mechanics
and electrodynamics, these transformations and their symmetries are combined into
Poincar\'e or Lorentz transformations, including time reversal T ($t\to -t$) and the
parity transformation P (${\vec x}\to -{\vec x}$).

The description of elementary-particle dynamics by quantum mechanics and quantum field
theory with three interactions obeys Poincar\'e and Lorentz symmetry as well, and in addition the symmetry 
C under the exchange of particles and antiparticles. Around 1956 experiments showed that two
interactions are fully symmetric and that the weak interaction is only symmetric
under Poincar\'e transformations excluding P and C. These two symmetries are maximally broken,
up to 1964 in such a way that CP and T were assumed to be perfect symmetries. Since C, P and CP are described
by unitary transformations, strong and electromagnetic interactions conserve the quantum
numbers C, P and CP, whereas the weak interaction conserves only the quantum number CP and
neither C nor P. The transformations T and CPT are antiunitary; therefore, there exist no conserved quantum
numbers for them in any of the three interactions. All three interactions could
be assumed to be symmetric under the transformations T and CPT.

In 1964 decays of neutral K mesons into two charged $\pi$ mesons at very late lifetimes indicated
that CP symmetry is broken in either the weak or in a new interaction. Since $ CP = CPT \times T$,
the symmetries T or CPT or both had also to be broken. Around 1968 it became clear that CP symmetry is dominantly broken
in $K^0\Kqz$ transitions and unmeasurably small in the decay amplitudes. Measurements of all four CP-violating 
parameters in $K^0\to\pi\pi$ decays and 
measured limits on CP violation in all other $K^0$ decays allowed in 1970 to conclude that only T symmetry is significantly 
violated in $K^0\Kqz$ transitions and that CPT symmetry therein is valid within errors.

The review describes the phenomenology of $K^0\Kqz$ transitions (Section 4) and all essential experimental results 
(Section 5) up to mid
2014. T violation and validity of CPT have been established with steadily increasing precision between 1970 and 2013.
T violation and CPT validity are also reviewed for $K^0\to\pi\pi$ decay amplitudes, as well as searches for T violation 
in $K^0\to\pi\ell\nu$ amplitudes and in transverse muon polarisation in $K\to\pi\mu\nu$ decays (Section 6).

The discoveries of transitions between $B^0$ and $\Bqz$ in 1987, between $D^0$ and $\Dqz$ in 2007, as well as detailed results 
on $B_s\Bqs$ transitions since 2006 led to intense searches for CP, T and CPT violation in also these three neutral-meson
systems. Apart from unconfirmed $3\,\sigma$ indications, there exist only upper limits; all results up to mid 2014 
are reported here in section 7.

T violation and the search for CPT violation in $B^0$ decays are discussed in detail in Section 8.

Throughout this review, we use the sign convention $p M^0 - q \Mqz$ for the heavier mass eigenstate of the $M^0$ system ($M=K,D,B,B_s$)
and $p M^0 + q \Mqz$ for the lighter mass eigenstate. The quantity $(|p/q|-1)/2$ is called $\reps$,
even if it should be $\reps/(1+|\epsilon|^2)$, i.~e.~both $\reps$ and the unobservable Im~$\epsilon$ are approximated
to be $\ll 1$. Two further conventions used are $\hbar = c = 1$.

\section{Prehistory}\label{Sec-Prehistory}

The first identified meson was a positive Kaon with a mass of $(500\pm 60)$ MeV, found 
in 1944 by L.~Leprince-Ringuet and M.~Lh\'eritier \cite{1944-LeprinceRinguet} in cosmic-ray
observations using a triggered cloud chamber, three years before the discovery of the 
charged $\pi$ meson \cite{1947-Lattes}. The first observation of a neutral Kaon was reported 
by G.~D.~Rochester and C.~C.~Butler \cite{1947-RochesterButler}, again using cosmic-ray
events in a cloud chamber. With the advance of experiments at proton accelerators after 1950, 
a puzzle appeared when comparing $\pi^+$, $\pi^0$, $\pi^-$ with masses around 140 MeV, 
and $K^+$, $K^0$, $K^-$ with masses around 500 MeV. The heavier mesons were called $\theta$ 
when decaying into two and $\tau$ when decaying into three pions. All these mesons were 
produced with comparably large rates in proton-proton collisions, and with the exception of
the $\pi^0$ they decayed with comparably long lifetimes. Since the lightest charged meson
cannot decay by strong interactions, its lifetime was understood to be similar to that of 
the muon owing to weak interactions. Puzzling was the observation that the K mesons also
decayed with long weak-interaction lifetimes in spite of their abundant production by strong interactions;
strong-interaction decays with very short lifetimes into two or three pions are
not forbidden by conservation of energy and momentum.

\begin{figure}[h]
\begin{minipage}{0.30\textwidth}
\includegraphics[width=0.90\textwidth]{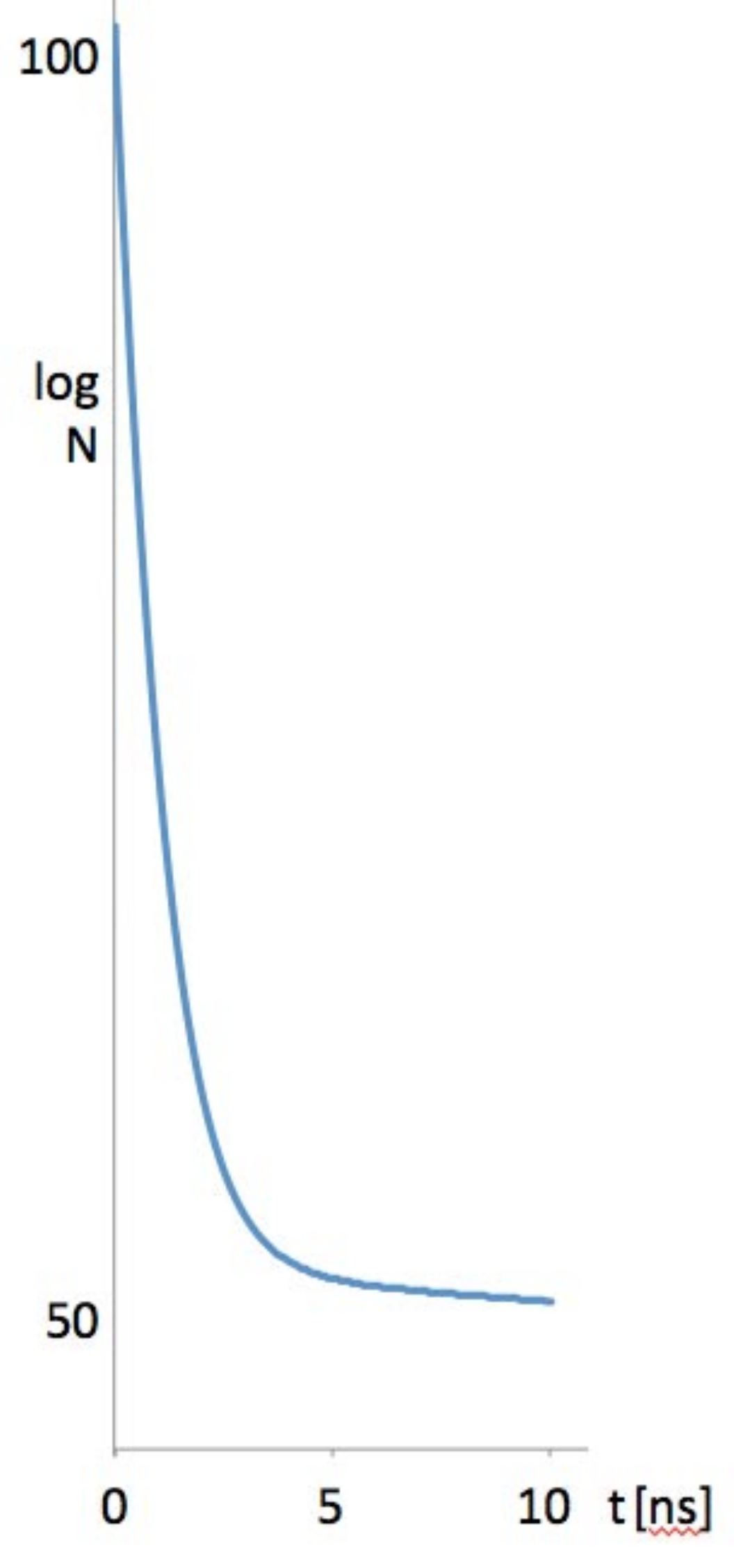}
\caption{The non-exponential decay law of neutral Kaons, $N(t)=N(0)\times(\re^{-\Gamma_1 t}+\re^{-\Gamma_2 t})/2$.}
\label{Fig-1} 
\end{minipage}\hfill
\begin{minipage}{0.65\textwidth} 
The solutions of the puzzle were found around 1953 by A.~Pais, M.~Gell-Mann and K.~Nishijima by
introducing the concept of ``associated production" \cite{1952-Pais}, e.~g.~$p p \to p p K^+ K^-$, 
and of the quantum number ``strangeness" \cite{1953-GellMann,1955-Nishijima}, 
e.~g.~$S(K^+)=+1, S(K^-)=-1$, which is conserved in strong and violated in weak interactions.

An important step in 1953 was the discovery that $\theta^+$ and $\tau^+$ are identical, also
$\theta^0$ and $\tau^0$, thanks to the invention of the Dalitz plot \cite{1953-Dalitz} demonstrating
that the three-pion state in $\tau^+$ decays had spin $J=0$. Parity conservation led to
$J^P=0^-$ for the $\tau^+$ and $J^P=0^+$ for the $\theta^+$, but the $\tau$ and $\theta$ masses 
were equal and
also their lifetimes. The increasing number of observed $\tau$ and $\theta$ decays led
T.~D.~Lee and C.~N.~Yang \cite{1956-LeeYang} to the hypothesis in 1956 that parity symmetry is
broken in weak interactions. This was quickly confirmed by two experiments at Columbia University
on 15 January 1957, that of L.~M.~Lederman's group \cite{1957-Lederman} in the decay chain $\pi^+\to
\mu^+\nu\to(e^+\nu{\overline\nu})\nu$ and that of C.~S.~Wu et al \cite{1957-Wu} in $\beta$
decays of polarized ${}^{60}\rm Co$ nuclei.

Neutral Kaons have a second very strange property, a non-exponential decay law, predicted in 1955
by M.~Gell-Mann and A.~Pais \cite{1955-GellMannPais}. Starting from the dominant decay
$K^0\to\pi^+\pi^-$ and its known rate $\Gamma_1$, they argued that C symmetry (charge conjugation)
requires the existence of the ${\overline K}{}^0$ meson and the decay ${\overline K}{}^0\to\pi^+\pi^-$ with 
the same rate $\Gamma_1$. Transitions $K^0\lra\pi\pi\lra{\overline K}{}^0$ with virtual
$\pi\pi$ pairs should  then occur with a transition rate comparable to $\Gamma_1$, and all 
linear superpositions 
\end{minipage}
\end{figure}

\be \Psi = \psi_1 K^0 + \psi_2 \Kqz \label{Eq-2-1}\ee
are possible states. One of these states,
\be K^0_1 = (K^0 + \Kqz)/\sqrt{2} \label{Eq-1}\ee
decays into two pions with rate $\Gamma_1$.
The orthogonal state,
\be K^0_2 = (K^0 - \Kqz)/\sqrt{2} \label{Eq-2}\ee
is not allowed to decay into two pions because of C conservation, but it can decay into other modes
like $3\pi$, $\pi e \nu$ or $\pi\mu\nu$ with a much smaller total rate $\Gamma_2$. Neutral Kaons
produced in the state $K^0$, e.~g.~in the reaction $n p\to p p K^0 K^-$, have a non-exponential decay
law, decaying dominantly into $\pi\pi$ at early times with a rate $\Gamma_1$ and later into the other
modes with $\Gamma_2\ll\Gamma_1$. With today's values \cite{2012-PDG} of $\Gamma_1$ and $\Gamma_2$, 
their decay law is shown in Fig.~\ref{Fig-1}.

This remarkably early prediction turned out to be true in spite of the fact that together with P symmetry
also C symmetry is broken in weak interactions. Soon after establishing that P is maximally violated, 
L.~Landau \cite{1956-Landau} found that C symmetry is maximally violated as well and that the
combined symmetry CP is unbroken in weak interactions. For the two neutral Kaon states, which are both
P eigenstates with $P=-1$, CP symmetry implies C symmetry, and the arguments of Gell-Mann and Pais
remain unchanged.

In 1956 the group of L.~M.~Lederman \cite{1956-Lande} observed 20 $K^0$ decays into charged final states 
incompatible with $\pi^+\pi^-$, living much longer and indicating $K^0_2$ states with $\Gamma_2/\Gamma_1<0.10$.
In 1958 M.~Baldo-Ceolin et al. \cite{1958-BaldoCeolin} reported a nuclear-emulsion experiment in a $K^+$ beam, 
where $K^0$ mesons
with $S=+1$ from a charge-exchange reaction such as $K^+ n \to K^0 p$ produced $\Lambda$ hyperons with
$S=-1$ in a reaction such as $\Kqz p \to \Lambda \pi^+$. This was one of the first clear 
demonstrations of the transition $K^0\to {\overline K}{}^0$.
In 1961 R.~H.~Good et al. \cite{1961-Good} further investigated this transition by passing a neutral Kaon beam
far from its production target through a metal plate and observing the time dependence of $K\to\pi^+\pi^-$
decays closely behind the plate. Owing to the two-state nature of neutral Kaons, the time-dependent transitions 
$K^0\lra {\overline K}{}^0$ cannot be described by a single ``transition rate". There are two coherent
contributions, the mass and width differences $\Delta m = m(K^0_2)-m(K^0_1)$ and $\Delta\Gamma = \Gamma_2
- \Gamma_1$, as described in detail in Section \ref{Sec-Description}. In addition to $|\Delta \Gamma| 
\approx 0.9\times\Gamma_1$
from Ref.~\cite{1956-Lande}, the metal-plate ``regeneration" experiment \cite{1961-Good} determined $|\Delta m|=(0.8\pm 0.2)
\times \Gamma_1$, so both contributions to the transition rate are indeed of order $\Gamma_1$ as predicted
by Gell-Mann and Pais \cite{1955-GellMannPais}.

Regeneration will be discussed in Section \ref{Sub-Regeneration}. Studying these 
transitions $K^0_2\to K^0_1$ in matter, the 
group of R.~Adair \cite{1963-Adair} found an effect in 1963 which they called ``anomalous regeneration".
At the limit of significance, a larger rate of $\pi^+\pi^-$ decays was observed in a beam of neutral Kaons 
behind a liquid-hydrogen regenerator than that expected from the estimated regeneration rate. The more precise and 
famous experiment of J.~H.~Christenson, J.~W.~Cronin, V.~L.~Fitch and R.~Turlay \cite{1964-FitchCronin}
discovered one year later with high significance that this larger rate occurs also in a near-to-vacuum
material, gaseous helium. The influence of regeneration was negligible; long-living Kaons in the assumed CP
eigenstate $K^0_2$ with $CP=-1$ decayed into the final state $\pi^+\pi^-$ with $CP=+1$. This evident 
non-conservation of CP symmetry could still have two origins: the decay dynamics of $K^0_2\to\pi\pi$
or the transition dynamics of $K^0\lra\Kqz$ generating a long-living state which is not a CP eigenstate.
In 1967 the experiment of J.~Steinberger's group \cite{1967-Steinberger} gave the answer that the second
origin is dominating, by observing that the decay rate of the long-living state into $\pi^-e^+\nu$ is larger
than that into $\pi^+e^-{\overline\nu}$.

The first observed meson was a $K^+$ in 1944, its neutral partner and the charged pion were discovered in
1947, $K^0\Kqz$ transitions were predicted in 1955 and observed since 1956, the study of $K^+\to\pi^+
\pi^+\pi^-$ decays in 1954 led to the discovery of P violation, and the 1964 observation of long-living neutral 
Kaons decaying into $\pi^+\pi^-$ was the discovery of CP violation. In the following years, we had the obligation to
find out experimentally if CPT symmetry is also broken, more specific: what are the contributions of T and of
CPT violation to the observed CP violation in $K^0\Kqz$ transitions? 

\section{The Symmetries T and CPT} \label{Sec-TSym}

The concept of time reversal in the basic laws of Nature is different from that of
the ``arrow of time", both in classical and in quantum physics. The omnipresent
arrow of time is a consequence of causality and of entropy increase in closed systems.
We only get older and never get younger, and we can never influence an event in the past.
However, a basic law can be either time-reversal (T) symmetric or T violating,
even if obeying causality and increasing entropy in both cases. The dynamics of weak
interactions can be T symmetric or T violating, but in both cases an ensemble of unstable particles shows the arrow of time 
by becoming a different ensemble composed of surviving particles and various decay states.
  
\subsection{Time Reversal in Classical Mechanics} \label{Sub-Classical}

The discussion can be kept on the level of equations of motion. In cases where the equation
\be m~\ddot{\r}= {\vec F}\ee
for the motion of a point-particle with mass $m$ and coordinate $\r$ contains a time-independent force $\vec F$,
the equation is invariant under the time-reversal transformation T,
\be t\to -t~.\label{Eq-3-1}\ee
The transformation here leads to time-reversal symmetry, called T symmetry in the following. As a consequence, we
observe "motion-reversal" symmetry, the symmetry between two motions, as illustrated in the following example. 
\begin{figure}[h]
\begin{minipage}[t]{0.30\textwidth}
\includegraphics[width=\textwidth]{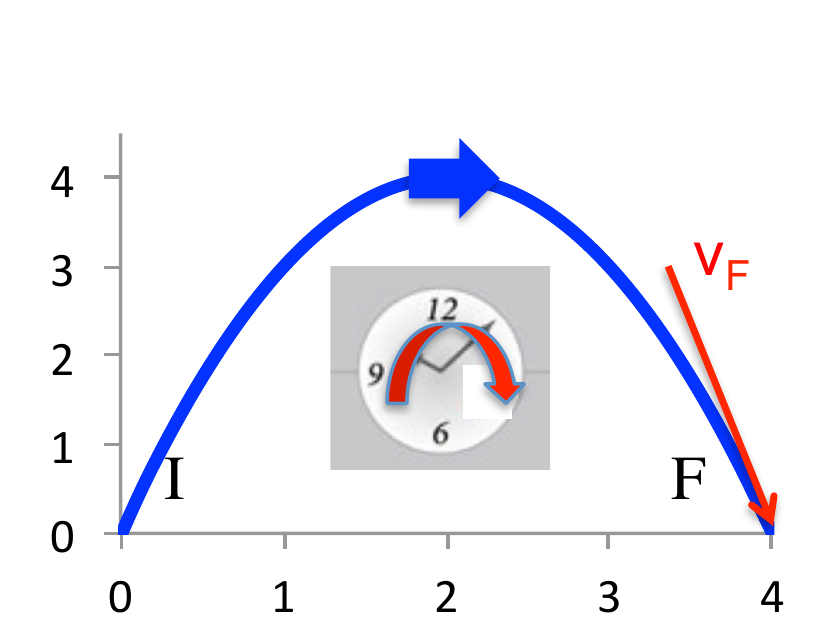}
\caption{Video recording of a motion with velocity ${\vec v}{}_F$ in the endpoint F together with a clock showing the running time.}
\label{Fig-2a} 
\end{minipage}\hfill
\begin{minipage}[t]{0.30\textwidth}
\includegraphics[width=\textwidth]{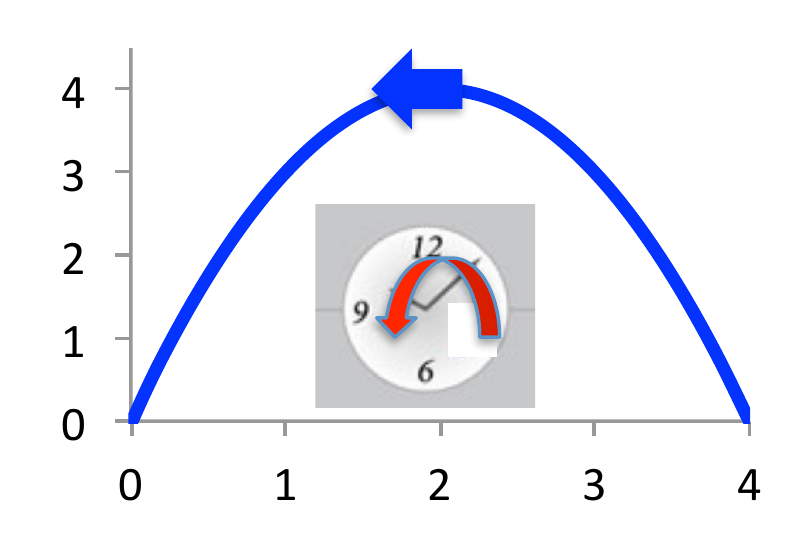}
\caption{Playback of the recorded video in the reversed time-direction.}
\label{Fig-2b} 
\end{minipage}\hfill
\begin{minipage}[t]{0.30\textwidth}
\includegraphics[width=\textwidth]{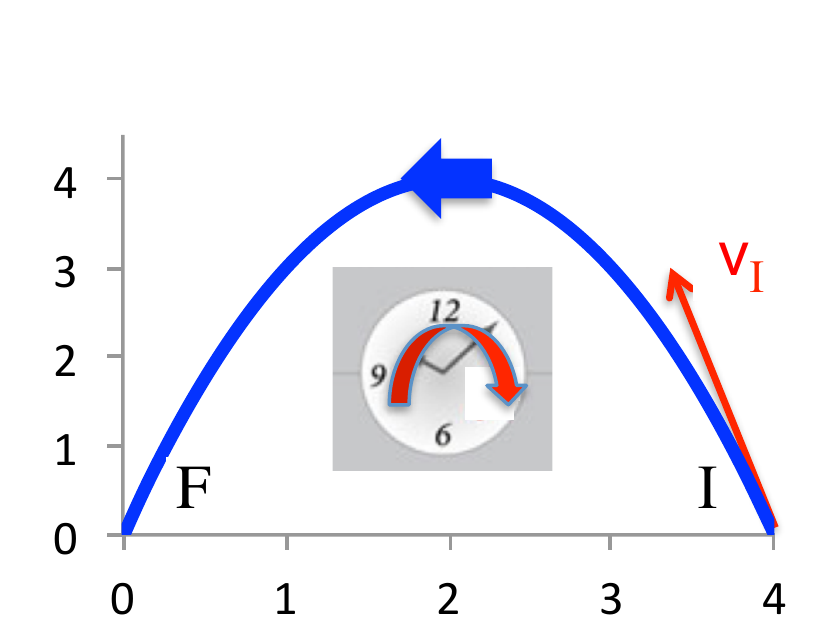}
\caption{The time-reversed motion with velocity ${\vec v}{}_I=-{\vec v}{}_F$ in the new start point I together with a clock
showing the forward-running time.}
\label{Fig-2c} 
\end{minipage}
\end{figure}
Fig.~\ref{Fig-2a} shows the orbit of the motion of a ball on the surface of the earth, in very good approximation
if velocity and distance are small.
This motion can be recorded together with a clock showing the time for the motion from the start to the end point.
A video recorder allows to replay the recorded movie in the backward direction, as shown in Fig.~\ref{Fig-2b}.
Even if the motion of the ball looks familiar to the viewer, the clock tells him that this process does not take place
in his world. The replayed movie shows an unobservable process; time does never run backwards. It is, however, possible
to observe the reversed motion in the real world: we have to start it at the end point of the original motion with a
velocity vector equal to the opposite of the final velocity of the original motion, as shown in Fig.~\ref{Fig-2c}.
This observable motion with forward-running time is called ``reversed motion". The operation with starting at the
end point and reversing the velocity is called ``motion reversal", and the comparison of Figs.~\ref{Fig-2c} and \ref{Fig-2a}
shows ``motion-reversal symmetry".

\begin{figure}[h]
\centering\includegraphics[width=0.60\textwidth]{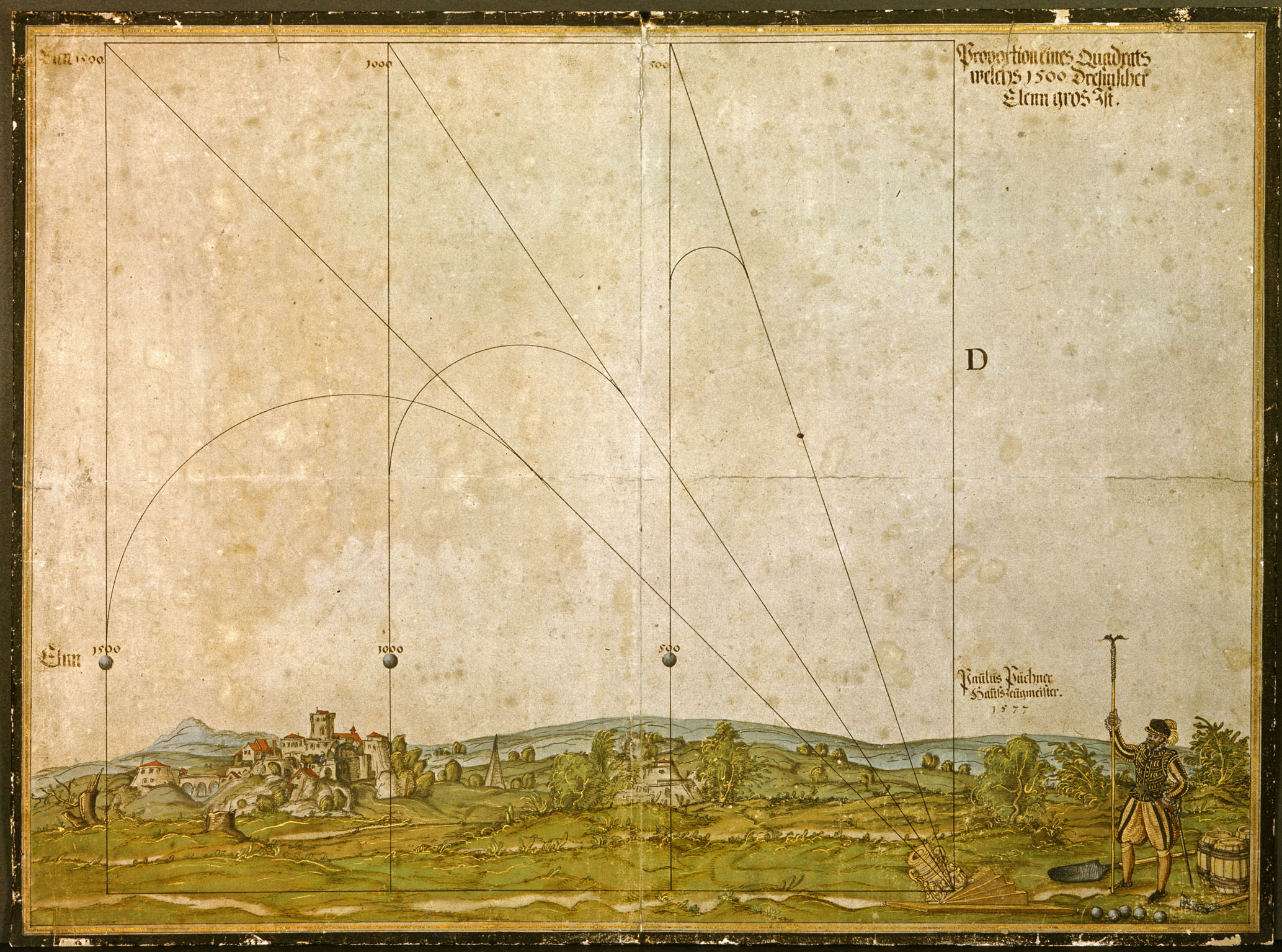}
\caption{P.~Puchner, Dresden 1577 \cite{2014-SKD}, Gun-ball orbits according to a theory of N.~Tartaglia \cite{1537-Tartaglia}.}
\label{Fig-Puchner}
\end{figure}

Motion-reversal symmetry is a consequence of T symmetry in the equation of motion, here
\be m~\ddot{\r}= m~{\vec g}~.\ee
Motion-reversal symmetry violation,
i.~e.~different orbits for the reversed and the original motion, proves T violation of the underlying
dynamics. This symmetry violation with balls of higher velocity in air is observed since long time, a 16th century
example is shown in Fig.~\ref{Fig-Puchner}. Motion reversal leads to a very different orbit here; we understand the reason
since the approximate equation of motion,
\be m~\ddot{\r} = m~{\vec g}-\eta~\dot{\r}~,\label{Eq-3-3}\ee
is not invariant under the T operation in Eq.~\ref{Eq-3-1}.
The example shows that we have two ways to prove T violation: either the ``direct" observation of motion-reversal violation,
here moving the gun to the target position and firing in the opposite direction, or the ``indirect" measurement of the
parameter $\eta$ in Eq.~\ref{Eq-3-3} finding that it is different from zero.

\begin{figure}[h]
\begin{minipage}{0.32\textwidth}
\includegraphics[width=\textwidth]{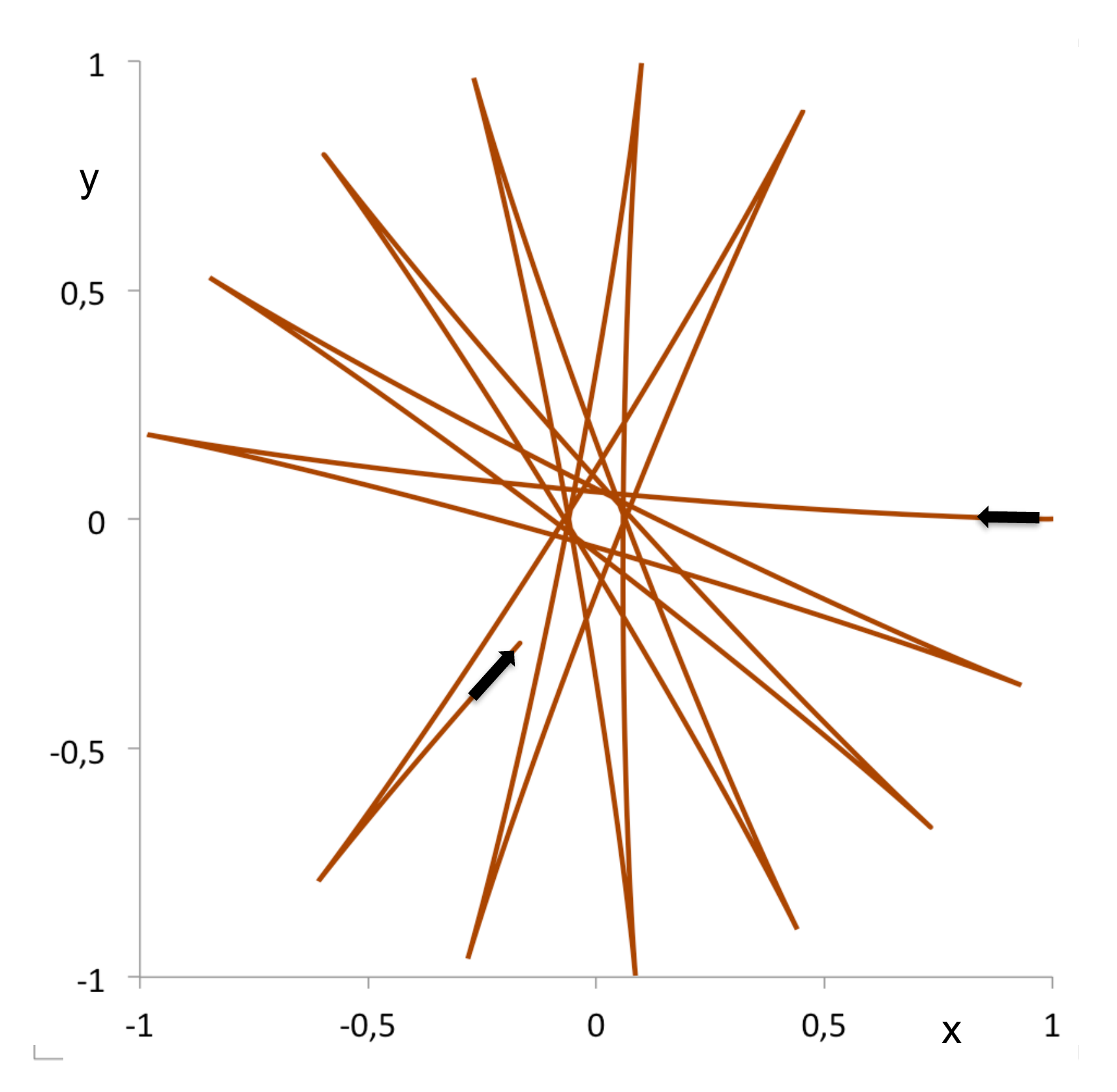}
\caption{Orbit of a pendulum on a turntable with arbitrary $\vec\Omega$.}
\label{Fig-Foucault} 
\end{minipage}\hfill
\begin{minipage}{0.64\textwidth}
Another example for motion-reversal symmetry violation, without dissipation, is the Foucault pendulum. With $x, y \ll \ell$,
where $\ell$ is the suspension length, its equation of motion
in the frame of the rotating earth with ${\vec\Omega}=2\pi {\vec e}_z/24~h$ is
\be m~\ddot{\r} = -m~\frac{g}{\ell}~\r+2~ m~ \dot{\r}\times {\vec\Omega}~.\label{Eq-3-4}\ee
This equation is not invariant under $t\to -t$, leading to rosette-like orbits like those in Fig.~\ref{Fig-Foucault}. In all parts of the orbit
we observe deflections to the right side. At every turning point the pendulum starts again to be deflected to the right side
instead of returning on the motion-reversed path. The basic laws of classical mechanics are T symmetric; motion reversal
symmetry would be restored if we would not only reverse the motion of the Foucault pendulum but also the rotation of the earth.
\end{minipage}
\end{figure}

\subsection{Time Reversal in Quantum Mechanics} \label{Sub-Quantum}

Time reversal in the laws of quantum mechanics was first discussed 1932 by E.~Wigner \cite{1932-Wigner}. Detailed discussions are presented 
in the textbooks of Sachs \cite{1987-Sachs} and Branco Lavoura Silva \cite{BrancoLavouraSilva}, summarized here in a few lines only. 
In the Schr\"odinger representation, the state vector $|\psi\rangle$ of a quantum system develops in time according to
\be \ri \,\frac {\partial \,|\psi(t)\rangle }{\partial\, t}={\cal H}\,|\psi(t)\rangle~,\label{Eq-3-2-1}\ee
and observables of the system are described by hermitean operators $Q$ with the expectation values 
\be \langle Q\rangle = \langle\psi|Q|\psi\rangle~.\ee
The T transformation $t\to t^\prime =-t$ transforms state vectors and observables using the $T$ operator
\be |\psi^\prime(t^\prime)\rangle = T\,|\psi(t)\rangle~,~~Q^\prime = T Q T^{-1}~.\ee 
Under the T transformation, the space and momentum observables $\vec x$ and $\vec p$ have the properties ${\vec x}{}^\prime =\vec x$ and 
${\vec p}{}^\prime =-\,\vec p$. Consequently, the quantum-mechanical commutation relations $[p_i,x_j]=-\ri\,\delta_{ij}$ require 
\be T\,\ri\,T^{-1}=-\,\ri~.\ee
Since $|\psi\rangle$ and $|\psi^\prime\rangle$ are in general states of the same system, the  operator $T$ must have the property
\be T = U\, K~,~~K\,z\,K^{-1} = z^*~,~~T|\psi\rangle = U|\psi^*\rangle~,\ee
where $z$ is any complex number, $U$ is unitary, $UU^\dag=\bf 1$, and $K$ is the
operator for complex conjugation, $K^2=\bf 1$, $K^{-1}=K$, $T^{-1}=K U^\dag$. 
The operator T is antilinear,
\be T(c_1|\psi_1\rangle + c_2|\psi_2\rangle)=U(c_1^*|\psi_1^*\rangle + c_2^*|\psi_2^*\rangle)~,\ee
and ``antiunitary". Antiunitarity means that the operator  $T^\dag=K^\dag U^\dag$ is different from $T^{-1}$. The operator $K^\dag$ must be
understood to complex-conjugate all numbers to its left side, i.~e.
\be \langle T\psi_2 |T\psi_1 \rangle = \langle \psi_2 |T^\dag T|\psi_1 \rangle = \langle \psi_2 |K^\dag U^\dag U K|\psi_1 \rangle 
     = \langle \psi_2^*|\psi_1^* \rangle = \langle \psi_1|\psi_2\rangle~.\label{Eq-T-on-scalar-products}\ee
With an arbitrary operator ${\cal O}$, the time-reversed matrix element of $\langle \psi_2|{\cal O}|\psi_1\rangle$ is given by
\bea \langle T \psi_2|T {\cal O} T^{-1}|T \psi_1\rangle &=& \langle \psi_2|K^\dag U^\dag U K {\cal O} K U^\dag U K|\psi_1\rangle\nonumber\\
             &=& \langle \psi_2^*|{\cal O}^*|\psi_1^*\rangle = \langle \psi_2|{\cal O}|\psi_1\rangle ^* = 
             \langle \psi_1|{\cal O}^\dag|\psi_2\rangle~.\label{Eq-T-on-operators}\eea

Invariance of Eq.~\ref{Eq-3-2-1} under a unitary transformation $U$, i.~e.~$U\,{\cal H}\,U^{-1}= {\cal H}$,
leads to a conservation law, an example is CP conservation. The same invariance under T or any other antiunitary transformation like CPT
does not lead to a conservation law. These symmetries do not lead to a conserved quantum number. There can be T or CPT symmetry, but never T or CPT
conservation. Violation of T symmetry can be called T violation, the same for CPT.

The asymptotic solutions of Eq.~\ref{Eq-3-2-1}, $\psi(t_2)=\re^{-\ri{\cal H}(t_2-t_1)}~\psi(t_1)$, are given by the operator $S$,
\be S = \lim_{t_1\to\infty}~\lim_{t_2\to\infty}~\re^{-\ri{\cal H}(t_2-t_1)}~,\ee
which is unitary, $S S^\dag = \bf 1$. With $T{\cal H}T^{-1}=\cal H$, i.~e.~energy conservation under time reversal, the $S$ operator
transforms like
\be T S T^{-1}= S^\dag~.\ee
It is usually separated into two parts,
\be S = {\bf 1}+\ri~D~,\label{Eq-DefD}\ee
where $D$ describes transitions and decays from $|\psi_1(t_1)\rangle$ into different states $|\psi_2(t_2)\rangle$; it transforms like
\be T D T^{-1}= D^\dag~.\ee
States containing particles with momenta ${\vec p}_i$ and spins ${\vec s}_i$ transform like
\be T~|{\vec p}_i,{\vec s}_i\rangle = U K |{\vec p}_i,{\vec s}_i\rangle = \re^{\ri\phi}|-{\vec p}_i,-{\vec s}_i\rangle~,\ee
with an arbitrary phase $\phi$.
Since $T S T^{-1}=S^\dag$, the S-matrix elements $\langle {\vec p}_f,{\vec s}_f |S|{\vec p}_i,{\vec s}_i\rangle$ transform like
\bea T(\langle {\vec p}_f,{\vec s}_f |S|{\vec p}_i,{\vec s}_i\rangle) &=& \re^{\ri\Delta\phi}
     \langle {-\vec p}_f,-{\vec s}_f |S^\dag|{-\vec p}_i,{-\vec s}_i\rangle\nonumber\\
     &=& \re^{\ri\Delta\phi}\langle {-\vec p}_i,-{\vec s}_i |S|{-\vec p}_f,{-\vec s}_f\rangle ^*\nonumber\\
 |T(\langle {\vec p}_f,{\vec s}_f |S|{\vec p}_i,{\vec s}_i\rangle) | &=& |\langle {-\vec p}_i,-{\vec s}_i |S|{-\vec p}_f,{-\vec s}_f\rangle |~,\eea
and T invariance means
\be |\langle {\vec p}_f,{\vec s}_f |S|{\vec p}_i,{\vec s}_i\rangle | = |\langle {-\vec p}_i,-{\vec s}_i |S|{-\vec p}_f,{-\vec s}_f\rangle |~,
    ~~|\langle {\vec p}_f,{\vec s}_f |D|{\vec p}_i,{\vec s}_i\rangle | = |\langle {-\vec p}_i,-{\vec s}_i |D|{-\vec p}_f,{-\vec s}_f\rangle |~.\ee
This invariance is often called principle of detailed balance. It manifests itself by equal rates for the two transitions $i\to f$ and
$f{\rm (momenta~and~spins~reversed)}\to i{\rm (momenta~and~spins~reversed)}$, equivalent to motion reversal in classical mechanics.   
  
\subsection{Examples for Motion Reversal in Particle Physics} \label{Sub-Examples}

In the following, we shortly discuss three particle-physics examples of motion reversal: comparing two reactions between stable nuclei, comparing
two transitions between different neutrino species, and comparing decay and formation of an unstable particle.

\subsubsection{Nuclear Reactions}

A test of detailed balance, i.~e.~motion-reversal symmetry, by measuring the energy-dependent cross sections of the nuclear reactions
$A+B\to C+D$ and $C+D\to A+B$ has been performed in 1967 at Heidelberg by W.~von Witsch et al.~\cite{1968-vonWitsch} using 
$\rm {}^{24} Mg +\alpha \rightleftharpoons {}^{27} Al+p$. 
At the same center-of-mass (*) energies and scattering angles, T symmetry requires for the differential cross sections
\be \frac{\rd\sigma ^\rightarrow}{\rd\Omega^*}=\frac{\rd\sigma ^\leftarrow}{\rd\Omega^*}\times\left(\frac{p_f^*}{p_i^*}\right)^2
    \frac{(2s_{1f}+1)(2s_{2f}+1)}{(2s_{1i}+1)(2s_{2i}+1)}~,\ee
where $p_f^*$ and $p_i^*$ are the momenta, and $s_{1i},s_{2i}$ and $s_{1f},s_{2f}$ the spins of the two particles in the initial and
final state of the reaction $\rightarrow$, respectively.

The sensitivity to T violation is especially high in reactions where many amplitudes contribute at the same energy. The chosen example has 
this property since the reactions proceed via the compound nucleus $\rm {}^{28}Si$ in an energy regime with large cross-section
fluctuations of the Ericson type.  Cross-section results are presented in several energy ranges at several scattering angles. The authors
conclude that no deviation from T symmetry is observed. The upper limit for the T-violating fraction of the reaction amplitude is quoted to be 
$3\times 10^{-3}$ with 85\% confidence.

\subsubsection{Neutrino Oscillations}

The standard phenomenology of neutrino mixing uses the $3\times 3$ unitary PMNS matrix
\cite{2013-NakamuraPetcov, 1962-MNS, 1967-Pontecorvo} relating the three mass eigenstates $\nu_j,~j=1,2,3$, to the three flavor eigenstates
$\nu_i,~i=e,\mu,\tau$, by
\be \nu_i = \sum_{j=1}^3 U_{ij}\nu_j~.\ee
At a given neutrino energy $E$ and distance $L$ between production and detection, the transition rates between $\mu$- and $e$-neutrinos are given by
\be P(\nu_e\to\nu_\mu)= \left |\sum_{j=1}^3 U_{ej} U_{\mu j}^*~
\re^{-\ri m_j^2L/2E}\right |^2~,~~P(\nu_\mu\to\nu_e)= \left |\sum_{j=1}^3 U_{\mu j} U_{e j}^*~\re^{-\ri m_j^2L/2E}\right |^2~.\ee
The comparison of the two rates will be difficult to measure, but the standard phenomenology allows different rates, i.~e.~T violation, if the 
Jarlskog invariant \cite{1985-Jarlskog} $J=J(U_{ij}) = {\rm Im}(U_{e1}U_{e2}^*U_{\mu 2}U_{\mu 1}^*)$ is different from zero. For the experimental difficulties, including matter effects, see e.~g.~Ref.~\cite{2004-Akhmedov}.

A comparison of the rates $P(\nu_e\to\nu_\mu)$ and $P({\overline\nu}{}_\mu\to{\overline\nu}{}_e)$ will be a test of CPT symmetry. The standard phenomenology
predicts the rate difference to be zero even if $J=J(U_{ij})\ne 0$. The underlying condition is CPT symmetry between the mixing matrices $U_{ij}$ of neutrinos
and ${\overline U}{}_{ij}$ of antineutrinos,
\be {\overline U}{}_{ij} = U_{ij}^*~.\ee
An observed CPT violation in the rates would imply that there is no CPT symmetry between the two mixing matrices,
i.~e.~${\overline U}{}_{ij} \ne U_{ij}^*$.

\subsubsection{Formation and Decay of the $J/\psi$ Meson}

The $J/\psi$ meson has been discovered in 1974 in two experiments, in the production reaction $p + N\to J/\psi + X,~J/\psi\to e^+ e^-$ \cite{1974-Ting} and in the formation
reaction $e^+ e^- \to J/\psi$ \cite{1974-Richter}. The decay rate $\Gamma=\Gamma_{tot}(J/\psi)\times {\cal B}(J/\psi\to e^+ e^-)$ and the formation cross section $\sigma
=\sigma(e^+ e^-\to J/\psi)$ are given by the same matrix element if T symmetry holds. A measurement of $\Gamma$ then predicts the formation cross section $\sigma$.
However, establishing motion-reversal symmetry by the two measurements would not be a test of T symmetry since CPT symmetry predicts exactly the same relation between
$\Gamma$ and $\sigma$.

\subsection{T-Symmetry in Decays}

The decay of an unstable particle allows perturbative descriptions if the decay proceeds much slower than the production. Therefore, the discussion here is limited
to decays induced by weak and electromagnetic interactions, including possible new physics in addition to the Standard interactions. These decays allow to test T symmetry of the decay 
dynamics without using motion-reversal experiments. This turns out to be useful since the reversal of a decay like $n\to p e^-{\overline\nu}{}_e$ is practically impossible.

Because of unitarity of the $S$ operator, the transition operator $D$ as defined in Eq.~\ref{Eq-DefD} has the property
\be D - D^\dag = \ri D^\dag D~.\label{Eq-DDdag}\ee
When there are no strong interactions involved, the couplings in $D$ are so small that the second-order transition on the right-hand side of Eq.~\ref{Eq-DDdag} is
negligibly small compared to the transitions produced by $D$ and $D^\dag$, effectively leading to hermiticity $D^\dag = D$. With $\langle f|D|i\rangle=\langle f|D~i\rangle$ and
Eq.~\ref{Eq-T-on-scalar-products}, T symmetry requires 
\be \langle f|D~i\rangle = \re^{\ri\Delta\phi}~\langle f|D^\dag~i\rangle ^*~,\ee
and hermiticity of $D$ leads to
\be \langle f|D|i\rangle = \re^{\ri\Delta\phi}~\langle f|D|i\rangle ^*~,\label{Eq-30}\ee
where the unobservable phase $\Delta\phi$ originates from the arbitrary phases of the states $|i\rangle$  and $|f\rangle$. The T transformation of
two different transitions $D_1$ and $D_2$, both without strong interactions, between $|i\rangle$  and $|f\rangle$ lead to the same
unobservable phase, i.~e.~T symmetry and hermiticity of $D_1$ and $D_2$ require that the phase between $\langle f|D_1|i\rangle$ and
$\langle f|D_2|i\rangle$ has to be 0 or $180^\circ$. Here follow two examples:

The beta decay of polarized neutrons allows observation of a T-odd triple product,
\be {\vec P}{}_n\cdot [{\vec p}{}_e \times {\vec p}{}_\nu]~~{\rm with}~~T\{{\vec P}{}_n\cdot [{\vec p}{}_e \times {\vec p}{}_\nu]\}=-{\vec P}{}_n
    \cdot [{\vec p}{}_e \times {\vec p}{}_\nu]~,\ee
since the polarization vector ${\vec P}{}_n$ of the neutron and the center-of-mass momenta ${\vec p}{}_e$ and ${\vec p}{}_\nu$ of electron and neutrino change sign under time reversal.
The hermitean effective Hamiltonian for neutron decay with only vector and axial-vector couplings contains a leptonic, ${\overline e} \gamma_\mu(1-\gamma_5)\nu_e$, and a hadronic
current, ${\overline p} \gamma_\mu(G_V+G_A\gamma_5)n$. T invariance of the Hamiltonian requires \cite{BrancoLavouraSilva}
\be T[{\overline e} \gamma_\mu(1-\gamma_5)\nu_e][{\overline p} \gamma_\mu(G_V+G_A\gamma_5)n]T^{-1} = 
    [{\overline e} \gamma_\mu(1-\gamma_5)\nu_e][{\overline p} \gamma_\mu(G_V^*+G_A^*\gamma_5)n] \ee
up to an unobservable arbitrary phase $\phi$, i.~e.
\be G_V+G_A\gamma_5 = \re^{\ri\phi}(G_V^*+G_A^*\gamma_5)~,~~{\rm Im}(G_A/G_V)=0~,~~\Phi_{AV}=0^\circ~~{\rm or}~~180^\circ~,\ee
where $\Phi_{AV}$ is the relative phase between $G_A$ and $G_V$.
A proposal of J.~D.~Jackson et al.~in 1957 \cite{1957-JacksonTreiman} relates $\Phi_{AV}$ to the coefficient $D$ in the observable partial rate
\be \frac{\rd\Gamma(n\to p e^-{\overline\nu})}{\rd E_e \rd\Omega_e\rd\Omega_\nu}=f_1(E_e,E_\nu)+D\times f_2(E_e,E_\nu)
     \times {\vec P}{}_n\cdot [{\vec p}{}_e \times {\vec p}{}_\nu]~,\ee
where the functions $f_1$ and $f_2$ do not only depend on the electron and neutrino energies but also on $|G_A/G_V|$. Measurements of $D$
with cold polarized neutrons have been performed since 1974 \cite{2012-PDG}. The most recent result, 
obtained in 2012 by T.~E.~Chupp et al.~\cite{2012-Chupp} with the emiT-II detector at NIST, is 
\be D= (-0.9 \pm 1.9 \pm 1.0)\times 10^{-4}~.\ee
The expectation from the CP- and T-violating Standard Model is $D\approx 10^{-12}$ \cite{2012-Chupp}, and final-state interactions
which could give $D\ne 0$ in spite of T symmetry, see the next Subsection, are estimated to produce $D\approx  10^{-5}$ \cite{2012-Chupp}.
Combined with earlier measurements \cite{2012-PDG}, the present best values are
\be D= (-1.2\pm 2.0)\times 10^{-4}~,~~\Phi_{AV}= (180.017\pm 0.026)^\circ~.\ee
The imaginary part of $G_A/G_V$ is compatible with zero, no T violation is seen in neutron decay.

The second example: Radiative decays of excited atoms can be used to test T symmetry of the underlying dynamics in ``forbidden" transitions,
i.~e.~when selection rules exclude E1 but allow E2 and M1 transitions. S.~P.~Lloyd has shown in 1951 \cite{1951-Lloyd} that T invariance
requires ${\rm Im}(A_{M1}/A_{E2})=0$ for the ratio of the two transition amplitudes. 

\subsection{Final State Interactions} \label{Sub-FinalStates}

Strong final-state interactions (FSI) between the decay products of a weak decay can be described by factorizing
\be \langle f | D | i \rangle = \langle f | D | i \rangle _{\rm weak} \times \re^{\ri\delta_{FSI}}~,\ee
where $\delta_{FSI}$ is a strong-interaction phase shift. Following Eq.~\ref{Eq-30}, T symmetry requires
\be \langle f | D | i \rangle = \langle f | D | i \rangle _{\rm weak}^* \times \re^{\ri(\Delta\phi+\delta_{FSI})}~.\ee
This does not lead to a deviation from the result that two transition amplitudes $A_1$ and $A_2$ between $|i\rangle$ and $|f\rangle$ have to
obey ${\rm Im}~(A_2/A_1)=0$ as long as the state $|f\rangle$ is a single eigenstate of the strong interaction. If this is not the case,
an example is the well defined single state $|f\rangle = |\pi^+\pi^-\rangle$ with two strong-interaction eigencomponents 
$|f_1\rangle =|\pi\pi, I=0\rangle$ and $|f_2\rangle =|\pi\pi, I=2\rangle$, the decay has to be described by two amplitudes
\be A = A_1 + A_2 =  a_1~\re^{\ri\delta_1}+a_2~\re^{\ri\delta_2}~~{\rm with}~~a_k = \langle f_k | D | i \rangle_{\rm weak}~,~k=1,2~.\ee 
T symmetry requires Im~$(a_2/a_1)=0$, and
\be {\rm Im}\left(\frac{A_2}{A_1}\right) = \frac{a_2}{a_1}\times\sin(\delta_2-\delta_1)~.\ee
This does not vanish if $\delta_2-\delta_1$ is different from 0 and $180^\circ$, i.~e.~final-state interactions can mimic T violation in decays
if two contributing transition amplitudes have different FSI phases. 

A well-known example in particle physics is the weak hyperon decay $\Lambda\to p\pi^-$. Without FSI, T symmetry requires that the partial decay
rate does not depend on the T-odd triple product ${\vec p}_p\cdot[{\vec\sigma}_\Lambda\times{\vec\sigma}_p]$ with proton momentum ${\vec p}_p$ and spins
$\vec\sigma$ of $\Lambda$ and $p$ \cite{2012-Commins}. The triple-product dependence of the rate is proportional to the phase difference $\Phi$
between the amplitudes $A_p = a_p\re^{\ri\delta_p}$ and $A_s = a_s\re^{\ri\delta_s}$ for p- and s-wave transitions with weak amplitudes
$a_j$ and FSI phases $\delta_j$, $j=p,s$. The present average \cite{2012-PDG} from three experiments is $\Phi=-6.5^\circ \pm 3.5^\circ$. The
expectation is $\Phi=\eta+\delta_p-\delta_s$, where $\eta$ is the T-violating phase between $a_p$ and $a_s$. The FSI phases have been measured
in pion-nucleon scattering assuming the isospin selection rule $\Delta I = 1/2$ \cite{1987-Sachs}, $\delta_p-\delta_s=-6.5^\circ\pm 1.5^\circ$.
This leads to $\eta = 0\pm 3.8^\circ$, no T violation is observed in $\Lambda\to p\pi^-$ decays.

\subsection{CPT Symmetry} \label{Sub-CPT}

The transformation CPT is the product of the three symmetry transformations C, P and T, interchanging particles and antiparticles, $\vec x$ and
$-\vec x$, $t$ and $-t$, respectively. Like T, the operator CPT is antiunitary. Any of the three symmetries C, P, T may be violated in Nature and in
quantum field theories. The CPT theorem, proven in 1954 by G.~L\"uders \cite{1954-Lueders} and 1955 by W.~Pauli \cite{1955-Pauli} shows that
local field theories with Lorentz invariance strictly obey CPT symmetry. Since it is not clear by how much the conditions of Lorentz invariance 
and locality are valid in e.~g.~string theory or quantized gravity, all experimental efforts for searching small
deviations from CPT symmetry are legitimate and well motivated as they explore uncharted territory.

Consequences of the CPT theorem are equal masses for particles and antiparticles, also opposite
charges, opposite magnetic moments, opposite gyromagnetic $g$ values, and equal lifetimes $\tau = 1/\Gamma_{tot}$,
\be \Gamma_{tot}({\overline i}) =\sum_{\overline f}|\langle {\overline f}|D|{\overline i}\rangle|^2 
           = \sum_f|\langle f|D|i\rangle|^2 =\Gamma_{tot}(i)~,\ee
where the sums run over all final states $f$ of an unstable particle $i$ and over all final states $\overline f$ of its antiparticle $\overline i$.
This equality of summed rates is also valid \cite{BrancoLavouraSilva} for each subset of final states which is not connected by final-state
interactions (FSI) to another subset. One example from K-meson physics: Conservation of G parity,  a consequence of C symmetry and
isospin symmetry neglecting electromagnetism, forbids FSI transitions between two-pion and three-pion states (four pions are forbidden because 
of $4 m_\pi > m_K$). The CPT theorem requires 
\be \Gamma(K^+\to\pi^+\pi^0) = \Gamma(K^-\to\pi^-\pi^0)~,\label{Eq-LOeYexample}\ee
i.~e.~there can be no CP violation here because of CPT symmetry. For neutral Kaons,
\be \Gamma(K^0\to\pi^+\pi^-) + \Gamma(K^0\to\pi^0\pi^0) =  \Gamma(\Kqz\to\pi^+\pi^-) + \Gamma(\Kqz\to\pi^0\pi^0)~.\ee
CP violation is possible here, $\Gamma(K^0\to\pi^+\pi^-)\ne\Gamma(\Kqz\to\pi^+\pi^-)$ and the same for the $\pi^0\pi^0$ channel. CPT symmetry
requires only absence of CP violation in the sum of both channels. The same holds for the three-pion states with
\bea \Gamma(K^+\to\pi^+\pi^+\pi^-) + \Gamma(K^+\to\pi^+\pi^0\pi^0) &=&  \Gamma(K^-\to\pi^-\pi^+\pi^-) + \Gamma(K^-\to\pi^-\pi^0\pi^0) ~,\nonumber\\ 
     \Gamma(K^0\to\pi^0\pi^+\pi^-)  + \Gamma(K^0\to\pi^+\pi^0\pi^0) &=& \Gamma(\Kqz\to\pi^0\pi^+\pi^-) + \Gamma(\Kqz\to\pi^0\pi^0\pi^0) ~.\eea
If a single final state is FSI-unconnected to any other final state such as the example in Eq.~\ref{Eq-LOeYexample}, it must have a single FSI phase,
\be \langle f_{\rm single}|D|i\rangle = \langle f_{\rm single}|D|i\rangle_{\rm weak}\times \re^{\ri\delta}~.\ee
Because of the C symmetry of the strong interaction, the CP-conjugate decay must also be a single final state with the same FSI phase \cite{1952-Watson}, 
\be \langle {\overline f}_{\rm single}|D| {\overline i}\rangle = \langle {\overline f}_{\rm single}|D| {\overline i}\rangle_{\rm weak}\times \re^{\ri\delta}~.\ee
In all such cases, CPT symmetry forbids CP violation,
\be |\langle f_{\rm single}|D|i\rangle|^2 = |\langle {\overline f}_{\rm single}|D| {\overline i}\rangle|^2~.\label{Eq-LeeOehmeYang}\ee
A proof has been given in 1957 by T.~D.~Lee, R.~Oehme and C.~N~.Yang \cite{1957-LeeOehmeYang} for weak decays of spinless particles:
The definition of T and Eqs.~\ref{Eq-T-on-scalar-products}, \ref{Eq-T-on-operators} give
\be \langle f|D|i\rangle_{\rm weak}^* = \langle f_T|T D T^{-1}|i_T\rangle_{\rm weak}~.\ee
For spin zero and in the rest frame of the states, we have $\langle i_T|=\langle i|, \langle f_T|=\langle f|$, and CPT symmetry requires $CP=T^{-1}$, leading to
\be \langle f|D|i\rangle_{\rm weak}^* = \pm \langle {\overline f}| D |{\overline i}\rangle_{\rm weak}~,\ee
where the sign is given by the parity change in the decay. Including the FSI phase, this leads to the result in Eq.~\ref{Eq-LeeOehmeYang}.
  
As closing remark, the CPT operator is the product $CPT=CP\times T$. For bosonic systems, $T^2 = {\bf 1}$. Therefore, multiplication with $T$ gives
\be CP = CPT \times T~.\ee
Whenever CP symmetry is violated, T or CPT or both must also be violated.

\section{Description of $K^0\Kqz$ Transitions} \label{Sec-Description} 

\subsection{Quasistable Particles and their Unitarity Relation} \label{Sub-Quasistable}

Elementary quantum mechanics describes the state $\psi$ of a stable particle with rest mass $m$ in its rest
frame by the wave function
\be \psi(t) = \re^{-\ri mt},\ee
with $|\psi|^2=const =1$ and solving the Schr\"odinger equation
\be \ri{\dot\psi}={\cal H}~\psi = m~\psi~,\ee
where the Hamilton operator ${\cal H}$ is hermitean since $m$ is real. This simple linear equation is not sufficient for 
describing strong interactions
with rapid particle production and decay. However, the time scale of weak interactions, where our electronic eyes are
fast enough to measure differences between production and decay time with high resolution, allows the effective concept
of quasistable particles. Their states obey again a linear Schr\"odinger equation
\be \ri{\dot\psi}={\cal H}_{eff}~\psi = \mu \psi = (m-\ri\frac{\Gamma}{2}) \psi~, \label{Eq-4-2}\ee
with the solution
\be\psi(t) =\re^{-\ri mt-\Gamma t/2}~,~~|\psi|^2 = \re^{-\ri\Gamma t}~.\ee
Their norm $|\psi|^2$ shows the exponential decay law of radioactivity with a mean life of $\tau = 1/\Gamma$. The
``effective" Hamiltonian ${\cal H}_{eff}=m-\ri\Gamma/2$ is not hermitean in spite of the hermiticity of the weak-interaction
Hamiltonian ${\cal H}_{weak}$. This Hamiltonian produces one or more final states $f_i ~(i=1,N)$ with decay rates
$\Gamma_i$, respectively. It acts in the full space of states $(\psi,f_1 \ldots f_N)$ where hermiticity guarantees unitarity, 
i.~e.~the sum of probabilities is conserved:
\be |\psi|^2 + \sum_{i=1}^N {|f_i|^2} = 1~.\label{Eq-4-1-5}\ee
The partial decay rates are given by
\be \Gamma_i = |\langle f_i|D|\psi\rangle|^2~,\ee
where $D$ is the transition operator produced by the Hamiltonian ${\cal H}_{weak}$. At $t=0$, unitarity requires
\be -\frac{d|\psi|^2}{dt}=\Gamma = \sum_{i=1}^N {|\langle f_i|D|\psi\rangle|^2}~. \label{Eq-4-6}\ee
The result in Eq.~\ref{Eq-4-6} is called a ``unitarity relation". It relates the observed mean life $\tau = 1/\Gamma$ with
squared decay matrix elements. In contrast to a one-state quasistable particle which has $N+1$ alternatives at any
time, to decay or not to decay, a $K^0$ meson has $N+2$ alternatives, to remain a $K^0$, to transform itself into
a $\Kqz$ or to decay into one of $N$ final states. Its effective Schr\"odinger equation, therefore, requires 
a modification of Eq.~\ref{Eq-4-2}  as described in the next subsection.

\subsection{$K^0\Kqz$ Transitions} \label{Sub-Transitions}

Weak-interaction transitions are again sufficiently well described by a Schr\"odinger equation. The state in 
Eq.~\ref{Eq-2-1}, $\Psi =\psi_1 |K^0\rangle + \psi_2 |\Kqz\rangle =(\psi_1,\psi_2)$ obeys the simplest linear equation 
$$ \ri \dot{\psi_i}=\mu_{ij}\psi_j =(m_{ij}-\ri\Gamma/2)\psi_j~,$$
\be 
\ri~\frac{\partial}{\partial t }\left(\begin{array}{c}\psi_1\\ \psi_2\end{array}\right)=
{\cal H}_{eff} \left(\begin{array}{c}\psi_1\\ \psi_2\end{array}\right) =
\left[\left(\begin{array}{cc}m_{11}&m_{12}\\ m_{12}^*&m_{22}\end{array}\right)
-\frac{\ri}{2}\left(\begin{array}{cc}\Gamma_{11}&\Gamma_{12}\\
\Gamma_{12}^*&\Gamma_{22}\end{array}\right)\right]
\left(\begin{array}{c}\psi_1\\\psi_2\end{array}\right)~,\label{Eq-4-2-1} 
\ee
where $m_{ij}$ and $\Gamma_{ij}$ are both hermitean matrices such as $m$ and $\Gamma$ are both real numbers in Eq.~\ref{Eq-4-2}.
The equation was derived in perturbation theory in 1930 by V.~F.~Weisskopf and E.~P.~Wigner \cite{1930-WeisskopfWigner} and in field 
theory in 1963 by R.~G.~Sachs \cite{1963-Sachs}.  Eq.~\ref{Eq-4-2-1} contains eight real parameters, but only seven of them are observables. 
Owing to arbitrary phase transformations of the states $|K^0\rangle $ and $|\Kqz\rangle$, the matrix elements  $m_{12}$ and $\Gamma_{12}$ 
have unobservable phases. However, the phase transformations change  $m_{12}$ and $\Gamma_{12}$ simultaneously, their relative phase is
phase-convention free. The seven observable real parameters in Eq.~\ref{Eq-4-2-1} are $m_{11}$, $m_{22}$, $\Gamma_{11}$, $\Gamma_{22}$, 
$|m_{12}|$, $|\Gamma_{12}|$ and $\phi(\Gamma_{12}/m_{12})$.
The equation is solved by diagonalisation, its two eigenstates are
\bea K^0_S(t) = [p\cdot\sqrt{1+2\delta}\cdot K^0 +q\cdot\sqrt{1-2\delta}\cdot\Kqz]\cdot \re^{-\ri m_S t-\Gamma_S t/2}~,\nonumber\\
     K^0_L(t) = [p\cdot\sqrt{1-2\delta}\cdot K^0 -q\cdot\sqrt{1+2\delta}\cdot\Kqz]\cdot \re^{-\ri m_L t-\Gamma_L t/2}~, \label{Eq-4-2-1a}  \eea
with arbitrary phases for $p$ and $q$ and normalized to 1 at $t=0$. The seven observable parameters of the solutions, following
unambiguously from the seven parameters in Eq.~\ref{Eq-4-2-1}, are $m_S$, $m_L$, $\Gamma_S$, $\Gamma_L$, $|p/q|$, Re\,$\delta$ and
Im\,$\delta$. Setting
\be \frac{p}{q}=\frac{1+\epsilon}{1-\epsilon}~,\ee
Im\,$\epsilon$ is unobservable and the observed values or limits for Re\,$\epsilon/\sqrt{1+|\epsilon|^2}$ and $|\delta|$ are of order 
$10^{-3}$ or smaller, see Section~\ref{Sec-Determination1}. Therefore, we use in all following discussions the approximations $|\epsilon|
\ll 1$, $|\delta|\ll 1$ and neglect all terms of order $|\epsilon|^2$, $|\delta|^2$ and smaller. Wherever $\reps$ appears in this text and the precision
is asked to be better than $10^{-6}$, then $\reps$ has to be replaced by $\reps/\sqrt{1+|\epsilon|^2}$. With this approximation and this notation, 
the two eigenstates are
\bea K^0_S(t) = [(1+\epsilon+\delta)\cdot K^0 +(1-\epsilon-\delta)\cdot\Kqz]\cdot \re^{-\ri m_S t-\Gamma_S t/2}/\sqrt{2}~,\nonumber\\
     K^0_L(t) = [(1+\epsilon-\delta)\cdot K^0 -(1-\epsilon+\delta)\cdot\Kqz]\cdot \re^{-\ri m_L t-\Gamma_L t/2}/\sqrt{2}~.
\label{Eq-eigenstates}\eea
The two states are normalized at $t=0$, but they are not orthogonal,
\be \langle K^0_S|K^0_S\rangle = \langle K^0_L|K^0_L\rangle =1~, ~~ \langle K^0_S|K^0_L\rangle = 2~{\rm Re}~\epsilon - 2\,\ri~{\rm Im}~\delta
    ~.\label{Eq-4-2-KSKL}\ee
The parameter $|p/q|$ is replaced by Re\,$\epsilon$, 
\be |p/q| = 1 + 2 Re\,\epsilon~,\ee
and by also introducing 
\be m = (m_S +m_L)/2~,~~\Delta m = m_L-m_S~,~~\Gamma=(\Gamma_S+\Gamma_L)/2~,~~\Delta\Gamma = \Gamma_L-\Gamma_S~,\label{Eq-4-2-DeltaM}\ee
the seven parameters of the eigenstates are given by the seven parameters of matrix $\mu_{ij}$: 
$$ m = (m_{11}+m_{22})/2~,~~\Gamma = (\Gamma_{11}+\Gamma_{22})/2~,$$
$$ |\Delta m| = 2|m_{12}|~,~~|\Delta\Gamma| = 2|\Gamma_{12}|~,~~\Delta m\cdot\Delta\Gamma = 4~{\rm Re}(m^*_{12}\Gamma_{12})~,$$
\be {\rm Re}~\epsilon =\frac{{\rm Im}(m^*_{12}\Gamma_{12})}{4|m_{12}|^2+|\Gamma_{12}|^2}~,~~\delta =\frac {(m_{22}-m_{11})
     -\ri(\Gamma_{22}-\Gamma_{11})/2}{2\Delta m-\ri\Delta\Gamma}~,\label{Eq-4-2-muParams}\ee
neglecting terms of order $|\epsilon|^2$ and $|\delta|^2$.  Note that we define the signs of $\Delta m$ and $\Delta\Gamma$ in such a way that
the observed values are positive for $\Delta m$ and negative for $\Delta\Gamma$. 
Only the relative signs of $\Delta m$ and $\Delta\Gamma$ are given by the matrix $\mu_{ij}$.
Expressions for $\Delta m$ and $\Delta\Gamma$ without the approximations $|\epsilon ,\delta|\ll 1$ can be found in Ref.~\cite{BrancoLavouraSilva};
a useful strict relation is
\be \frac{q}{p} =  \sqrt{ \frac{2 m_{12}^*-\ri\,\Gamma_{12}^*}{2 m_{12}-\ri\,\Gamma_{12}}  }~.\label{strictq/p}\ee 
The last result in Eq.~\ref{Eq-4-2-muParams} shows that $\delta$ is observable in modulus and phase, free of $|K^0\rangle$ and $|\Kqz\rangle$ phase conventions.

All solutions of the Schr\"odinger equation and their time dependence are given by linear superpositions of the two eigenstates
which are the only two states with an exponential decay law.
A Kaon in the initial state $\Psi_K(0)=K^0$ is found to have the solution
\be \Psi_K (t) = \frac{1}{2}\{[(1+2\delta)K^0+(1-2\epsilon)\Kqz]\re^{-\ri m_S t-\Gamma_S t/2}
                             +[(1-2\delta)K^0-(1-2\epsilon)\Kqz]\re^{-\ri m_L t-\Gamma_L t/2}\}~,\label{Eq-Kevolution}\ee
and with the initial condition $\Psi_{\overline K}(0)=\Kqz$ it has the solution
\be \Psi_{\overline K} (t) = \frac{1}{2}\{[(1+2\epsilon)K^0+(1-2\delta)\Kqz]\re^{-\ri m_S t-\Gamma_S t/2}
                             -[(1+2\epsilon)K^0-(1+2\delta)\Kqz]\re^{-\ri m_L t-\Gamma_L t/2}\}~.\label{Eq-Kbarevolution}\ee
The two results give the four time-dependent transition probabilities $K^0,\Kqz\to K^0,\Kqz$
\bea P(K^0\to K^0) = |(1+2\delta)\re^{-\ri m_S t-\Gamma_S t/2}+ (1-2\delta)\re^{-\ri m_L t-\Gamma_L t/2}|^2/4~,\nn\\
     P(\Kqz\to\Kqz) = |(1-2\delta)\re^{-\ri m_S t-\Gamma_S t/2}+ (1+2\delta)\re^{-\ri m_L t-\Gamma_L t/2}|^2/4~,\nn\\
     P(K^0\to\Kqz) = |(1-2\epsilon)\re^{-\ri m_S t-\Gamma_S t/2}- (1-2\epsilon)\re^{-\ri m_L t-\Gamma_L t/2}|^2/4~,\nn\\
     P(\Kqz\to K^0) = |(1+2\epsilon)\re^{-\ri m_S t-\Gamma_S t/2}- (1+2\epsilon)\re^{-\ri m_L t-\Gamma_L t/2}|^2/4~.\eea
These results show the symmetry conditions for $\epsilon$ and $\delta$. CPT symmetry in the transitions requires
\be P(\Kqz\to\Kqz) = P(K^0\to K^0) ~\Longleftrightarrow~ \delta = 0~,\label{Eq-4-2-CPT}\ee
and T symmetry requires
\be P(K^0\to\Kqz) = P(\Kqz\to K^0) ~\Longleftrightarrow~ {\rm Re}\,\epsilon = 0~.\ee
The explicit expressions for the probabilities,
$$   P(K^0\to K^0) = (\frac{1}{4} + {\rm Re}~\delta)\re^{-\Gamma_S t}+(\frac{1}{4}- {\rm Re}~\delta)\re^{-\Gamma_L t}
                     + (\frac{1}{2}\cos\Delta m t -2~{\rm Im}~\delta~\sin\Delta m t)\re^{-\Gamma t} ~,$$
$$  P(\Kqz\to\Kqz) = (\frac{1}{4} - {\rm Re}~\delta)\re^{-\Gamma_S t}+(\frac{1}{4}+ {\rm Re}~\delta)\re^{-\Gamma_L t}
                     + (\frac{1}{2}\cos\Delta m t +2~{\rm Im}~\delta~\sin\Delta m t)\re^{-\Gamma t} ~,$$
\bea P(K^0\to\Kqz) =  (\frac{1}{4} - {\rm Re}~\epsilon)~
                       (\re^{-\Gamma_S t}+\re^{-\Gamma_L t}-2\cos\Delta mt\cdot\re^{-\Gamma t}) ~,\nn\\
     P(\Kqz\to K^0) =  (\frac{1}{4} + {\rm Re}~\epsilon)~
                       (\re^{-\Gamma_S t}+\re^{-\Gamma_L t}-2\cos\Delta mt\cdot\re^{-\Gamma t})~,\label{Eq-4-2-1b}\eea
show that the CPT asymmetry 
\be A_{CPT} = \frac {P(\Kqz\to\Kqz)-P(K^0\to K^0)}{P(\Kqz\to\Kqz)+P(K^0\to K^0)}\label{Eq-KasymCPT}\ee
depends on time, whereas the T asymmetry
\be A_{T} = \frac {P(\Kqz\to K^0)-P(K^0\to\Kqz)}{P(\Kqz\to K^0)+P(K^0\to\Kqz)} = 4~{\rm Re}~\epsilon~,\label{Eq-KasymT}\ee
is time-independent. A measurement of the CPT asymmetry determines ${\rm Re}~\delta$ and ${\rm Im}~\delta$ from a fit
to the time dependence, and the time-independent T asymmetry determines ${\rm Re}~\epsilon$, the only T-symmetry violating
parameter in $K^0\Kqz$ transitions. For the determination of the sign of Im\,$\delta$, remember that $\Delta m$ was defined as 
$m_L - m_S$ in Eq.~\ref{Eq-4-2-DeltaM}. 

\subsection{The CPT-symmetric and CP-symmetric Limits} \label{Sub-CPTsymm}

CPT symmetry in $K^0\Kqz$ transitions requires $\delta =0$, see Eq.~\ref{Eq-4-2-CPT}, i.~e.~$m_{22}=m_{11}$ and 
$\Gamma_{22}=\Gamma_{11}$ in the transition matrix $\mu_{ij}$, see Eq.~\ref{Eq-4-2-muParams}. The two eigenstates and
the four transition probabilities are given in the previous subsection, setting $\delta =0$.

CP symmetry requires CPT and T symmetries, i.~e.~\,$\delta=0$ and Re\,$\epsilon =0$. The latter condition requires a phase of 0 or $\pi$
between $m_{12}$ and $\Gamma_{12}$ since Re\,$\epsilon\propto {\rm Im}(m^*_{12}\Gamma_{12})$, see Eq.~\ref{Eq-4-2-muParams}. The phases of $m_{12}$ and $\Gamma_{12}$
are unobservable, but all properties of $K^0\Kqz$ transitions in the CP-conserving limit can be discussed in the convention that
all four parameters in
\be 
\ri~\frac{\partial}{\partial t }\left(\begin{array}{c}\psi_1\\ \psi_2\end{array}\right)=
\left[\left(\begin{array}{cc}m & m_{12}\\ m_{12}&m\end{array}\right)
-\frac{\ri}{2}\left(\begin{array}{cc}\Gamma&\Gamma_{12}\\
\Gamma_{12}&\Gamma\end{array}\right)\right]
\left(\begin{array}{c}\psi_1\\\psi_2\end{array}\right) 
\ee
are real. Its two eigenstates are the ``historical" states $K^0_1$ and $K^0_2$ in Eqs.~\ref{Eq-1} and \ref{Eq-2},
multiplied by $\re^{-\ri m_1 t -\Gamma_1 t/2}$ and $\re^{-\ri m_2 t-\Gamma_2 t/2}$, respectively. 
There are two different sign conventions for the short- and long-living CP eigenstates $K^0_1$ and $K^0_2$. Keeping the convention of
Eqs.~\ref{Eq-1} and \ref{Eq-2}, $(K^0+\Kqz)/\sqrt{2}$~=~short-living, we have 
\be \Gamma_{12}>0~~{\rm and}~~m_{12}<0~.\ee 

\subsection{Two more Sets of Unitarity Relations} \label{Sub-Unitarity}

Including all decay states $f_1 \ldots f_N$ of $K^0$ and $\Kqz$, unitarity in the space $(\psi_1,\psi_2,f_1\ldots f_N)$
requires
\be |\psi_1|^2 + |\psi_2|^2 + \sum_{i=1}^N {|f_i|^2} = const. = 1~,\ee
as in Eq.~\ref{Eq-4-1-5} for a single-state quasistable particle. Applying
\be -\frac{d|\Psi|^2}{dt}(t=0) = \sum_{i=1}^N |\langle f_i|D|\Psi\rangle|^2 \ee
to the two states $\Psi = K^0_1$ and $K^0_2$ in the CP-symmetric limit, we obtain
\bea \Gamma_1 = \Gamma + \Gamma_{12}=\sum_{i=1}^N |\langle f_i|D|K^0_1\rangle|^2 
      = \frac{1}{2}\sum_{i=1}^N |\langle f_i|D|K^0\rangle +\langle f_i|D|\Kqz\rangle |^2~,\nn\\
     \Gamma_2 = \Gamma - \Gamma_{12}=\sum_{i=1}^N |\langle f_i|D|K^0_2\rangle|^2 
      = \frac{1}{2}\sum_{i=1}^N |\langle f_i|D|K^0\rangle  -\langle f_i|D|\Kqz\rangle |^2~.\label{Eq-here2}\eea
Adding both equations gives
\be 2 \Gamma = \sum_{i=1}^N |\langle f_i|D|K^0\rangle|^2 +\sum_{i=1}^N |\langle f_i|D|\Kqz\rangle |^2~.\label{Eq-here}\ee
CP symmetry requires $|\langle f_i|D|K^0\rangle|^2=|\langle {\overline f}{}_i|D|\Kqz\rangle|^2$, and all states ${\overline f}{}_i$ are
contained in the set of states $f_i$. Therefore, both sums on the right-hand side of Eq.~\ref{Eq-here} are equal, leading to the
unitarity relation
\be \Gamma = \sum_{i=1}^N |\langle f_i|D|K^0\rangle|^2 = \sum_{i=1}^N |\langle f_i|D|\Kqz\rangle|^2~.\ee
Subtracting the two equations \ref{Eq-here2} leads to the second unitarity relation for the CP-symmetric limit, 
\be \Gamma_{12} = \sum_{i=1}^N \langle f_i|D|K^0\rangle^* \langle f_i|D|\Kqz\rangle~.\label{Eq-here3}\ee
Because of the large observed difference between $\Gamma_1=\Gamma+\Gamma_{12}$ and $\Gamma_2=\Gamma-\Gamma_{12}$,
the relation in Eq.~\ref{Eq-here3} tells us that the total decay rates of $K^0$ and $\Kqz$ 
are dominated by decay channels which are common to $K^0$ and $\Kqz$.

We now derive, following Ref.~\cite{BrancoLavouraSilva}, the unitarity relations given by the full Hamiltonian in
Eq.~\ref{Eq-4-2-1} without imposing CP, CPT or T symmetry. The state $\Psi(0)=\alpha K^0_S+\beta K^0_L$ has the evolution
\be \Psi(t)=\alpha~\re^{-\ri m_S t-\Gamma_S t/2} K^0_S + \beta~\re^{-\ri m_L t-\Gamma_L t/2} K^0_L \ee
with the norm
\be |\Psi(t)|^2 = |\alpha|^2~\re^{-\Gamma_S t} + |\beta|^2~\re^{-\Gamma_L t}+2~{\rm Re} [\alpha^*\beta~\re^{(-\Gamma_S -\Gamma_L
    -2\ri\Delta m)t/2}\langle K^0_S|K^0_L\rangle] \ee
and its derivative at $t=0$
\be -\frac{d|\Psi(t)|^2}{dt} = |\alpha|^2~\Gamma_S + |\beta|^2~\Gamma_L +{\rm Re} [\alpha^*\beta~(\Gamma_S +\Gamma_L
    +2\ri\Delta m)~\langle K^0_S|K^0_L\rangle]~. \label{Eq-here-x}\ee
Unitarity requires
\be -\frac{d|\Psi(t)|^2}{dt}(0) = \sum_{i=1}^N|\langle f_i|D|\Psi (0)\rangle|^2~=~\sum_{i=1}^N|\alpha \langle f_i|D|K^0_S\rangle
    +\beta \langle f_i|D|K^0_L\rangle|^2~. \label{Eq-here-y}\ee
The right-hand sides of Eqs.~\ref{Eq-here-x} and \ref{Eq-here-y} have to be equal for all values of $\alpha$ and $\beta$,
resulting in
$$ \Gamma_S =  \sum_{i=1}^N|\langle f_i|D|K^0_S\rangle|^2~,~~\Gamma_L =\sum_{i=1}^N|\langle f_i|D|K^0_L\rangle|^2~,$$
\be\left(\frac{\Gamma_S+\Gamma_L}{2}+\ri\Delta m\right)\langle K^0_S|K^0_L\rangle~=~\sum_{i=1}^N
     \langle f_i|D|K^0_S\rangle^* \langle f_i|D|K^0_L\rangle ~.\ee
Using the result for $\langle K^0_S|K^0_L\rangle$ in Eq.~\ref{Eq-4-2-KSKL}, the third relation can be written as
\be {\rm Re}\,\epsilon-\ri~{\rm Im}\,\delta =\frac {\sum_{i=1}^N\langle f_i|D|K^0_S\rangle^* \langle f_i|D|K^0_L\rangle }
    {\Gamma_S+\Gamma_L+2\ri\Delta m}~.\label{Eq-BellSteinberger}\ee
It was derived in 1966 by J.~S.~Bell and J.~Steinberger \cite{1966-BellSteinberger}
and allows to determine the two $K^0\Kqz$ transition parameters ${\rm Re}\,\epsilon$ and ${\rm Im}\,\delta$ using measurable
decay properties, as described in the next Section.  

\section{Determination of $K^0\Kqz$ Transition Parameters} \label{Sec-Determination1} 

The seven transition parameters of the eigenstates in Eq.~\ref{Eq-eigenstates} are directly related to the seven parameters of the
effective Hamiltonian in Eq.~\ref{Eq-4-2-1}, as given in Eqs.~\ref{Eq-4-2-muParams}
neglecting terms of order $|\epsilon|^2$ and $|\delta|^2$. In this Section we concentrate on the determination of the three CP-violating
parameters Re\,$\delta$, Im\,$\delta$ and Re\,$\epsilon$, directly related to $m_{22}-m_{11}$, $\Gamma_{22}-\Gamma_{11}$ and
Im\,$(m_{12}/\Gamma_{12})$. For the four CP-conserving parameters, we will only present their present 
best values  in the first subsection and cite the most recent experiments for their determination.

\subsection{The Parameters $\Gamma$, $\Delta\Gamma$, $m$ and $\Delta m$}

The Review of Particle Properties \cite{2012-PDG} lists
\bea \tau_S = 1/\Gamma_S & = & (0.8954\pm 0.0004)\times 10^{-10}~{\rm s}~,\nn\\
     \tau_L = 1/\Gamma_L & = & (5.116\pm 0.021)\times 10^{-8}~{\rm s}~,\nn\\
                       m & = & (497.614\pm 0.024)~{\rm MeV}~,\nn\\
                \Delta m & = & +(0.5289\pm 0.0010)\times 10^{10}/{\rm s}~.\eea
For both $\tau_S$ and $\Delta m$ it lists two values, with and without CPT symmetry. The differences are insignificant in both cases.
Therefore, we prefer to give here only the value with the (slightly) larger uncertainty.
The most recent and most precise results for the mean life $\tau_S$ have been obtained in 2011 by the experiments KLOE \cite{2011-KLOEtauS}
and KTeV \cite{2011-KTeV}. For $\tau_L$ see the 2006 KLOE result \cite{2006-KLOEtauL}. For the mass $m$ see the 2007 KLOE result
\cite{2007-KLOEmK0}, for $\Delta m$ see KTeV 2011 \cite{2011-KTeV}. 

Early evidence for the positive sign of $\Delta m$, i.\,e.\ $m_L>m_S$,
has been found in 1966 by Meisner et al.~\cite{1966-Meisner} and in 1968 by Mehlhop et al.~\cite{1968-signDmK} by observing the interference 
between $K^0$ states and regenerated $K^0_S$ states behind a regenerator close to the production target. 

\subsection{Regeneration}  \label{Sub-Regeneration}

Regeneration has been an efficient tool for observing interferences between the two decay amplitudes $\langle \pi\pi|D|K^0_S,K^0_L\rangle$.  
It uses the different strong-interaction properties of $K^0$ and $\Kqz$ mesons in matter, as first proposed in 1955 by A.~Pais and O.~Piccioni
\cite{1955-PaisPiccioni}. At the ``low" accelerator energies between 1955 and 1975, the absorption cross sections of $\Kqz$  are much higher 
than those of $K^0$, owing to strong-interaction reactions such as $\Kqz p\to \Lambda\pi^+$ which are not possible for $K^0$ mesons because of strangeness 
conservation. The optical theorem, relating the forward scattering amplitude $f(0)$ for the laboratory-frame momentum $p=\beta\gamma m$ to the total 
cross section $\sigma_{tot}$ by Im\,$f(0)=p\cdot\sigma_{tot}/4\pi$, then requires
\be {\rm Im}\,{\overline f} > {\rm Im}\,f~,~~ {\overline f} \ne f~,\ee
where $f$ and $\overline f$ are the scattering amplitudes of $K^0$ and $\Kqz$ at the forward angle $\theta = 0$.  
After a thin layer of matter with
thickness $L\ll \beta\gamma\tau_S$, Kaons in the state $K^0_L$ leave the layer in the state
\be K^0_L + \rho K^0_S~,~~\rho = \ri~\pi~ \frac{f-{\overline f}}{p}~n~L~,\label{Eq-rho}\ee
in a narrow cone around the original direction, where $n$ is the number of target nuclei per volume. The opening angle of the cone \cite{1976-Kleinknecht} 
is of order $10^{-7}$\,rad. In addition to this coherent regeneration, where the $K^0_S$ intensity increases proportional to $L^2$, there is diffraction
regeneration from the incoherent contribution of all un-broken nuclei outside the narrow forward cone, proportional to $L$, and incoherent regeneration
from reactions where the nuclei break up and the $K^0_S$ mesons appear at angles outside the diffraction cone. In the following, we discuss only
coherent regeneration, though corrections from the two other contributions must be taken into account carefully in experiments. In thick regenerators 
\cite{1957-Good,1976-Kleinknecht} of length $L$, the coherent-regeneration amplitude $\rho$ in Eq.~\ref{Eq-rho} is given by
\be \rho = \ri~\pi~ \frac{f-{\overline f}}{p}~n~\beta\gamma\tau_S~\frac{\re^{(\ri\Delta m-\Gamma_S/2)T}-1}{\ri\Delta m\tau_S-1/2} ~,\label{Eq-Regenerator}\ee
where $T=L/\beta\gamma$ is the flight time through the regenerator. The regeneration phase is usually expressed by the sum
$\phi_\rho=\phi_f+\phi_L$, where $\phi_f$ is the single-nucleus contribution $\phi[i(f-{\overline f})]$ and $\phi_L$ is the phase of
$(\re^{\ri\Delta m-\Gamma_S/2)L/\beta\gamma}-1)/(\ri\Delta m\tau_S-1/2)$ which takes into account phase differences from interactions along the 
Kaon path in the regenerator between $z=0$ and $z=L$.

Modulus and phase of $\rho$ can be determined from time-dependent $K_{\ell 3}$ decays after regeneration. Since the ``$\Delta Q\ne\Delta S$" decays 
$K^0\to\pi^+\ell^-{\overline\nu}$ and $\Kqz\to\pi^-\ell^+ \nu$ are essentially absent, see Section \ref{Sec-Determination2}, and CPT symmetry
requires the same rates $|A_\ell|^2$ for the two ``$\Delta Q = \Delta S$" decays $K^0\to\pi^-\ell^+\nu$ and $\Kqz\to\pi^+\ell^-{\overline\nu}$,
the two $K_{\ell 3}$ decays have the rates
\bea N(\pi^-\ell^+\nu) &=& |A_{\ell}|^2~|\re^{-\ri m_L t-\Gamma_L t/2}+\rho~\re^{-\ri m_S t-\Gamma_S t/2}|^2/2\nn\\
               &=& |A_{\ell}|^2~[\re^{-\Gamma_L t}+|\rho|^2~\re^{-\Gamma_S t}+2|\rho|\cos(\Delta m t+\phi_\rho)~\re^{-\Gamma t}]/2\nn\\
     N_(\pi^+\ell^-{\overline\nu}) &=& |A_{\ell}|^2~|\re^{-\ri m_L t-\Gamma_L t/2}-\rho~\re^{-\ri m_S t-\Gamma_S t/2}|^2/2\nn\\
               &=& |A_{\ell}|^2~[\re^{-\Gamma_L t}+|\rho|^2~\re^{-\Gamma_S t}-2|\rho|\cos(\Delta m t+\phi_\rho)~\re^{-\Gamma t}]/2~.\eea
Fits of the data to these rates are sensitive to $\Gamma_S$, $\Gamma_L$, $\Delta m$ and to modulus and phase of $\rho$. If $\phi_\rho$
is close to zero, the rates are not sensitive to the sign of $\Delta m$. However, when a regenerator is placed closely behind the production target, a
produced $K^0$ with the state $\Psi(0) = K^0_L + K^0_S$ is in the state $\Psi(t) = K^0_L(t) + K^0_S(t) +\rho K^0_S(t-t_2)$ for
$t>t_2$, where $t_2$ is the time at the exit of the regenerator. This leads to an interference term that contains $\cos\Delta m t$ and
$\sin\Delta m t$ even for $\phi_\rho = 0$.

Decays into a CP eigenstate $f$ from the state $K^0_L +\rho K^0_S$ with amplitudes
\be A = \langle f|D|K^0_S\rangle~,~~A_{L} =  \langle f|D|K^0_L\rangle\ee
need a CP-violating description. Introducing the CP-violating parameter
\be \eta =\frac {\langle f|D|K^0_L\rangle}{\langle f|D|K^0_S\rangle}~,\ee
they have the rate
\bea N_f & = & |A_L~ \re^{-\ri m_L t-\Gamma_L t/2}+\rho A ~\re^{-\ri m_S t-\Gamma_S t/2}|^2\nn\\
         & = & |A|^2~|\eta ~\re^{-\ri m_L t-\Gamma_L t/2}+\rho ~\re^{-\ri m_S t-\Gamma_S t/2}|^2\nn\\
         & = & |A|^2~[|\eta|^2\re^{-\Gamma_L t}+|\rho|^2\re^{-\Gamma_S t}
               + 2 |\eta||\rho| \cos(\Delta m t+\phi_\rho-\phi_\eta)~\re^{-\Gamma t}]~.\label{Eq-Regen}\eea
The time dependence of the rate allows to determine the CP-violating amplitude ratio $\eta$ in
modulus and phase. The phase determination requires knowledge of $\phi_\rho$, either from semileptonic decay
rates or from analyses of K-nucleon scattering. 

\newpage

\subsection{Two-pion Decays, the Dominant Contribution to Unitarity} \label{Sub-TwoPi}

The CP-violation discovery experiment \cite{1964-FitchCronin} found ${\cal B}(K^0_L\to\pi^+\pi^-)=1.6\times 10^{-3}\pm 20\%$
assuming a value of 0.8 for the ratio of ${\rm charged ~decays / all~ decays}$. Here and in all following discussions, $\cal B$ denotes the decay fraction.
The result gives the modulus of the parameter 
\be \eta_{+-} =\eta(\pi^+\pi^-) = \frac {\langle \pi^+\pi^-|D|K^0_L\rangle}{\langle \pi^+\pi^-|D|K^0_S\rangle}~,\label{Eq-5-eta+-}\ee
\be |\eta_{+-}|= \sqrt{\frac{\Gamma_L\times{\cal B}(K^0_L\to\pi^+\pi^-)}{\Gamma_S\times{\cal B}(K^0_S\to\pi^+\pi^-)}}=(2.0\pm 0.2)\cdot 10^{-3}~.\ee

The phase of $\eta_{+-}$ is found from the time-dependent rate of $\pi^+\pi^-$ decays behind a regenerator, as given in Eq.~\ref{Eq-Regen}.
Early experiments at Princeton \cite{1965-Fitch}, Brookhaven \cite{1966-Firestone} and CERN \cite{1966-AlffSteinberger, 1967-BottBodenhausen}
and a few more led to a 1968 mean value \cite{1968-Rosenfeld} of
\be \phi_{+-} = \phi(\eta_{+-}) = 65^\circ \pm 11^\circ~.\ee

\begin{figure}[h]
\begin{minipage}{0.32\textwidth}
\includegraphics[width=\textwidth]{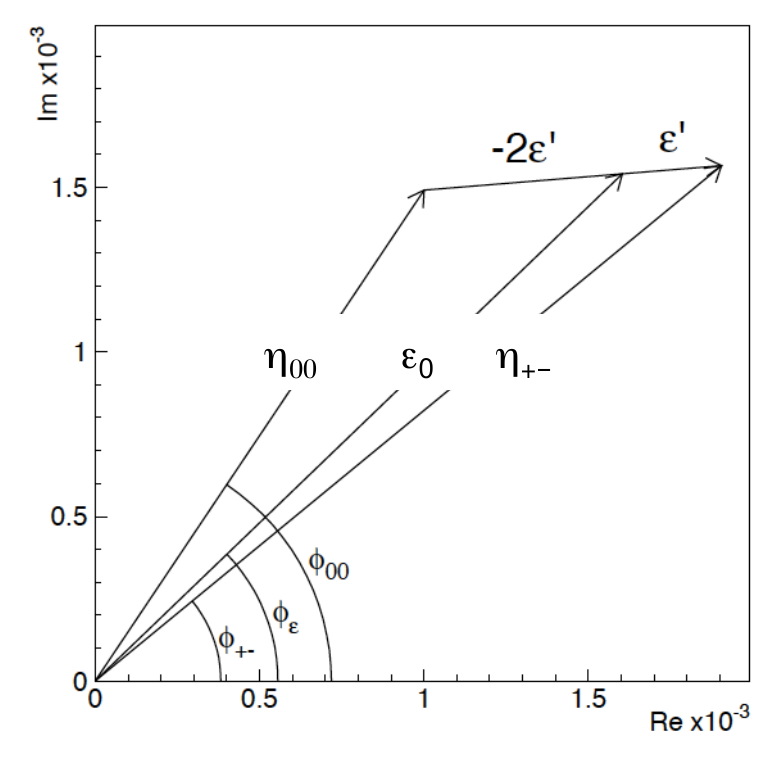}
\caption{The Wu-Yang triangle \cite{1964-WuYang} relating the parameters $\eta_{+-}$, $\eta_{00}$, $\epsilon_0$ and $\epsilon^\prime$.} 
\label{Fig-WuYang} 
\end{minipage}\hfill
\begin{minipage}{0.64\textwidth}
CP violation in the decay $K^0_L\to\pi^0\pi^0$ was observed in 1967 by J.~M.~Gaillard et al.~at CERN \cite{1967-Gaillard} and by J.~W.~Cronin 
et al.~at Princeton \cite{1967-Cronin}. These and other early experiments suffered from high background; the
1968 best value  for the modulus of
\be \eta_{00} =\eta(\pi^0\pi^0) = \frac {\langle \pi^0\pi^0|D|K^0_L\rangle}{\langle \pi^0\pi^0|D|K^0_S\rangle}~,\label{Eq-5-eta00}\ee
was approximately $(4\pm 1)\times 10^{-3}$. No information on 
\be\phi_{00} = \phi(\eta_{00})\ee
was available until 1970.
The standard phenomenology, following T.~T.~Wu and C.~N.~Yang \cite{1964-WuYang}, relates $\eta_{+-}$
and $\eta_{00}$ to the parameters $\epsilon_0$ and $\epsilon^\prime$ defined by the isospin-0 and isospin-2 states of the $\pi\pi$ system:
\end{minipage}\end{figure}

\be \epsilon_0 =\frac {\langle \pi\pi ,I=0|D|K^0_L\rangle}{\langle \pi\pi ,I=0|D|K^0_S\rangle}~,
    ~~\epsilon^\prime = \frac{1}{\sqrt{2}}\,\frac {\langle \pi\pi ,I=2|D|K^0_L\rangle}{\langle \pi\pi ,I=0|D|K^0_S\rangle}~.\label{Eq-here8}\ee
The ``experimental" phenomenology \cite{BrancoLavouraSilva} uses
\be \epsilon_0 =\frac {2\eta_{+-}+\eta_{00}}{3}~,~~\epsilon^\prime = \frac {\eta_{+-}-\eta_{00}}{3}~,~~\eta_{+-}=\epsilon_0+\epsilon^\prime
     ~,~~\eta_{00}=\epsilon_0-2\,\epsilon^\prime~,\label{Eq-EpsilonExp}\ee
and both definitions are identical if $\omega = {\langle \pi\pi ,I=2|D|K^0_S\rangle}/{\langle \pi\pi ,I=0|D|K^0_S\rangle}=0$. Since
$|\omega|\approx 0.045$, see Section \ref{pipiParams}, the definitions differ only slightly, and we will use Eq.~\ref{Eq-EpsilonExp} 
for all discussions in this Review. The relation between $\epsilon_0$, $\epsilon^\prime$, $\eta_{+-}$ and $\eta_{00}$ is shown by
the ``Wu-Yang triangle" in Fig.~\ref{Fig-WuYang}.   

Measurements of the four quantities $|\eta_{+-}|$, $\phi_{+-}$, $|\eta_{00}|$ and $\phi_{00}$ determine modulus and phase of $\epsilon_0$.
If the two $\pi\pi$ modes dominate the sum $\sum_{i=1}^N\langle f_i|D|K^0_S\rangle^* \langle f_i|D|K^0_L\rangle$, the phase $\phi(\epsilon_0)$
must be $\sim 45^\circ$ or $225^\circ$ in the case of CPT symmetry and $\sim 135^\circ$ or $315^\circ$ in the case of T symmetry. This consequence 
of the Bell-Steinberger relation is easily demonstrated: Introducing $\alpha_i$ by
\be \langle f_i|D|K^0_S\rangle^* \langle f_i|D|K^0_L\rangle = \alpha_i\times \Gamma_S~,\label{Eq-alpha}\ee
i.~e.~$\alpha_i=\eta_i\times {\cal B}(K^0_S\to f_i)$ for the states $\pi^+\pi^-$ and $\pi^0\pi^0$, 
and approximating ${\cal B}(K^0_S\to\pi^+\pi^-)=2/3$, ${\cal B}(K^0_S\to\pi^0\pi^0)=1/3$ and ${\cal B}(\rm all~other~modes) = 0$,
we obtain for Eq.~\ref{Eq-BellSteinberger}:
\be \sum \alpha_i = (2\eta_{+-}+\eta_{00})/3 =\epsilon_0~,\ee 
\be {\rm Re}\,\epsilon - \ri\,{\rm Im}\,\delta = \frac{\epsilon_0}{1+\Gamma_L/\Gamma_S+2\ri\Delta m/\Gamma_S}~.\ee
Since $\Gamma_L\ll\Gamma_S$, the requirement of Re\,$\epsilon=0$ for T symmetry leads to
\be \epsilon_0 =-\ri\,{\rm Im}\,\delta(1+2\ri\Delta m/\Gamma_S)~,~~ \phi(\epsilon_0)={\rm atan}(-\Gamma_S/2\Delta m)~=133^\circ~{\rm or}~313^\circ~,\ee
and the requirement of $\delta =0$ for CPT symmetry to
\be \epsilon_0 ={\rm Re}\,\epsilon(1+2\ri\Delta m/\Gamma_S)~,~~ \phi(\epsilon_0)={\rm atan}(+2\Delta m/\Gamma_S)~=43^\circ~{\rm or}~223^\circ~.\ee

Determination of $\phi(\epsilon_0)$ requires both $\eta_{+-}$ and $\eta_{00}$ with both modulus and phase.  As already stated above, no conclusion on the validity of 
CPT or T symmetry could be drawn from the data around 1968 without a measurement of $\phi_{00}$, as quantitatively shown by R.~C.~Casella \cite{1968-69-Casella}.

\subsection{The First Measurement of $\phi_{00}$}

\begin{figure}[h]
\begin{minipage}{0.55\textwidth}
The phase $\phi_{00}$ of $\eta_{00}$ has been determined by J.~C.~Chollet et al.~at CERN \cite{1970-Chollet} in an experiment with optical spark chambers  using a neutral beam with 
$K^0_L$ momenta between 1.5 and 3 GeV/c and a Cu regenerator of 12 cm thickness. About 54 000 events with four converted photons were found by scanning the $\sim 240~000$  
recorded photographs. All these events were measured on hand-operated image-plane digitizers determining conversion point, direction and energy of each shower. 
After kinematical fits of the $K\to\pi^0\pi^0$ hypothesis to the measured events, all events within wide selection criteria were re-scanned by physicists. After the final selection
criteria, about 600 events remained with measured lifetimes between $-2$ and $16~\tau_S$. The lifetime resolution was $0.8\,\tau_S$~\cite{1969-Chollet}, 
and the measured-lifetime distribution is shown in Fig.~\ref{Fig-Chollet}. A fit of the expected distribution in Eq.~\ref{Eq-Regen}, modified for efficiency and resolution, 
gives $|\eta_{00}/\rho|$ and $\phi_{00}-\phi_\rho$. Because of uncertainties in the regeneration amplitude, ref.\,\cite{1970-Chollet} quotes only $|\eta_{00}|=
(1.23\pm 0.24)\times 10^{-3}\times (f-{\overline f})/p_0$ at $p_0= 2$\,GeV/c. \end{minipage}\hfill
\begin{minipage}{0.40\textwidth}
\includegraphics[width=\textwidth]{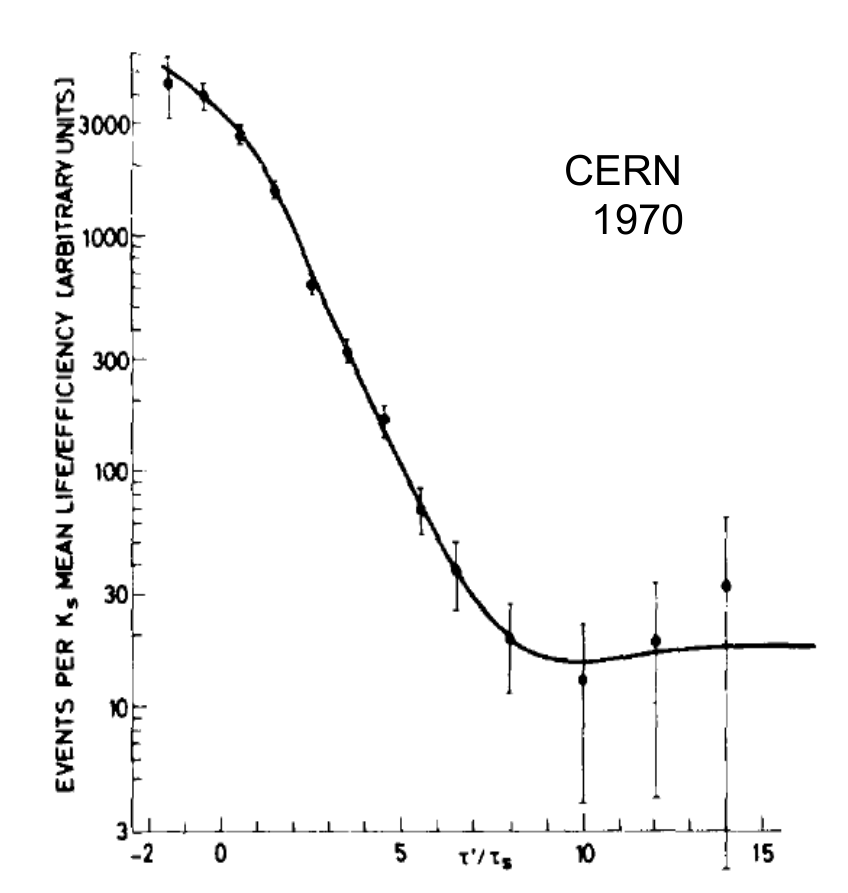}
\caption{Lifetime distributions of $\pi^0\pi^0$ events from neutral Kaons behind a Cu regenerator \cite{1970-Chollet}.} 
\label{Fig-Chollet}
\end{minipage}\end{figure}

The part $\phi_L$ of the regeneration phase is precisely known for $L=12$\, cm 
and the nuclear part had been measured to be $\phi_f=(-46.5\pm 4.4)^\circ$.
The fit result for the phase of $\eta_{00}$ is
\be \phi_{00}= 51^\circ \pm 30^\circ~.\ee
A preliminary result of the $\phi_{00}$ experiment was presented in January 1969 \cite{1969-Chollet}, together with the evidence for $\rm Re\,\epsilon\ne 0$, i.e. for
T violation in $K^0\Kqz$ transitions. The question remained if CPT symmetry is also broken and how big its contribution to CP violation could be. The answer was given
by using the Bell-Steinberger unitarity relation, and the group performed this analysis in the time between the preliminary and the final result.

\subsection{The First Bell-Steinberger Analysis}

Starting from the unitarity relation in Eq.~\ref{Eq-BellSteinberger} and using the parameters $\alpha_i$ as defined in Eq.~\ref{Eq-alpha}, the analysis 
for determining the T- and CPT-violating contributions to CP violation in $K^0\Kqz$ transitions has to use
\be {\rm Re\,}\epsilon -\ri\,{\rm Im\,}\delta = \frac {\sum \alpha_i}{1+\Gamma_L/\Gamma_S+2\ri\Delta m/\Gamma_S}~,\label{Eq-BellSt}\ee
where the sum runs over all final states which are reached from both $K^0_S$ and $K^0_L$. Taking into account all essential final states, $\pi^+\pi^-$, $\pi^0\pi^0$,
$3 \pi$ and $\pi^\mp\ell^\pm\nu$ with $\ell = \rm e~or~\mu$, results for ${\rm Re\,}\epsilon$ and  ${\rm Im\,}\delta$ have been obtained in 1970 
by the CERN group \cite{1970-Schubert} measuring $\phi_{00}$. 
In addition to their measurement $\phi_{00}=51^\circ\pm 30^\circ$, they used $|\eta_{00}|=(3.3\pm 0.6)\,10^{-3}$,
$|\eta_{+-}|=(1.92\pm 0.05)\,10^{-3}$ and $\phi_{+-}=44^\circ\pm 5^\circ$. As already discussed in Section \ref{Sub-TwoPi}, these four experimental results
dominate the inputs for the Bell-Steinberger relation; the two uncorrelated inputs for $\alpha_{\pi\pi}$ were
\bea  {\rm Re\,}\alpha_{\pi\pi}+{\rm Im\,}\alpha_{\pi\pi} &=& (3.36\pm 0.32)\,10^{-3}~,\\
      {\rm Re\,}\alpha_{\pi\pi}-{\rm Im\,}\alpha_{\pi\pi} &=& (-0.13\pm 0.73)\,10^{-3}~,\\
                       {\rm with}~~\alpha_{\pi\pi} &=& \alpha(\pi^+\pi^-) + \alpha(\pi^0\pi^0)~.\eea
All other $\alpha_i$ values were compatible with zero, but some of their uncertainties are large.
For $K_{3\pi}$ decays with $f= \pi^+\pi^-\pi^0$ and $3\,\pi^0$, an experiment \cite{1969-Webber3pi} with 71 decays $K^0,\Kqz\to\pi^+\pi^-\pi^0$ had determined
\be \langle \pi^+\pi^-\pi^0|D|K^0_S\rangle/\langle \pi^+\pi^-\pi^0|D|K^0_L\rangle = 0.05\pm 0.30 + \ri\,(-0.15\pm 0.45)~.\ee
Assuming that all $3\pi$ decays ($\pi^+\pi^-\pi^0$ and $3\pi^0$) are in the state with isospin $I=1$, i.~e.~no decays with $\Delta I=5/2$ into $3\pi$ with
$I=3$, and using $\Gamma(K^0_L\to{\rm all}\,3\pi)/\Gamma_S=5.5\times 10^{-4}$, leads to
\be \alpha_{3\pi}=(0.03\pm 0.17) 10^{-3}+\ri\,(0.08\pm 0.25) 10^{-3}~.\ee
For $K_{\ell 3}$ decays, the group re-analyzed two experiments \cite{1969-Bennet, 1969-WebberL3} without imposing CPT invariance by using four parameters related
to the  real and imaginary parts of $x_+$ and $x_-$, introduced later, see Section \ref{Sec-Determination2}, resulting in
\be \alpha_{\pi\ell\nu}=(-0.24\pm 0.21) 10^{-3}+\ri\,(0.09\pm 0.34) 10^{-3}~.\ee
Adding the $3\pi$ and $\pi\ell\nu$ contributions and using $2\Delta m/\Gamma_S=0.94$ gives
\bea      \sum\alpha &=& \alpha_{\pi\pi} + (-0.21\pm 0.27)\,10^{-3} +\ri\,(0.17\pm 0.42)\,10^{-3}\ ,\nn\\
  {\rm Re}\,\epsilon &=&  0.53~{\rm Re}\sum\alpha + 0.50~{\rm Im}\sum\alpha = (1.68 \pm 0.30)\,10^{-3}~,\nn\\
    {\rm Im}\,\delta &=&  0.50~{\rm Re}\sum\alpha - 0.53~{\rm Im}\sum\alpha = (-0.30 \pm 0.45)\,10^{-3}~. \label{Eq-KRSresult}\eea
The result for ${\rm Re}\,\epsilon$ was a first proof of T violation in particle physics, with a significance of about 5~$\sigma$. The result for ${\rm Im}\,\delta$ 
showed that there is no CPT violation in $K^0\Kqz$ transitions, with an uncertainty of about one third of the observed violation of CP and T. 

Since $\rm Im\,\epsilon$ is not an observable, it should be noted that the (historical) quantity in the analysis was defined by 
``$\rm Im\,\epsilon$"~$=\rm Im\,(\epsilon_0+\delta)$. The observable parameter $\rm Re\,\delta$ could not be determined since the overlap $\langle K^0_S|K^0_L\rangle$
in the unitarity relation is only sensitive to $\rm Im\,\delta$. Instead of $\rm Re\,\delta$, the analysis determined the real part of 
\be {\tilde\delta} = \delta + \frac{|\Aq_0|-|A_0|}{|\Aq_0|+|A_0|}~~{\rm with}~~A_0=\langle \pi\pi,\,I=0|D|K^0\rangle,~\Aq_0=\langle \pi\pi,\,I=0|D|\Kqz\rangle~.\ee
The result was $\rm Re\,{\tilde\delta}=( 0.07\pm 0.43)\,10^{-3}$,
showing that this sum of two phase-convention independent and CPT-symmetry testing quantities was compatible with zero.
The first determination of $\rm Re\,\delta$ independent of $|\Aq_0/A_0|$ was presented by CPLEAR in 1999 as discussed later. 

\subsection{Bell-Steinberger Updates and Present Precision Results}

With more and more precise measurements appearing, the Bell-Steinberger unitarity analysis has been repeated many times. The following overview is not complete but
presents some milestones for the increasing precision of $\rm Re\,\epsilon$ and $\rm Im\, \delta$.

\subsubsection{Cronin 1980} \label{Subsub-Cronin}

In his Nobel lecture \cite{1980-Cronin}, J.~Cronin updated the determination of  $\rm Re\,\epsilon$ and $\rm Im\, \delta$ with four considerably improved
input values for $\alpha_{\pi\pi}$. Three of them, $|\eta_{+-}| =(2.27\pm 0.02)\,10^{-3}$, $|\eta_{00}| =(2.32\pm 0.09)\,10^{-3}$ and $\phi_{+-}
 = 44.7^\circ\pm1.2^\circ$, were the result of a large number of experiments between 1970 and 1980, whereas the fourth, $\phi_{00}= 56^\circ\pm 6^\circ$,
was essentially the result of a single experiment \cite{1979-Christenson}. The progress in the precision of $\alpha_{3\pi}$ and $\alpha_{\pi\ell\nu}$ was
smaller, the sum of both had reached $(0.14\pm 0.19)\,10^{-3}+\ri (-0.19\pm 0.25)\,10^{-3}$. The results of the analysis were given as linear
combinations of real and imaginary parts of $\epsilon$ and $\tilde\delta$, allowing to extract 
\be \rm Re\,\epsilon = (1.61\pm 0.20)\,10^{-3}~,~~Im\,\delta = (-0.08\pm 0.17)\,10^{-3}~.\ee

\subsubsection{Moscow-Padova 1983} \label{Subsub-Okun}

A Moscow-Padova group \cite{1983-Barmin} succeeded in 1983 to measure
\be \eta_{000} =\frac{\langle 3\pi^0|D|K^0_S\rangle}{\langle 3\pi^0|D|K^0_L\rangle} =-0.08\pm 0.18 +\ri(-0.05\pm 0.27) \ee
using a Xenon bubble chamber in a $K^+$ beam at the ITEP-Moscow proton synchrotron. The result was obtained from the time distribution of 632 identified $3\pi^0$
decays from initial $K^0$ states produced by charge exchange. This was an essential input for a new Bell-Steinberger analysis, published by the group together
with L.~B.~Okun \cite{1983-BaldoCeolin} in the same year. The separately measured CP-violation limits $\eta_{000}$  and $\eta_{+-0}$ made the
analysis independent of the (well-motivated) assumption that $\Delta I =5/2$ decays into the $3\pi (I=3)$ state are negligible. Together with the 1983 best value
\be \eta_{+-0} =\frac{\langle \pi^+\pi^-\pi^0|D|K^0_S\rangle}{\langle \pi^+\pi^-\pi^0|D|K^0_L\rangle} =0.05\pm 0.07 +\ri(0.26\pm 0.13)~, \ee
the $3\pi$ contribution was now \cite{1984-Barmin} 
\be \alpha_{3\pi}=(-0.02\pm 0.07)\,10^{-3}+\ri (-0.04\pm 0.10)\,10^{-3}~.\ee
The lepton contribution was found to be \cite{1984-Barmin} $\alpha_{\pi\ell\nu}=\ri(0.01\pm 0.06)\,10^{-3}$, and their final result was
\be \rm Re\,\epsilon = (1.62\pm 0.05)\,10^{-3}~,~~Im\,\delta = (-0.11\pm 0.10)\,10^{-3}~.\ee

\subsubsection{CPLEAR 1999} \label{Subsub-CPLEAR}

The CPLEAR experiment \cite{2003-CPLEARreport} at CERN took data from 1992 to 1996 producing initial $K^0$ states in the reaction 
${\overline p}p\to K^-\pi^+K^0$ and $\Kqz$ states in ${\overline p}p\to K^+\pi^-\Kqz$. About $2\times 10^8$ decays into $\pi^+\pi^-$, $\pi^0\pi^0$,
$\pi e\nu$, $\pi^+\pi^-\pi^0$ and $3\pi^0$ final states have been collected. The analysis of $\pi^+\pi^-$ time distributions gave precision results for
$|\eta_{+-}|$, $\phi_{+-}$ and $\Delta m$. The analyses of $\pi e\nu$ decays determined results for $\rm Re\,\epsilon$,  $\rm Re\,\delta$, $\rm Im\,\delta$
and all five lepton-decay parameters $y$, ${\rm Re}\,x_+$, ${\rm Im}\,x_+$, ${\rm Re}\,x_-$ and ${\rm Im}\,x_-$, as presented in more detail in 
Section \ref{Sub-piellnu}. The analyses of the $3\pi$ decays determined 
\be \eta_{+-0}= (-2\pm 8)\,10^{-3}+\ri(-2\pm 9)\,10^{-3}~,~~ \eta_{000}= 0.18\pm 0.15 +\ri(0.15\pm 0.20)~,\ee
and it also determined the fraction of CP-allowed decays into the $\pi^+\pi^-\pi^0$ final state with $I=0$, ${\cal B}=(2.5{+1.3\atop -1.0}\pm 0.6)\,10^{-7}$. 
In contrast to the $A_T$ asymmetry in Eq.~\ref{Eq-KasymT}, the measured asymmetry with $\pi \ell\nu$ decays, 
\be A^{exp}_T = \frac{N(\Kqz\to\pi^- e^+\nu)-N(K^0\to\pi^+ e^-\nu)}{N(\Kqz\to\pi^- e^+\nu)+N(K^0\to\pi^+ e^-\nu)} = 4 ({\rm Re\,}\epsilon-y-{\rm Re\,}x_-)
                + f(t; {\rm Re\,}x_-,{\rm Im\,}x_+)~,\label{Eq-5-631}\ee
is time-dependent, with vanishing $f$ for large times $t$, and 
including a contribution of $-2(y+{\rm Re\,}x_-)$ from experimentally necessary normalisation factors.
Assuming CPT symmetry in the $\pi e\nu$ decay amplitudes, $y=x_-=0$, two results were given in 1998 \cite{1998-CPLEAR-T}, first the average for times from
1 to $20\,\tau_S$,
\be \langle A^{exp}_T (t)\rangle = 4 {\rm Re\,}\epsilon + \langle f(t;{\rm Im\,}x_+) \rangle = (6.6\pm 1.3 \pm 1.0)\, 10^{-3}~,\label{Eq-here-44}\ee
and second the result from a two-parameter time-dependent fit,
\be 4 {\rm Re\,}\epsilon = (6.2\pm 1.4\pm 1.0)\,10^{-3}~,~~{\rm Im\,}x_+  = (1.2\pm 1.9\pm 0.9)\, 10^{-3}~.\label{Eq-here-45}\ee
Since ${\rm Im\,}x_+$ is also T-violating, the result in Eq.~\ref{Eq-here-44} violates T symmetry with a significance of 4 to 5 standard deviations, and the 
value of ${\rm Re\,}\epsilon $ in Eq.~\ref{Eq-here-45} was in perfect agreement with the so-far reached values from the Bell-Steinberger analyses.
The CPLEAR group called their results ``First direct observation of time-reversal non-invariance in the neutral-kaon system", using ``direct T violation"
for motion-reversal violation as defined in Subsections \ref{Sub-Quantum} and \ref{Sub-Examples} of this Review. 

The CPLEAR $\pi e\nu$ data were also used in 1998 for the first determination of ${\rm Re\,}\delta $ \cite{1998-CPLEAR-CPT} independent of possible CPT violations in
$\pi\pi$ and $\pi e \nu$ decay amplitudes. This will be presented in Subsection \ref{Sub-FirstReDelta}.

The Bell-Steinberger analysis of CPLEAR, using the world averages of 1999 for all input values, was presented in Ref.~\cite{1999-CPLEAR-BellSt}. Since
\be \alpha_{\pi\ell\nu}= 2({\rm Re\,}\epsilon +{\rm Im\,}\delta-y-\ri\,{\rm Im \,}x_+)\times {\cal B}(K^0_L\to\pi\ell\nu)\,\Gamma_L/\Gamma_S~,\label{Eq-piellnu}\ee
the analysis performed a combined fit to the unitarity relation in Eq.~\ref{Eq-BellSt}, to the CPLEAR $\pi e\nu$ data and to the 1999 best value of the
asymmetry
\be\Delta_L=\frac {N(K^0_L\to\pi^-\ell^+\nu)-N(K^0_L\to\pi^+\ell^-\nu)}{N(K^0_L\to\pi^-\ell^+\nu)+N(K^0_L\to\pi^+\ell^-\nu)}
         = (3.27\pm 0.12)\times 10^{-3}=2\,{\rm Re\,}(\epsilon-\delta-y-x_-)~,\label{Eq-DeltaL}\ee
resulting in 
\bea {\rm Re\,}\epsilon &=& (1.649\pm 0.025)\times 10^{-3}~,\nn\\
      {\rm Im\,}\delta &=&  (0.024 \pm 0.050)\times 10^{-3}~,\nn\\
      {\rm Re\,}\delta &=&  (0.24\pm 0.27)\times 10^{-3}~.\label{Eq-ReDelta1}\eea
The result on ${\rm Re\,}\delta$ is slightly more precise than that obtained from the $\pi e\nu$ data alone, see Subsection \ref{Sub-FirstReDelta}.
The fit also gave $y+{\rm Re\,}x_- =(-0.2\pm 0.3)\,10^{-3}$, showing that the CPT-violating contributions to the $A^{exp}_T$ asymmetries in Eqs.
\ref{Eq-here-44} and \ref{Eq-here-45} are negligible.

\subsubsection{KLOE 2006} \label{Subsub-KLOE}

The KLOE experiment at Frascati took data from 2000 to 2006. In the reaction $e^+ e^-\to\phi\to \rm two~entangled~neutral~Kaons$, 
they collected $2.5\times 10^{9}$ $K^0\Kqz$ pairs with decays into all essential final states. Since the entangled state from the $\phi$-meson decay is
\be \frac{K^0\Kqz-\Kqz K^0}{\sqrt{2}} = \frac{K^0_S K^0_L - K^0_L K^0_S}{\sqrt{8}pq}= \frac{K^0_S K^0_L - K^0_L K^0_S}{\sqrt{2}}~\rm within~  
     o(|\epsilon|^2)~,\ee
the experiment could examine very rare $K^0_S$ decays by tagging with $K^0_L$ decays. An example is the 2013 search for $K^0_S\to 3\pi^0$ decays
\cite {2013-KLOE-3pi0} with the so-far best limit on $\eta_{000}$. Precision results were obtained for essentially all decay modes of $K^0_S$ and $K^0_L$
including $\pi e \nu$ and $\pi\mu\nu$ decay parameters and form factors. 

The 2006 Bell-Steinberger update of the KLOE group together with G.~D'Ambrosio and G.~Isidori \cite{2006-KLOE-BellSt} profited from many of KLOE's precision
results. Their unitarity relation including their definition of $\alpha_i$, 
\be \left(\frac{\Gamma_S+\Gamma_L}{\Gamma_S-\Gamma_L}+\ri\frac{2\Delta m}{\Gamma_S-\Gamma_L}\right)\,\left(\frac{\reps}{1+|\epsilon|^2}-\ri\,\idel\right)=
     \frac{\sum_i \alpha_i\,\Gamma_S}{\Gamma_S-\Gamma_L}~,\label{Eq-BellStKloe}\ee
is identical with that of all earlier analyses described in this review. The only difference is the
explicit appearance of $1+|\epsilon|^2$ which becomes relevant when the error of $\reps$ gets smaller than $10^{-8}$; see the remark above 
Eq.~\ref{Eq-eigenstates}. 

The dominant two-pion contributions, 
\be \alpha_f=\sqrt{{\cal B}(K^0_S\to f)\,{\cal B}(K^0_L\to f)\,\Gamma_L/\Gamma_S}\,\re^{\ri\phi_f}~,\ee
with $f= \pi^+\pi^-(\gamma)$ including bremsstrahlung photons with an energy below 20 MeV and $f=\pi^0\pi^0$ were determined with the 2006 best values for
$\Gamma_S$, ${\cal B}(K^0_L\to \pi^0\pi^0)/{\cal B}(K^0_L\to \pi^+\pi^-)$, $\phi_{+-}$ and $\phi_{00}$ and with all other input values from KLOE,
resulting in
\bea \alpha_{\pi^+\pi^-} &=& [(1.115 \pm 0.015) + \ri\,(1.055 \pm 0.015)]\times 10^{-3}~,\nn\\
     \alpha_{\pi^0\pi^0} &=& [(0.489 \pm 0.007) + \ri\,(0.468 \pm 0.007)]\times 10^{-3}~.\eea
The $\alpha$ value for $\pi^+\pi^-\gamma$ with direct photon emission was estimated to be below $10^{-6}$ using published measurements and a Monte-Carlo extrapolation.
For $f=\pi^+\pi^-\pi^0$ and $3\,\pi^0$, the $\eta_{+-0}$ value was taken from CPLEAR  and $|\eta_{000}|$ from KLOE's own 2006 value.
For $\alpha_{\pi\ell\nu}$, see Eq.~\ref{Eq-piellnu}, the contribution of $\reps$ and $y$ was replaced by $\reps-y = (\Delta_L +\Delta_S)/4$, where $\Delta_{L,S}$ 
are the semileptonic charge asymmetries, as defined in Eq.~\ref{Eq-DeltaL} for $K^0_L$ and analogously for $K^0_S$. The contribution of ${\rm Im}\,x_+$ was taken from CPLEAR, and
the contribution of $\idel$ has to be moved to the left-hand side of the unitarity relation in Eq.~\ref{Eq-BellSt},
leaving ${\tilde\alpha}_{\pi\ell\nu}$ on the right-hand side,  
\be {\tilde\alpha}_{\pi\ell\nu}=\alpha_{\pi\ell\nu}-2\,\idel\times{\cal B}(K^0_L\to \pi\ell\nu)\,\Gamma_L/\Gamma_S~,\label{Eq-5-tilde}\ee 
The results of KLOE's analysis are
\be \reps =  (159.6 \pm 1.3) \times 10^{-5}~,~~\idel = (0.4 \pm 2.1) \times 10^{-5}~,\ee
with a weak correlation of $\rho=-0.17$. The analysis determines also $\rdel$ fitting time distributions and constraining by $\Delta_S-\Delta_L$. The result,
$\rdel = (0.23 \pm 0.27) \times 10^{-3}$, is essentially equal to that of the CPLEAR analysis, see Eq.~\ref{Eq-ReDelta1}. The results for $\reps$ and $\idel$ show
considerable improvement.

\subsubsection{Particle Data Group 2013} \label{Subsub-PDG2012}

\begin{figure}[h]
\begin{minipage}{0.20\textwidth}
\caption{Likelihood contours for $-\delta m$ vs.~$-\delta\Gamma$ \cite{2012-PDG}. The outer two and inner two dashed lines correspond to Eq.~\ref{Eq-delMdelG-outer}
          (no constraint on $\delta\Gamma$) and Eq.~\ref{Eq-delMdelG-inner} ($\delta\Gamma=0$), respectively.}\label{Fig-9-delMdelG}
\end{minipage}\hfill
\begin{minipage}{0.32\textwidth} 
\includegraphics[width=\textwidth]{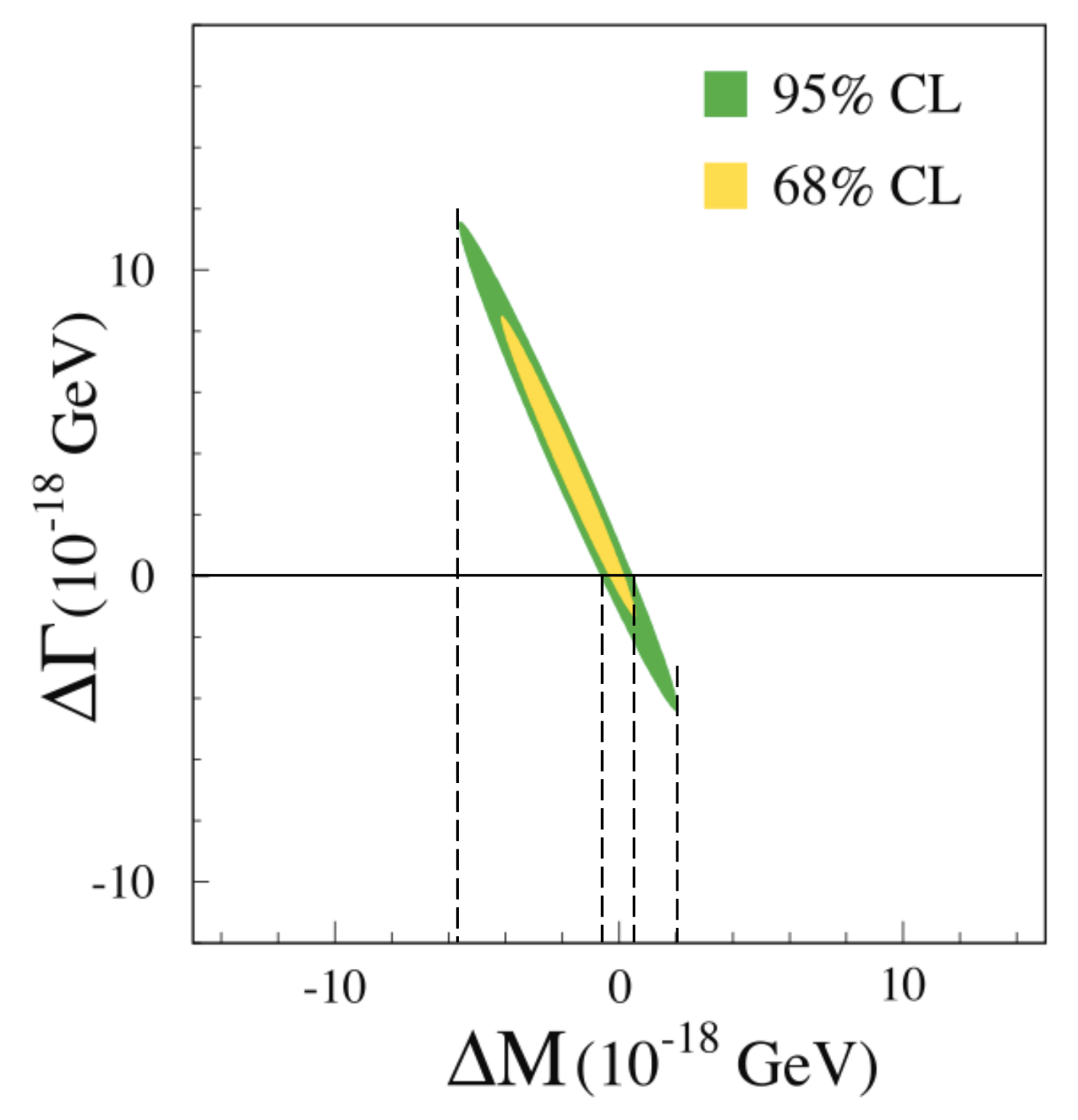}
\end{minipage}\hfill
\begin{minipage}{0.40\textwidth}
\begin{center} 
Table 1: Inputs for the 2013 Bell-Steinberger analysis \cite{2012-PDG}. The last line gives ${\tilde\alpha}_{\pi\ell\nu}$ as defined in Eq.~\ref{Eq-5-tilde}.\\[4mm] 
\begin{tabular}{|r|r|r|}
\hline
$f_i$ & ${\rm Re}\,\alpha_i~[10^{-6}]$ & ${\rm Im}\,\alpha_i~[10^{-6}]$\\
\hline
$\pi^+\pi^-$ & $1112 \pm 10$ & $1061\pm 10$ \\
$\pi^0\pi^0$ & $493 \pm 5$ & $471\pm 5$ \\
$\pi^+\pi^-\pi^0$ & $0 \pm 2$ & $0\pm 2$ \\
$\pi^0\pi^0\pi^0$ & $0 \pm 1$ & $0\pm 1$ \\
$\pi\ell\nu$ & $-2 \pm 5$ & $1\pm 5$ \\
\hline\end{tabular}
\end{center}
\end{minipage}
\end{figure}

For the Review of Particle Properties by the Particle Data Group \cite{2012-PDG}, M.~Antonelli and G.~D'Ambrosio have updated the 2006 Bell-Steinberger analysis
\cite{2006-KLOE-BellSt} using 2013 values for all inputs. The most important changes concern $\alpha_{\pi\pi}$ from KTeV at FNAL \cite{2011-KTeV}, see
Section \ref{Sec-Determination2}, and $\alpha_{3\pi^0}$ from the 2013 result of KLOE \cite{2013-KLOE-3pi0}. The complete list of inputs is given in Table 1.
With these updates, the authors obtain
\bea \reps &=& (161.1\pm 0.5)\times 10^{-5}~,\nn\\ \idel &=& (-0.7\pm 1.4)\times 10^{-5}~,\label{Eq-137}\eea
with a correlation of $\rho =+0.09$. The error on $\reps$ improved by a factor of 2.6, on $\idel$ by 1.5. Also
\be \rdel = (0.24\pm 0.23)\times 10^{-3}~,\ee
shows a slight improvement. Using the definition in Eq.~\ref{Eq-4-2-muParams}, results for the CPT-testing differences $\delta m = m(\Kqz)-m(K^0)$ and
$\delta \Gamma = \Gamma(\Kqz)-\Gamma(K^0)$ can be obtained from $\rdel$, $\idel$ as shown in Fig.~\ref{Fig-9-delMdelG} with $\Delta M =-\delta m$ and $\Delta\Gamma=-\delta\Gamma$.
The correlation between $\delta m$ and
$\delta\Gamma$ is large since the errors on $\rdel$ and $\idel$ differ by a large factor of 16. From the figure, we can read
\be -2\times 10^{-18}<\delta m< +6\times 10^{-18}~\rm GeV~,~~-12\times 10^{-18}<\delta \Gamma< +4\times 10^{-18}~\rm GeV~~at~95\%~C.L.\label{Eq-delMdelG-outer}\ee
The authors quote only the result with the constraint $\delta\Gamma = 0$.
\be -4\times 10^{-19}<\delta m< +4\times 10^{-19}~\rm GeV~~at~95\%~C.L.\label{Eq-delMdelG-inner}\ee

\subsection{The First Determination of $\rm Re\,\delta$} \label{Sub-FirstReDelta}

When presenting $A_T$ with the determination of $\reps$ in 1998 \cite{1998-CPLEAR-T}, see Subsection \ref{Subsub-CPLEAR},
the CPLEAR group also published their analysis of $A_{CPT}$ with the first
determination of $\rdel$ independent of any other CPT- or T-violating parameters \cite{1998-CPLEAR-CPT}. The asymmetry $A_{CPT}$, as introduced in Eq.~\ref{Eq-KasymCPT}, is
time-dependent and sensitive to $\rdel$ and $\idel$. For large times $t\gg\tau_S$, $A_{CPT}=4\,\rdel$. Using their $\pi e\nu$ events, CPLEAR could have measured
\be A^{exp}_{CPT}=\frac{N(\Kqz\to\pi^+ e^-\nu)-N(K^0\to\pi^- e^+\nu)}{N(\Kqz\to\pi^+ e^-\nu)+N(K^0\to\pi^- e^+\nu)} = 4\,\rdel+2\,y+2\,{\rm Re\,}x_-
                + f(t)~,\ee
but this depends on $y$ and $x_-$ in addition to $\rdel$. They, therefore, determined the quantity 
\bea A_{\delta} &=&\frac{N(\Kqz\to\pi^- e^+\nu)-N(K^0\to\pi^+ e^-\nu)(1+4\,{\rm Re}\,\eta_{+-})}{N(\Kqz\to\pi^- e^+\nu)+N(K^0\to\pi^+ e^-\nu)(1+4\,{\rm Re}\,\eta_{+-})}\nn\\ 
            &+& \frac{N(\Kqz\to\pi^+ e^-\nu)-N(K^0\to\pi^- e^+\nu)(1+4\,{\rm Re}\,\eta_{+-})}{N(\Kqz\to\pi^+ e^-\nu)+N(K^0\to\pi^- e^+\nu)(1+4\,{\rm Re}\,\eta_{+-})}
            = 8\,\rdel+ f(t)~,\eea
with ${\rm Re}\,\eta_{+-}=\reps-\rdel$ and $f$ vanishing for $t\gg\tau_S$. For large values of $t$, this quantity is strictly independent of all parameters with the exception 
of $\rdel$, and CPLEAR found, using their own $\pi^+\pi^-$ data for ${\rm Re}\,\eta_{+-}$,
\be \rdel = (0.30\pm 0.33\pm 0.06)\times 10^{-3}~.\ee 
The error is slightly larger than that in Eq.~\ref{Eq-ReDelta1} obtained one year later using unitarity. 

\section{$K_{\pi\pi}$ and $K_{\pi\ell\nu}$ Decay Parameters} \label{Sec-Determination2}
In this Section we discuss the results and symmetry properties of all parameters for the description of 
decays $K^0,\Kqz\to\pi\pi$ and $K^0,\Kqz\to\pi\ell\nu$.  

\subsection{Decays into $\pi\pi$} \label{pipiParams}

The CP-violating amplitude ratios $\eta_{+-}$ and $\eta_{00}$ for $\pi^+\pi^-$ and $\pi^0\pi^0$ decays have been introduced in 
Eqs.~\ref{Eq-5-eta+-} and \ref{Eq-5-eta00}.
Eqs.~\ref{Eq-here8} and \ref{Eq-EpsilonExp} relate them to the amplitude ratios $\epsilon_0$ and $\epsilon^\prime$ for $\pi\pi\, ,I=0$ and 
$\pi\pi\, ,I=2$ decays, respectively. Before reporting the most precise experimental results, let us look at the symmetry properties of $\epsilon_0$ and 
$\epsilon^\prime$ using the $\pi\pi$ amplitudes for $I=0$ and 2, 
\be A_I=\langle \pi\pi,\,I|D|K^0\rangle = a_I\times \re^{\ri\,\delta_I}~,~ ~ \Aq_I= \langle \pi\pi,\,I|D|\Kqz\rangle = a^*_I\times \re^{\ri\,\delta_I}~,\ee
where $a_I$ are the weak amplitudes and $\delta_I$ the scattering phases of final-state interactions \cite{1952-Watson}.

For the phase of $\epsilon_0$, CPT symmetry requires
\be \phi(\epsilon_0)=\arctan(-2\Delta m/\Delta\Gamma)=\phi_W=43.45^\circ\pm 0.06^\circ~.\label{Eq-6-1phiW}\ee
This phase is historically called ``superweak phase" following L.~Wolfenstein's hypothesis \cite{1964-Wolfenstein} of a superweak explanation of
CP violation. Since this hypothesis has  been falsified, it could be called ``unitarity phase" or Wolfenstein phase $\phi_W$. The proof for the
condition in Eq.~\ref{Eq-6-1phiW} follows from the unitarity relation $\Gamma_{12}=A_0^*\Aq_0$, Eq.~\ref{Eq-here3} with dominance of $A_0$ and $\Aq_0$.
The full expression for $\Gamma_{12}$ gives the same result within the quoted errors.
Inserting the $K^0_L$ and $K^0_S$ states into the definition of $\epsilon_0$, we obtain
\be \epsilon_0 = \frac{1-\delta-(1+\delta)\lambda_0}{1+\delta+(1-\delta)\lambda_0}~{\rm with}~\lambda_0=\frac{q\Aq_0}{pA_0}~.\ee
CPT symmetry requires $|\Aq_0/A_0|=1$ and $\delta =0$. The two conditions lead to
\be \Aq_0/A_0 = \sqrt{\Aq_0A_0^*/(A_0\Aq_0^*)}=\sqrt{\Gamma_{12}/\Gamma_{12}^*}~,\ee
and with $q/p$ from Eq.~\ref{strictq/p}
\be \lambda_0 = \sqrt{\frac{2 m_{12}^*\Gamma_{12}-\ri~\Gamma_{12}^*\Gamma_{12}}{2 m_{12}\Gamma_{12}^*-\ri~\Gamma_{12}\Gamma_{12}^*}}=
   \sqrt{\frac{-2(m_{12}/\Gamma_{12})^*+\ri}{-2(m_{12}/\Gamma_{12})+\ri}}~.\ee
Setting
\be -2\,m_{12}/\Gamma_{12}=R\,\re^{\ri\phi}=R(1+\ri\,\phi)~~{\rm with}~~\phi\ll 1~,\ee
we obtain
\bea \lambda_0 &=& \sqrt{\frac{R(1-\ri\,\phi)+\ri}{R(1+ü\ri\,\phi)+\ri} }= 1-\phi\,\frac{\ri\, R}{R+\ri}~,~~ 
     \epsilon_0 = \frac{1-\lambda_0}{1+\lambda_0}=\frac{\phi}{2}\,\frac{\ri\, R}{R+\ri}~,\nn\\
     \phi(\epsilon_0) &=& \phi(1+\ri\,R)=\arctan(2\,|m_{12}|/|\Gamma_{12}|) =\arctan(-2\Delta m/\Delta \Gamma)~.\eea

The requirement of CPT symmetry for the phase of $\epsilon^\prime$ is obtained from the definition in Eq.~\ref{Eq-here8} using $\delta=0$ and the
approximation $\lambda_0=1+o(10^{-3})\approx 1$  \cite{BrancoLavouraSilva}:
\bea     \epsilon^\prime &=& \frac{1}{\sqrt{2}}\,\frac{p\, a_2-q\, a_2^*}{p\, a_0+q\, a_0^*}\,\frac{\re^{\ri\delta_2}}{\re^{\ri\delta_0}}
                          = \frac{\re^{\ri(\delta_2-\delta_0)}}{\sqrt{2}}\,\frac{a_2}{a_0}\,\frac{1-\lambda_2}{1+\lambda_0}\nn\\
                         &=& \frac{\re^{\ri(\delta_2-\delta_0)}}{\sqrt{2}}\,\left(\frac{a_2}{2\,a_0}-\frac{a_2^*}{2\,a_0^*}\right)
                          = \frac{\ri}{\sqrt{2}}\,{\rm Im}\,\frac{a_2}{a_0}\,\re^{\ri(\delta_2-\delta_0)}~,\nn\\
   \phi(\epsilon^\prime) &=& \pi/2+\delta_2-\delta_0 + n\,\pi= 45^\circ\pm 6^\circ~{\rm or}~225^\circ\pm6^\circ~,\label{Eq-6-phiepsprime}\eea
using $a_2\lambda_2/a_0=\lambda_0 a_2^*/a_0^*\approx a_2^*/a_0^*$, and taking $\delta_2-\delta_0$ from Ref.~\cite{1991-Gasser}.

A third complex amplitude ratio in $\pi\pi$ decays of Kaons is
\be \omega = \frac{\langle \pi\pi,I=2|D|K^0_S\rangle}{\langle \pi\pi,I=0|D|K^0_S\rangle}~.\ee
Neutral-Kaon decays are only sensitive to ${\rm Re}\,\omega$. Including isospin-breaking effects and with $|\omega|\ll 1$, the branching ratio
${\cal B}_{+-}/{\cal B}_{00}=|\langle \pi^+\pi^-|D|K^0_S\rangle/\langle \pi^0\pi^0|D|K^0_S\rangle|^2$ is calculated to be \cite{BrancoLavouraSilva} 
\be {\cal B}_{+-}/{\cal B}_{00} = 1.97 (1+3\sqrt{2}\,{\rm Re}\,\omega)~.\ee
The present best values from Eq.~\ref{Eq-6-BFpipi} for the branching fractions ${\cal B}_{+-}$ and ${\cal B}_{00}$  give 
\be {\rm Re}\,\omega=0.033~.\ee
Assuming CPT symmetry, the phase of $\omega$ is $\delta_2-\delta_0$ \cite{BrancoLavouraSilva}, but this is not needed in the further discussion. The $\pi\pi$
input for the Bell-Steinberger analyses is
\be \alpha_{\pi\pi} = {\cal B}_{+-}\, \eta_{+-} + {\cal B}_{00}\, \eta_{00} 
       =({\cal B}_{+-} + {\cal B}_{00})\times\epsilon_0 + ({\cal B}_{+-} -2 {\cal B}_{00})\times\epsilon^\prime~,\ee
with ${\cal B}_{+-} + {\cal B}_{00} \approx 0.999$ and ${\cal B}_{+-} -2 {\cal B}_{00}\approx 0.078$, demonstrating the strong dominance of the $I=0$ contribution.

Precision measurements of $\eta_{+-}$, ${\cal B}_{+-}$, $\eta_{00}$ and ${\cal B}_{00}$ test CPT symmetry of the $\pi\pi$ decay amplitudes. 
The most precise determination of the branching fractions is obtained by KLOE \cite{2006-KLOEc,2012-PDG},
\be {\cal B}_{+-}/{\cal B}_{00}= 2.255\pm 0.005~,~~{\cal B}_{+-}=(69.20\pm0.05)\%~,~~{\cal B}_{00}=(30.69\pm 0.05)\%~.\label{Eq-6-BFpipi}\ee
\begin{figure}[h!]
\begin{minipage}{0.40\textwidth}
\includegraphics[width=\textwidth]{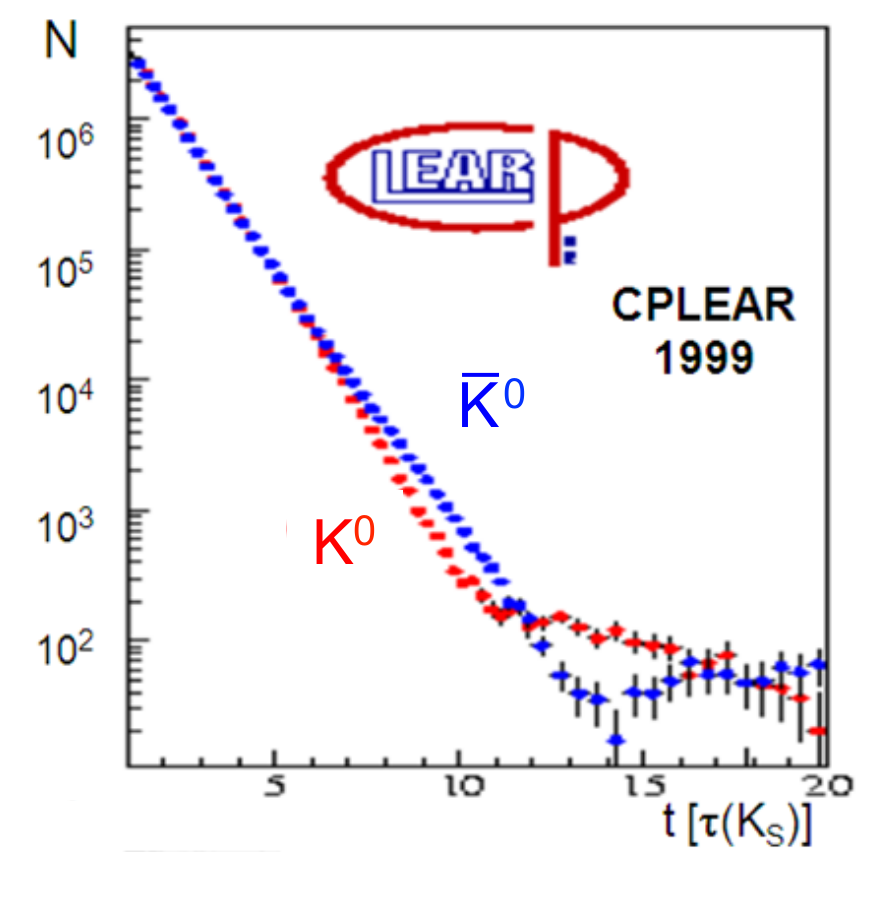}
\caption{Lifetime distributions of $\pi^+\pi^-$ decays from initial $K^0$ and $\Kqz$ states \cite{1999-CPLEAReta+-}.} 
\label{Fig-CPLEAR}
\end{minipage}\hfill
\begin{minipage}{0.55\textwidth}
For $|\eta_{+-}|$, the three experiments KTeV \cite{2004-KTeV}, KLOE \cite{2006-KLOEf} and NA48 \cite{2007-NA48} have nearly equal precision. Their results
dominate the present best value \cite{2012-PDG},
\be |\eta_{+-}|=(2.232\pm 0.011)\times 10^{-3}~.\ee
The most precise measurement of $\phi_{+-}$ is that of CPLEAR 1999 \cite{1999-CPLEAReta+-}. The time dependences of $70 \times 10^6$ $\pi^+\pi^-$ decays 
from  initial $K^0$ and $\Kqz$ states
produced in the reaction $p{\overline p}\to K^\mp\pi^\pm K^0,\Kqz$ are shown in Fig.~\ref{Fig-CPLEAR}. 
The opposite-sign interference term between 
$K^0_L$ and $K^0_S$ decays in the two decay curves does not only give the phase of $\eta_{+-}$, it is also a beautiful example for the demonstration of
CP violation: Particles and antiparticles have different decay laws, seen without any use of quantum mechanics. CPLEAR obtained 
$\phi_{+-}=(42.9\pm 0.6\pm 0.3)^\circ$, the present best average \cite{2012-PDG} is
\be \phi_{+-}=(43.4\pm 0.5)^\circ~.\ee
\end{minipage}\end{figure}

For $|\eta_{00}|$, the two competing experiments NA48 at CERN \cite{2002-NA48} and KTeV at FNAL \cite{2011-KTeV} have reached a precision which definitively
allows to conclude that $\epsilon^\prime\ne 0$. For minimizing systematic errors, both measure the double ratio
\be R=\frac{\Gamma(K^0_L\to\pi^0\pi^0)}{\Gamma(K^0_S\to\pi^0\pi^0)}\left/\frac{\Gamma(K^0_L\to\pi^+\pi^-)}{\Gamma(K^0_S\to\pi^+\pi^-)}\right.
=\left|\frac{\eta_{00}}{\eta_{+-}}\right|^2=1-6\,{\rm Re}\frac{\epsilon^\prime}{\epsilon_0}~.\ee
NA48 uses two parallel beams with $K^0_S$ from a near and $K^0_L$ from a far production target, and a common detector with a magnetic spectrometer
for the charged and a liquid-Krypton calorimeter for the neutral decays. Their result, combining data from several runs between 1997 and 2001, is
\be  {\rm Re}\,(\epsilon^\prime/{\epsilon_0})=(1.47\pm 0.22)\times 10^{-3}~.\ee
NA 48 does not measure the phase of $\eta_{00}$. The KTeV experiment determines $|\eta_{00}/\eta_{+-}|$ and the phases $\phi_{+-}$ and $\phi_{00}$.
It uses two parallel $K^0_L$
beams, one without and one with a lead-scintillator regenerator, and a common detector with a magnetic spectrometer for the charged and a Cesium-Iodide
crystal calorimeter for the charged decays. From the time dependences of the two decay modes, they determine
\bea  {\rm Re}\,(\epsilon^\prime/{\epsilon_0}) &=& (2.11\pm 0.34)\times 10^{-3}~,\\
                          \phi_{00} &=& (44.06\pm 0.68)^\circ~,~~\phi_{+-} = (43.76\pm 0.64)^\circ~,\eea
and, taking into account correlations,
\bea \phi(\epsilon_0) &=& (43.86\pm 0.63)^\circ~,\label{Eq-6-phi0}\\ \phi_{00} -\phi_{+-} &=& (0.30\pm 0.35)^\circ~.\label{Eq-6-phiprime}\eea
They separately determine ${\rm Re}\,({\epsilon^\prime}/{\epsilon_0})$ imposing CPT symmetry, 
i.~e.~$\phi(\epsilon_0)=\arctan(-2\Delta m/\Delta\Gamma)$. The result is
\be {\rm Re}(\epsilon^\prime/\epsilon_0)_{CPT} = (1.92\pm 0.21)\times 10^{-3}~.\ee
\begin{figure}[h]
\begin{minipage}{0.3\textwidth}
\includegraphics[width=\textwidth]{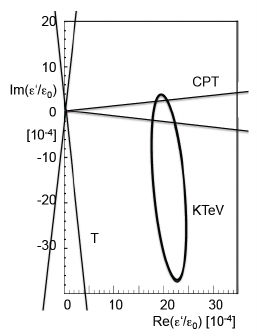}
\caption{$\Delta\chi^2=1$  contour for $\epsilon^\prime/\epsilon_0$ measured by KTeV \cite{2011-KTeV}. The two bands marked CPT and T are the allowed regions 
$(\pm 1\,\sigma)$ for CPT symmetry and T symmetry of $\epsilon^\prime$, respectively. T is violated with $\approx 6\sigma$, CPT is valid within $1\sigma$.} 
\label{Fig-6-KTeV} 
\end{minipage}\hfill
\begin{minipage}{0.65\textwidth}
The result for $\phi(\epsilon_0)$ in Eq.~\ref{Eq-6-phi0} agrees with the CPT prediction $\phi_W$ in Eq.~\ref{Eq-6-1phiW} within one standard deviation,
$\epsilon_0$ is CP-violating, CPT-symmetric and T-violating. The prediction for T symmetry is not unique and depends on the value of $\rm Im\,\delta$.

The phase difference in Eq.~\ref{Eq-6-phiprime} determines the phase of $\epsilon^\prime$. 
As derived in Eq.~\ref{Eq-6-phiepsprime}, CPT symmetry requires
$\phi(\epsilon^\prime)=45^\circ \pm 6^\circ$ which accidentally coincides with $\phi(\epsilon_0)$, thus $\epsilon^\prime/\epsilon_0$ has to be real.
The measured phase difference in Eq.~\ref{Eq-6-phiprime} shows that it is indeed real, since
\be \rm Im\,(\epsilon^\prime/\epsilon_0)=(\phi_{+-}-\phi_{00})/3 = (1.7 \pm 2.0)\ 10^{-3}~.\ee
The amplitude $\langle \pi\pi,\,I=2|D|K^0\rangle$ violates CP symmetry, $\epsilon^\prime \ne 0$. It does not violate CPT, so it must violate T.
Quantitatively, this is shown in Fig.~\ref{Fig-6-KTeV} with only the KTeV data. T symmetry does not predict the phase of $\epsilon^\prime$ since this depends
on $\delta$ and on $a_2=(|\Aq_2|-|A_2|)/(|\Aq_2|+|A_2|)$ which is real. Since $\delta$, CPT violation in mixing, has been measured to be very small, we assume
$|\delta|\ll|a_2|$ and obtain that the T-symmetry expectation is 
\be \phi(\epsilon^\prime) = \delta_2 -\delta_0~,\ee
shown as the vertical ``T" band in the Figure. Note that this band shows the ratio of a T-symmetric, CPT-violating value of $\epsilon^\prime$ and the measured
value of $\epsilon_0$. The KTeV result \cite{2011-KTeV} is about $6\,\sigma$ away from T-symmetry of $\epsilon^\prime$. 
\end{minipage}\end{figure}

\subsection{Decays into $\pi\ell\nu$} \label{Sub-piellnu}

The symmetry properties of $\pi e \nu$ and $\pi\mu\nu$ decays depend crucially on the ``$\Delta Q/\Delta S$" rule, i.~e.~$K^0$ states decay only into positive 
leptons and $\Kqz$ states only into negative leptons. The rule was first discussed 1957 for hyperon decays by R.~P.~Feynman and
M.~Gell-Mann \cite{1958-FeynmanGellmann}. If it is valid, the only way to break CP symmetry in $\pi\ell\nu$ decays  would be CPT violation with
\be y =\frac{|\langle\pi^+\ell^-\nu|D|\Kqz\rangle|-|\langle\pi^-\ell^+\nu|D|K^0\rangle|}{|\langle\pi^+\ell^-\nu|D|\Kqz\rangle|+|\langle\pi^-\ell^+\nu|D|K^0\rangle|}   
      =\frac{1}{2}\, \frac{\Gamma(\Kqz\to\pi^+\ell^-\nu)-\Gamma(K^0\to\pi^-\ell^+\nu) }{\Gamma(\Kqz\to\pi^+\ell^-\nu)+\Gamma(K^0\to\pi^-\ell^+\nu) }~,\label{Eq-6-y}\ee
where $|y|\ll 1$ is assumed.
In addition to $y$, there are four more decay parameters, ${\rm Re}\,x_+$, ${\rm Im}\,x_+$, ${\rm Re}\,x_-$ and ${\rm Im}\,x_-$, 
\bea x=\frac{\langle \pi^-\ell^+\nu|D|\Kqz\rangle}{\langle \pi^-\ell^+\nu|D|K^0\rangle} &,& {\overline x}=\frac{\langle \pi^+\ell^-\nu|D|K^0\rangle^*}
{\langle \pi^+\ell^-\nu|D|\Kqz\rangle^*}~,\nn\\[3mm]
     x_+ =(x+{\overline x})/2 &,& x_- =(x-{\overline x})/2~,\eea
where all matrix elements are integrated over the full three-body phase space. The $y$ parameter is
$\Delta Q/\Delta S$ conserving and CPT-violating, the four $x$ parameters are $\Delta Q/\Delta S$ violating, $x_+$ is CPT-symmetric, $x_-$ CPT-violating,
the two real parts are T-symmetric, and the two imaginary parts are T-violating. Only ${\rm Re}\,x_+$ is CP-conserving, but it violates the $\Delta Q/\Delta S$ rule.
All five parameters have been determined for $\ell=e$ by CPLEAR from the time dependences of $e^+$ and $e^-$ final
states from initial $K^0$ and $\Kqz$ states. Lepton universality, which is well tested in many
other observations, allows to assume equal parameter values for $\pi\mu\nu$ decays. All five parameters are treated to be small, and explicit expressions 
for the four time-dependent decay rates are given e.~g.~in Ref.~\cite{2003-CPLEARreport}, 

\bea R_1=R(K^0\to\pi^-\ell^+\nu)=  R_0/4 \times \{(1+2{\rm Re\,}x +4{\rm Re\,}\delta-2y)\re^{-\Gamma_S t}
                                                 + (1-2{\rm Re\,}x -4{\rm Re\,}\delta-2y)\re^{-\Gamma_L t} \nn\\
         + [(2-4y)\cos(\Delta m t)-(8{\rm Im\,}\delta+4{\rm Im\,}x)\sin(\Delta m t)]\re^{-(\Gamma_S+\Gamma_L)t/2}\}~,\nn\\[3mm]
R_2=R(\Kqz\to\pi^+\ell^-\nu) = R_0/4 \times \{(1+2{\rm Re\,}{\overline x} -4{\rm Re\,}\delta+2y)\re^{-\Gamma_S t}
                                                 + (1-2{\rm Re\,}{\overline x} +4{\rm Re\,}\delta+2y)\re^{-\Gamma_L t} \nn\\
  + [(2+4y)\cos(\Delta m t)+(8{\rm Im\,}\delta+4{\rm Im\,}{\overline x})\sin(\Delta m t)]\re^{-(\Gamma_S+\Gamma_L)t/2}\},\nn\\[3mm]
R_3=R(K^0\to\pi^+\ell^-\nu) = R_0/4 \times \{(1+2{\rm Re\,}{\overline x} -4{\rm Re\,}\epsilon+2y)\re^{-\Gamma_S t}
                                        + (1-2{\rm Re\,}{\overline x} -4{\rm Re\,}\epsilon+2y)\re^{-\Gamma_L t} \nn\\
 - [(2-8{\rm Re\,}\epsilon+4y)\cos(\Delta m t)+4{\rm Im\,}{\overline x}\,\sin(\Delta m t)]\re^{-(\Gamma_S+\Gamma_L)t/2}\}~,\nn\\[3mm]
R_4=R(\Kqz\to\pi^-\ell^+\nu) = R_0/4 \times \{(1+2{\rm Re\,}x +4{\rm Re\,}\epsilon-2y)\re^{-\Gamma_S t}
                                                 + (1-2{\rm Re\,}x +4{\rm Re\,}\epsilon-2y)\re^{-\Gamma_L t} \nn\\
 - [(2+8{\rm Re\,}\epsilon-4y)\cos(\Delta m t)-4{\rm Im\,}x\,\sin(\Delta m t)]\re^{-(\Gamma_S+\Gamma_L)t/2}\}~,~\eea 
where $R_0=[\Gamma(K^0\to\pi^-\ell^+\nu)+\Gamma(\Kqz\to\pi^+\ell^-\nu)]/2$.
CPLEAR uses ${\rm Re\,}y$ instead of $y$. Since CPT symmetry requires only the equality of the rates in Eq.~\ref{Eq-6-y}, $y$ has to be real.
${\rm Im\,}y$ is unobservable and does not appear in the four time-dependent rates. 

From their $1.3\times 10^6$ $\pi e\nu$ decays, only one external input ($\Delta_\ell$, see Eq.~\ref{Eq-DeltaL}), and the Bell-Steinberger unitarity
constraint (Section \ref{Subsub-CPLEAR}), CPLEAR determined all five parameters. The real part of $x_+$ and the imaginary part of $x_-$ were obtained
in 1998 together with $\Delta m$ from the asymmetry \cite{1998-CPLEAR-Rex+,2003-CPLEARreport}
\be A_{\Delta m}=\frac{(R_1+R_2)-(R_3+R_4)}{(R_1+R_2)+(R_3+R_4)}=\frac{[2\cos(\Delta m t)-4\,{\rm Im\,}x_-\,\sin(\Delta m t)]\re^{-(\Gamma_S+\Gamma_L)t/2}}
       {(1+2\,{\rm Re\,}x_+)\re^{-\Gamma_S t}+(1-2\,{\rm Re\,}x_+)\re^{-\Gamma_L t}}~.\ee
The results are \cite{2003-CPLEARreport}
\be  {\rm Re\,}x_+ = (-1.8\pm 6.1)\ 10^{-3}~,~~{\rm Im\,}x_- = (-0.8\pm 3.5)\ 10^{-3}~.\label{Eq-6-Imx}\ee
There is a strong correlation between 
$\Delta m$ and ${\rm Im\,}x_-$, the value in Eq.~\ref{Eq-6-Imx} was obtained with $\Delta m =(530.1\pm 1.4)\,10^7$~/s. 
The three other decay parameters are given in 1999 in Ref.~\cite{1999-CPLEAR-BellSt},
\be  y=(0.3\pm 3.0)\,10^{-3}~,~~{\rm Re\,}x_- = (-0.5\pm 3.0)\ 10^{-3}~,~~{\rm Im\,}x_+ = (-2.0\pm 2.6)\ 10^{-3}~,\label{Eq-6-yetc}\ee
with partially strong correlations as shown in Table 2 of Ref.~\cite{1999-CPLEAR-BellSt}. 

All four $x$ values are well compatible with zero. This is not a surprise because of the big success of the quark model; Standard-Model estimations 
\cite{1990-DibGuberina} for the breakdown of the $\Delta Q/\Delta S$ rule are on a level of $10^{-7}$. On the level of the CPLEAR precision, T violation
cannot be seen in $\pi\ell\nu$ decays. The only possible CP violation is CPT violation. The CPT result for $y$ excludes this on a level of $3\times 10^{-3}$
for the ratio of CPT-violating and CPT-symmetric amplitudes.

\subsection{Muon Polarisation in $\pi\mu\nu$ Decays}

As proposed in 1958 by J.~J.~Sakurai \cite{1958-Sakurai}, a measurement of the muon polarisation transverse to the decay plane in $K\to\pi\mu\nu$ decays
is a test of T symmetry. The matrix element for the decay $K\to\pi\ell\nu$ consists of a leptonic and a hadronic current; the latter has two independent form factors
$f_+(q^2)$ and $f_-(q^2)$. T invariance requires $\xi(q^2)=f_-(q^2)/f_+(q^2)$ to be real, and the transverse component of the lepton polarization is
\be P_T=\frac{{\vec\sigma_\ell}\cdot({\vec p}_\pi\times{\vec p}_\ell)}{|{\vec p}_\pi\times{\vec p}_\ell|}={\rm Im}\,\xi\,\frac{m_\ell}{m_K}\,f(|{\vec p}_\mu|,\theta)~,\ee
where ${\vec\sigma}_{\ell}$ is the spin of the lepton, ${\vec p}_i$ are momenta and $\theta$ is the angle between $\ell$ and $\nu$. There are two reasons for using muons
instead of electrons, the factor $m_\ell/m_K$ and the sequential P-violating decay $\mu\to e\nu\nu$ which allows to measure the polarisation. 
There is also a reason for choosing positive muons, they can be stopped and kept in an absorber until they decay. Two recent experiments have determined 
$\rm Im\,\xi$ with high precision, in 1980 by W.~M.~Morse et al. \cite{1980-Morse} at BNL using $K^0_L$ mesons, and in 2004 by M.~Abe et al. \cite{2004-AbeMuPolK+} at KEK
using $K^+$ mesons. Since there is only one hadron in the final state, no strong final-state
interactions can mimic T violation; electromagnetic final-state interactions are estimated to produce ${\rm Im}\,\xi\approx 0.008$ \cite{1973-GinsbergSmith}.
The BNL experiment \cite{1980-Morse} determines 
\be {\rm Im}\,\xi(K^0_L\to\pi^-\mu^+\nu) = 0.009 \pm 0.030~,\ee
and the KEK experiment \cite{2004-AbeMuPolK+}
\be {\rm Im}\,\xi(K^+\to\pi^0\mu^+\nu) = -0.005 \pm 0.008~.\ee
Both experiments find T symmetry within one standard deviation. Both test T violation in New Physics, since Standard-Model estimations give ${\rm Im}\,\xi=o\,(10^{-7})$
\cite{2000-BigiSanda}.

\section{$D^0\Dqz$, $B^0\Bqz$ and $B_s\Bqs$ Transitions} \label{Sec-Dzero}

Only four neutral-meson systems can show flavor transitions, $K^0\Kqz$, $D^0\Dqz$, $B^0\Bqz$ and $B_s\Bqs$, and all four transitions have been observed, 
Table \ref{Tab-7-1} gives an overview. 

\begin{table}[h]
\setlength{\belowcaptionskip}{2mm}
\begin{center}
\caption{Historical overview of flavor transitions. Column 2 gives the year of first observation, the next columns give the years when the five transition parameters
have been determined; L denotes existing limits, and (*) indicates present conflicting results.} 
\label{Tab-7-1}
\begin{tabular}{|c|c|c|c|c|c|c|}
\hline
& seen & $\Delta m$ & $\Delta\Gamma$ & $\rm Re\,\epsilon$ & $\rm Re\,\delta$ & $\rm Im\,\delta$ \\
\hline
$K^0\Kqz$ & 1956 & 1961 & 1956 & 1970 & L & L \\
$D^0\Dqz$ & 2007 & 2007 & 2007 & L &  &  \\
$B^0\Bqz$ & 1987 & 1987 & L & L & L & L \\
$B_s\Bqs$ & 1992 & 2006 & 2010 & $\rm L^{(*)}$ & & \\
\hline\end{tabular}
\end{center}\end{table}

\begin{figure}[t]
\centering\includegraphics[width=0.70\textwidth]{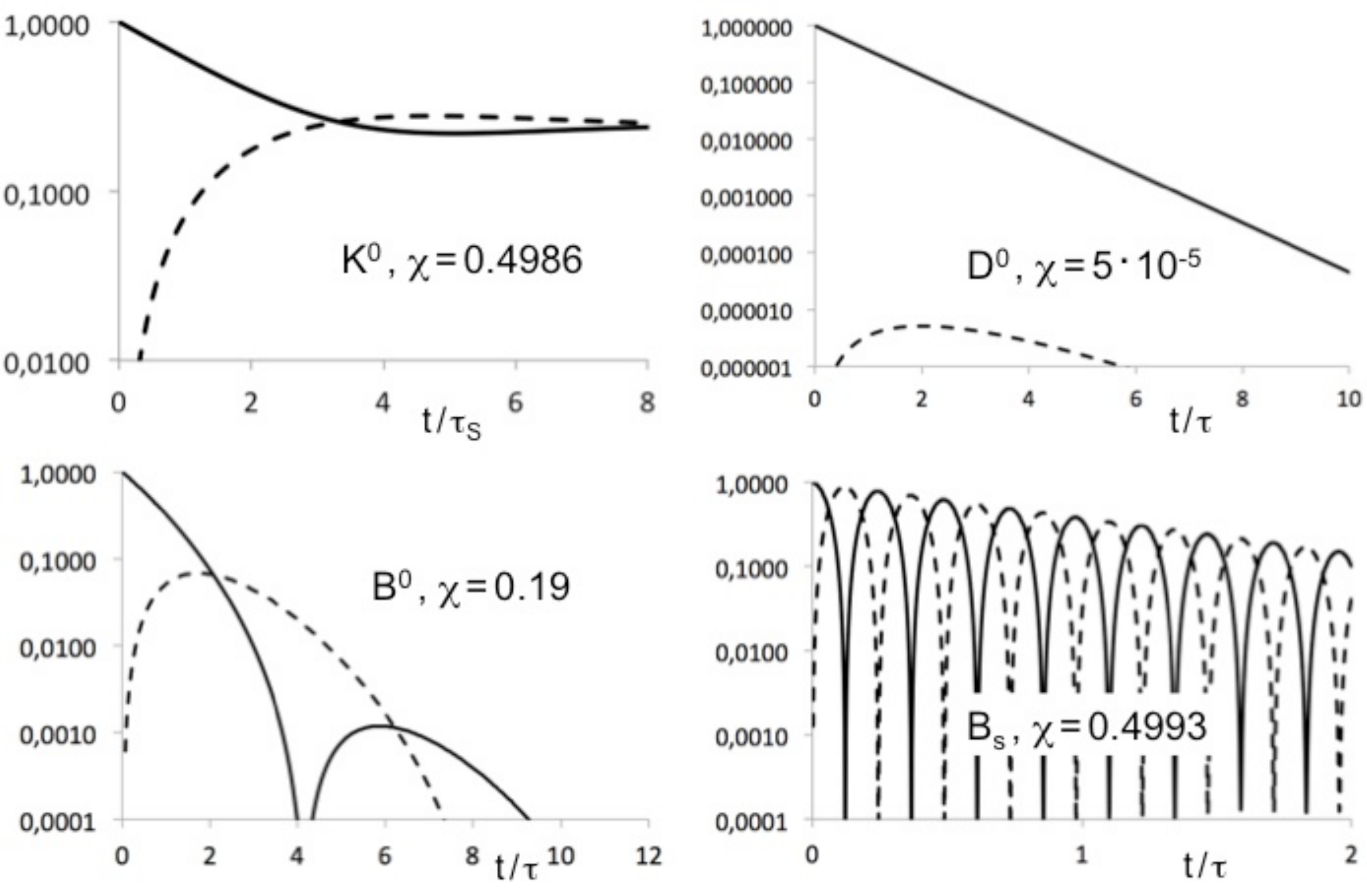}
\caption{Time dependences for $M^0$ decays (solid curves) and $\Mqz$ decays (dashed curves) from initial $M^0$ states for all four meson systems with their
transition parameters as measured; $\chi$ is the time-integrated fraction of $\Mqz$ decays. The vertical scales show $N(t)/N(0)$, the horizontal scales
use $\tau_S = \tau(K^0_S)$ and for the three heavier systems $\tau=1/\Gamma$.}
\label{Fig-4Mixings} 
\end{figure}

{\noindent The dynamics is the same in all four systems $M\,\overline M$: $M$ and $\overline M$ have non-exponential decay laws, the same lowest-order linear Schr\"odingier
equation (Eq.~\ref{Eq-4-2-1}) describes the transitions completely, its two eigenstates (Eq.~\ref{Eq-4-2-1a}) have exponential decay laws, and the evolutions of initial
$M$ and $\overline M$ states are given by the same expressions (Eqs.\,\ref{Eq-Kevolution},\,\ref{Eq-Kbarevolution},\,\ref{Eq-4-2-1b}). 
The eigenstates of the three heavy-flavor systems are called 
$M_h$ (heavy) and $M_\ell$ (light) in contrast to $K^0_L$ and $K^0_S$ where the sign of $\Delta m$ was determined much later than the sign of $\Delta\Gamma$.
The seven parameters are $m_h$, $m_\ell$, $\Gamma_h$, $\Gamma_\ell$, $|q/p|$, $\rm Re\,\delta$ and $\rm Im\,\delta$. For $\Delta m$ and $\Delta\Gamma$ we use the 
same convention for all four systems,
\be \Delta m = m_h - m_\ell > 0~, ~~ \Delta\Gamma = \Gamma_h -\Gamma_\ell~,\ee
and $\Gamma=(\Gamma_h+\Gamma_\ell)/2$, ${\rm Re\,}\epsilon =(1-|q/p|)/2$. The parameter values are very different for the four systems, leading to the 
evolution pictures in Fig.~\ref{Fig-4Mixings}.}

CP and T violations have been seen in the $K^0\Kqz$ system, in transitions and in decays. Since both violations are small, the two mass eigenstates are 
``almost CP eigenstates", the heavier $K^0_L$ has $CP\approx -1$. No CP violation has been established in the $D^0\Dqz$ system so far; the expected small
effects and the already achieved limits allow the same approximation, the heavier state $D^0_h$ has $CP\approx +1$. In the $B^0\Bqz$ system, no CP
violation has been observed in transitions, but there are large CP violations in decays and in the interplay of transitions and decays. Therefore,
the two mass eigenstates are no CP eigenstates, not even approximately. The state decaying into $\pi^+\pi^-\,(CP=+1)$ is different from the state decaying
into $J/\psi\, K^0_L\,({\rm also}~CP=+1)$, and both are not mass eigenstates. The $B_s\Bqs$ system
resembles the first two  systems, LHCb \cite{2012-LHCb-DelGamBs} recently determined $CP\approx -1$ for the heavier eigenstate. 

\subsection{The $D^0$ System} \label{Sub-D0}

Transitions between $D^0=c\,\overline u$  and $\Dqz=u\,\overline c$ have been found by the ``B-Meson Factory" experiments BABAR \cite{2007-BABAR-Dmix} at SLAC 
and Belle \cite{2007-Belle-Dmix} at KEK. A combination of all present observations leads to \cite{2013-HFAG}
\bea x = (m_h-m_\ell)/\Gamma &=& (4.1\pm 1.4)\times 10^{-3}~,\nn\\
     y = (\Gamma_h-\Gamma_\ell)/2\Gamma &=& (6.3\pm 0.7)\times 10^{-3}~,\nn\\
      {\rm Re}\,\epsilon &=& +0.04 \pm 0.05~,\eea
where $1/\Gamma = 0.41 \times 10^{-12}$ s.
Since transitions are very low in rate, the precision on T violation  is still very modest, and no
tests of CPT symmetry have been reported so far.

\subsection{The $B^0$ System} \label{Sub-B0}

Transitions between $B^0={\overline b}\,d$ and $\Bqz=b\,{\overline d}$ have first been seen in 1987 by ARGUS \cite{1987-ARGUS-Bmix} at the $e^+\,e^-$ storage ring DORIS-II
at DESY, producing entangled $B^0\Bqz$ pairs from $\Upsilon(4S)$ decays. They combined three observations, like-sign lepton pairs, events with 
$\ell^\pm$ and $D^{*\mp}\ell^\pm\nu$, and one event with
two fully reconstructed $B^0\to D^{*-}\mu^+\nu$ decays. Since no time between production and decay of the $B$ mesons were measured, $\Delta m$ and $\Delta\Gamma$
could not be determined, only the time-integrated ratio
\be \chi =\frac{N(B^0\to\Bqz\to\ell^-\nu X)}{N(B^0\to\ell^\pm\nu X)}= \frac{(\Delta m)^2+(\Delta\Gamma/2)^2}{2\Gamma^2+2(\Delta m)^2}
         =\frac{x^2+y^2}{2+2x^2}= 0.17\pm 0.05~.\label{Eq-7-3B}\ee
The first time-dependent measurement, showing the interference pattern of the transitions, was reported in 1993 by ALEPH at the $e^+e^-$ storage ring LEP 
\cite{1993-ALEPH-DmB0}.  Their result, $\Delta m = (0.52\pm 0.11)\,10^{12}/$s,
was in agreement with the $\chi$ value of ARGUS in Eq.~\ref{Eq-7-3B} if $\Delta\Gamma\ll\Delta m$. The presently most precise result,
$\Delta m = (0.516\pm 0.006)\,10^{12}/$s was obtained in 2013 by LHCb \cite{2013-LHCb-DmB0}, and the present world average \cite{2012-PDG} is
\be \Delta m(B^0) = (0.510\pm 0.003)\,10^{12}/\rm s~.\ee
In the decade between 2001 and 2010, the majority of the $B^0$-system results were obtained by the two experiments BABAR at SLAC and Belle at KEK. 
They used entangled $B^0\Bqz$ pairs produced on the $\Upsilon(4S)$ resonance in the energy-asymmetric $e^+\,e^-$ storage rings PEP-II and KEK-B, respectively. 
For $\Delta\Gamma$, they found upper limits in 2004 \cite{2004-BabarCPTprd} and 2012 \cite{2012-BelleCPT}. The present average, including results
from LHCb and D0 is \cite{2013-HFAG}
\be \Delta\Gamma/\Gamma = (0.1\pm 1.0)\,10^{-2}~.\label{Eq-7-DeltaGammaB0}\ee
The Standard-Model estimation is $(4\pm 1)\,10^{-3}$ \cite{2007-LenzNierste}.
The BABAR and Belle results have been obtained in the same analyses which determine the three CP-violation parameters in $B^0\Bqz$ transitions, 
as discussed in the following. With small $\Delta\Gamma/\Gamma$, it is convenient to rewrite Eqs.~\ref{Eq-4-2-1b} as
\bea P(B^0\to B^0) = \re^{-\Gamma t}\,[\frac{1}{2}\cosh(\Delta\Gamma t/2)+2{\rm Re\,}\delta\sinh(\Delta\Gamma t/2)+\frac{1}{2}\cos(\Delta m t)
                     -2{\rm Im\,}\delta\sin(\Delta m t)]~,\nn\\
    P(\Bqz\to\Bqz) = \re^{-\Gamma t}\,[\frac{1}{2}\cosh(\Delta\Gamma t/2)-2{\rm Re\,}\delta\sinh(\Delta\Gamma t/2)+\frac{1}{2}\cos(\Delta m t)
                     +2{\rm Im\,}\delta\sin(\Delta m t)]~,\nn\\[3mm]
    P(B^0\to\Bqz) = (\frac{1}{2}-2{\rm Re\,}\epsilon)\,\re^{-\Gamma t}\,[\cosh(\Delta\Gamma t/2)-\cos(\Delta m t)]~,\nn\\
    P(\Bqz\to B^0) = (\frac{1}{2}+2{\rm Re\,}\epsilon)\,\re^{-\Gamma t}\,[\cosh(\Delta\Gamma t/2)-\cos(\Delta m t)]~,\label{Eq-6-FourRates}\eea
since $\cosh(\Delta\Gamma t/2)\approx 1$ and $\sinh(\Delta\Gamma t/2)\approx \Delta\Gamma t/2$.

Assuming the validity of a ``$\Delta Q/\Delta b$" rule for semileptonic decays ($B^0$ decays into $\ell^+\nu X$ and $\Bqz$ into $\ell^-\nu X$ with strict absence of 
the opposite-sign decays) and in addition CPT symmetry in these decays, $\Gamma(B^0\to\ell^+\nu X) = \Gamma(\Bqz\to\ell^-\nu X)$, $\rm Re\,\epsilon$ can be 
obtained from the asymmetry $A_T$ as given in Eq.~\ref{Eq-5-631} for Kaons with $y=x_+=x_-=0$. From entangled $B^0\Bqz$ pairs we have
\be A_T = \frac{P(\Bqz\to B^0)-P(B^0\to\Bqz)}{P(\Bqz\to B^0)+P(B^0\to\Bqz)}=\frac{N(\ell^+\ell^+)-N(\ell^-\ell^-)}{N(\ell^+\ell^+)+N(\ell^-\ell^-)}
        = 4 {\rm Re\,}\epsilon = 4 \times (0.2 \pm 1.4)\times 10^{-3}~,\ee 
where the result is the mean of the 2006 measurements of BABAR \cite{2006-BABAR-AT} and Belle \cite{2006-Belle-AT}. Including two newer results, 
from D0 \cite{2012-D0} in 2011 in untagged $p\overline p\to D^{*\mp}\mu^{\pm} X$ events and from BABAR \cite{2013-BABAR} in 2013 with 
$(D^{*\mp}\ell^\pm\nu)\,(K^\pm X)$ events, the present best value for T violation in $B^0\Bqz$ transitions is
\be {\rm Re}\,\epsilon = (0.7\pm 0.7)\times 10^{-3}~.\label{Eq-6-epsB}\ee
The prediction of the Standard Model \cite{2007-LenzNierste} is 
\be {\rm Re}\,\epsilon = (-1.2\pm 0.3)\,10^{-4}~, \label{Eq-LenzNierste}\ee
one order of magnitude below the sensitivity of present experiments. 

Lepton pairs with opposite charges have been used for testing CPT symmetry. Instead of $\delta$, BABAR and Belle use $z=-2\,\delta$ for these tests. 
With $|\Delta\Gamma/\Delta m|\ll 1$, we obtain from Eq.~\ref{Eq-4-2-muParams}
\be {\rm Re\,}\delta = -\frac{{\rm Re\,}z}{2}=\frac{m_{22}-m_{11}}{2\,\Delta m}~,
~~{\rm Im\,}\delta = -\frac{{\rm Im\,}z}{2}=-\frac{\Gamma_{22}-\Gamma_{11}}{4\,\Delta m}~.\label{Eq-6-CPTtest}\ee
If the first $B$ in an entangled $B^0\Bqz$ pair decays into $\ell^+\nu X$, it prepares a $\Bqz$ state and a later $\ell^-\nu X$ decay has the rate $P(\Bqz\to\Bqz)$ with
$t= t(\ell^-~{\rm decay})-t(\ell^+~{\rm decay})$. The analogous argument holds for $\ell^-\nu X$ as first decay with $t=t(\ell^+~{\rm decay})-t(\ell^-~{\rm decay})$. 
The CPT asymmetry, defined as in Eq.~\ref{Eq-KasymCPT} for the $K^0$ system, is then given by
\be A_{CPT} = \frac{P(\Bqz\to\Bqz)-P(B^0\to B^0)}{P(\Bqz\to\Bqz)+P(B^0\to B^0)}=\frac{N^{+-}-N^{-+}}
     {N^{+-}+N^{-+}}=4\,\frac{-{\rm Re\,}\delta\sinh(\Delta\Gamma t/2)+{\rm Im\,}\delta\sin(\Delta mt)}{\cosh(\Delta\Gamma t/2)+\cos(\Delta m t)}~,\label{Eq-ACPTB0}\ee
where $N^{+-}$ and $N^{-+}$ are the number of events with $\ell^+$ and $\ell^-$ in the first decay, respectively.
From this asymmetry, BABAR \cite{2006-BABAR-AT} determined in 2006
\be {\rm Im\,}\delta=(+7.0\pm 3.7\pm 1.6)\,10^{-3}~.\ee 
${\rm Re\,}\delta$ could not be determined in this analysis since $A_{CPT}$ contains only the product ${\rm Re\,}\delta\times\Delta\Gamma$.
Sensitivity on both real and imaginary parts of $\delta$ was obtained in a combined analysis of $(D^{*\pm}\ell^\mp\nu,\,\ell^\pm\nu X)$ and 
$(J/\psi K_{S,L},\,\ell^\pm\nu X)$, the same analysis which led to
the $\Delta\Gamma$ result in Eq.~\ref{Eq-7-DeltaGammaB0}. The time-dependent rates for initial $B^0$ and $\Bqz$ decaying into flavor-specific
states are the same as given in Eq.~\ref{Eq-6-FourRates}, and for decays into CP eigenstates they are given in Section \ref{Sub-BtoF}, Eqs.~\ref{Eq-rateBtoF}.
Results for ${\rm Re\,}\delta$ and ${\rm Im\,}\delta$ from this combination have been obtained by BABAR \cite{2004-BabarCPTprd} in 2004, 
\be {\rm Re\,}\delta \times \cos\phi_\lambda = (-7\pm 18\pm 17)\times 10^{-3}~,~~{\rm Im\,}\delta = (-19\pm 15\pm 13)\times 10^{-3}~,~\ee
where $\cos\phi_\lambda$ is now known to be 0.73, and by Belle \cite{2012-BelleCPT} in 2012, 
\be {\rm Re\,}\delta =(-10\pm 19\pm 17)\,10^{-3}~,~~{\rm Im\,}\delta = (+2.9\pm 1.7\pm 1.7)\,10^{-3}~.~\ee
The three results for the imaginary part and the two for the real part lead to an average of
\be {\rm Re\,}\delta =(-10\pm 20)\,10^{-3}~,~~{\rm Im\,}\delta = (+3.7\pm 2.1)\,10^{-3}~.\label{Eq-7-deltaB}\ee 

With $\delta m =m_{22}-m_{11}$ and $\delta\Gamma=\Gamma_{22}-\Gamma_{11}$, Eqs.~\ref{Eq-6-CPTtest} allow the conclusions
\be \delta m/m = (-1.3\pm 2.6)\times 10^{-15}~,~~\delta\Gamma /\Gamma = (-11\pm 6)\times 10^{-3}~,\ee
in comparison to $K^0\Kqz$ transitions where $\delta m/m = (4\pm 8)\times 10^{-18}$, $\delta\Gamma /\Gamma = (-1\pm 2)\times 10^{-3}$.

\subsection{The $B_s$ System} \label{Sub-Bs}

The transitions between $B_s=s\,\overline b$ and $\Bqs = b\,\overline s$ have the highest transition probability, $\chi(B_s)\approx 0.5$. In a ``Search for $B^0 \Bqz$ 
oscillations" in 1987, UA1 \cite{1987-UA1} at the $\rm Sp{\overline p}S$ storage rings at CERN found a 2.9\,$\sigma$ excess of like-sign muon pairs in $p\overline p$ collisions. They reported
\be \chi = f_0 \,\chi(B^0) +f_s\,\chi(B_s) = 0.121 \pm 0.047~,\ee
where $f_0$ and $f_s$ are the production fractions of $B^0$ and $B_s$, and the $\chi$ values are their transition probabilities as defined in Eq.~\ref{Eq-7-3B}.
The conclusion was that the result is compatible with $\chi(B^0)=0$ and $\chi(B_s)=0.5$. 

At LEP, many events with wrong-sign decays of $B_s$ mesons have been observed, but the oscillation period $2\pi/\Delta m$ was too short for being determined. 
The experimental limits  improved from $\Delta m > 1.8\times 10^{12}$/s in 1994 \cite{1994-BUSCULIC94B} to $> 10.9\times 10^{12}$/s in 2003 \cite{2003-HEISTER03E}. 
CDF \cite{2006-ABULENCIA06G} at the Tevatron at FNAL succeeded in 2006 to determine 
\be \Delta m =(17.31{+0.33 \atop -0.18}\pm 0.07)\times 10^{12}/{\rm s}~,~~ x=\frac{\Delta m}{\Gamma} =26~,~~\chi=\frac{x^2}{2+2x^2}=0.4993~.\ee
The present world average is \cite{2013-HFAG}
\be \Delta m =(17.761\pm 0.022)\times 10^{12}/{\rm s}~,~~ x=26.85\pm 0.13~,~~\chi=0.499311\pm 0.000007~,\ee
including the latest result from LHCb \cite{2012-AAIJ12I}. 

The parameters $\Gamma=(\Gamma_h+\Gamma_\ell)/2$ and $\Delta\Gamma=\Gamma_h-\Gamma_\ell$ are obtained from the three time dependences of decays into 
(approximate) CP eigenstates 
with $CP\approx+1$ (e.\,g.~$K^+K^-$ or $J/\psi \phi$), into those with $CP\approx -1$ (e.\,g.~$J/\psi f_0$), and into summed flavor eigenstates (e.\,g.~the sum of
${\overline D}{}_s\ell^+ X$ and $D_s\ell^- X$). The data \cite{2012-PDG} show a clearly longer lifetime for the $CP\approx -1$ than for the $CP\approx +1$ states, and the
summed flavor eigenstates show a lifetime in between as expected by their decay law $\re^{-\Gamma t}\cosh(\Delta\Gamma t/2)$. 
Determining the phase-shift differences in final-state interactions of $K^+K^-$ pairs in s- and p-waves as function of the $K^+K^-$ mass, LHCb \cite{2012-LHCb-DelGamBs}
found in 2012 that $\Delta\Gamma=\Gamma_h-\Gamma_\ell<0$. A fit to all three time dependences gives \cite{2013-HFAG} 
\be \Gamma=(0.6598\pm 0.0034)\times 10^{12}/{\rm s}~,~~\Delta\Gamma =(-0.091\pm 0.009)\times 10^{12}/{\rm s}
   ~,~~\Delta\Gamma/\Gamma=-0.137\pm 0.013~.\ee
Measurements of T violation in $B_s\Bqs$ transitions are controversial. Using $D_s\mu X$ events, ${\rm Re\,\epsilon}=A_T/4$ is measured to be 
$(-2.8 \pm 1.9\pm 0.4)\times 10^{-3}$ by D0 in 2013 \cite{2013-D0-PRL110-011801} and $(-0.2 \pm 1.3\pm 0.9)\times 10^{-3}$ by LHCb in 2014 \cite{2014-LHCb-PLB728-607}.
Both results are compatible with each other and with no T violation. However, D0 has also measured several linear combinations of ${\rm Re\,\epsilon}(B_s)$ and
${\rm Re\,\epsilon}(B^0)$ from single-muon and likesign-dimuon events. A combined fit to these data in 2014 \cite{2014-D0-PRD89-012002} 
deviates from the Standard-Model expectations by $3.6\,\sigma$. The expectation for the $B^0$ system was given below Eq.~\ref{Eq-6-epsB}. For the $B_s$ system,
A.~Lenz and U.~Nierste \cite{2007-LenzNierste} obtain an even smaller value of $\rm Re\,\epsilon = (0.52\pm 0.14)\times 10^{-5}$. More precise data are clearly needed. 

\section{$B^0$ Decays and their Symmetry Properties} \label{Sec-BDecays}

Any difference in the time-dependent rates of $B^0$ and $\Bqz$ decays into a CP eigenstate is a violation of CP symmetry. Decays into the CP eigenstates $\jpsi K^0_S$
and $\jpsi K^0_L$ show large CP violation of this type since 2001 \cite{2001-BABAR,2001-Belle}. It can have four contributions, CPT violation in $B^0\Bqz$ transitions, 
T violation therein, CPT violation in the decay, and T violation therein. In this Section, we discuss how to determine the four fractions in the observed
CP violation. 
Decays into a final state $f$ are described by 
\be A_f = \langle f|D|B^0\rangle~,~~\Aq_f = \langle f|D|\Bqz\rangle~,\label{Eq-196}\ee
and we first develop the formalism for the time-dependent rates of decays into any state $f$ that can be reached from both $B^0$ and $\Bqz$, not necessarily a CP eigenstate.

\subsection{Time Dependence of $B^0$ and $\Bqz$ Decays into a Common Final State} \label{Sub-BtoF}

As light and heavy mass eigenstates of the $B^0$ system we use the same states as introduced in Section \ref{Sec-Dzero}, neglecting terms of order $|\epsilon|^2$
and $|\delta|^2$,
\bea B^0_\ell(t) = [(1+\epsilon+\delta)\, B^0 +(1-\epsilon-\delta)\,\Bqz]\times \re^{-\ri m_\ell t-\Gamma_\ell t/2}/\sqrt{2}~,\nonumber\\
     B^0_h(t) = [(1+\epsilon-\delta)\, B^0 -(1-\epsilon+\delta)\,\Bqz]\times \re^{-\ri m_h t-\Gamma_h t/2}/\sqrt{2}~.\eea
At $t=0$, this leads to
\bea B^0 & = & [(1-\epsilon+\delta)\, B^0_\ell +(1-\epsilon-\delta)\, B^0_h]/\sqrt{2}~,\nonumber\\
     \Bqz & = & [(1+\epsilon-\delta)\, B^0_\ell-(1+\epsilon+\delta)\, B^0_h ]/\sqrt{2}~.\eea
Like Kaons in Eqs.~\ref{Eq-Kevolution} and \ref{Eq-Kbarevolution}, the initial states $\Psi_B(0) = B^0$ and $\Psi_\Bq(0)=\Bqz$ have the evolutions
\bea
\Psi_B (t) = \frac{1}{2}\{[(1+2\delta)B^0+(1-2\epsilon)\Bqz]\,\re^{-\ri m_\ell t-\Gamma_\ell t/2}
                             +[(1-2\delta)B^0-(1-2\epsilon)\Bqz]\,\re^{-\ri m_h t-\Gamma_h t/2}\}~,\nn\\
\Psi_\Bq (t) = \frac{1}{2}\{[(1+2\epsilon)B^0+(1-2\delta)\Bqz]\,\re^{-\ri m_\ell t-\Gamma_\ell t/2}
                             -[(1+2\epsilon)B^0-(1+2\delta)\Bqz]\,\re^{-\ri m_h t-\Gamma_h t/2}\}~.\label{Eq-8-1} \eea
Using $1+2\epsilon =p/q$ and the amplitude definitions in Eq.~\ref{Eq-196}, the decays $\Psi_B\to f$ and $\Psi_\Bq\to f$ have the time-dependent amplitude 
\bea a_f(t) = \frac{\re^{-\Gamma t/2}}{2}\{[(1+2\,\delta)\,A_f+q\,\Aq_f/p]\,\re^{-\ri m_\ell t+\Delta\Gamma t/4}+[(1-2\,\delta)A_f-q\,\Aq_f/p]
              \,\re^{-\ri m_h t-\Delta\Gamma t/4}\}\, ,\nn\\
     \aq_f(t) = \frac{\re^{-\Gamma t/2}}{2}\{[p\,A_f/q+(1-2\,\delta)\,\Aq_f]\,\re^{-\ri m_\ell t+\Delta\Gamma t/4}-[p\,A_f/q-(1+2\delta)\Aq_f]
              \,\re^{-\ri m_h t-\Delta\Gamma t/4}\}\, .\eea
With the definitions
\be \frac{q \Aq_f}{p A_f}=\lambda_f~,~~\frac{p A_f}{q \Aq_f}=\frac{1}{\lambda_f}={\overline\lambda}{}_f~,\label{Eq-Deflambda}\ee
their time-dependent decay rates are
\bea N_{Bf}(t) = |a_f(t)|^2  = \frac{|A_f|^2\,\re^{-\Gamma t}}{4}\left|( 1+2\,\delta+\lambda_f)\,\re^{\ri\Delta m t}\,\re^{\Delta\Gamma t/4}
                                      +(1- 2\,\delta-\lambda_f)\,\re^{-\Delta\Gamma t/4}\right|^2,\nn\\
   N_{\Bq f}(t) = |\aq_f(t)|^2  =  \frac{|\Aq_f|^2\,\re^{-\Gamma t}}{4}\left|( 1-2\,\delta+{\overline\lambda}{}_f)\,\re^{\ri\Delta m t}\,\re^{\Delta\Gamma t/4}
                                      +(1+ 2\,\delta-{\overline\lambda}{}_f)\,\re^{-\Delta\Gamma t/4}\right|^2.\eea
In the approximation $|\lambda_f|^2\approx 1$ with $||\lambda_f|^2-1|$ as small as $|\epsilon|$ and $|\delta|$, the explicit dependences on all parameters are given by the 
following expressions:
\newpage 
\bea N_{Bf}(t) &=& |A_f|^2\,\re^{-\Gamma t}\left\{\left(\frac{|\lambda_f|^2+1}{2}+2\,{\rm Re\lambda_f Re\delta}+2\,{\rm Im\lambda_f Im\delta}\right)
                \,\cosh\frac{\Delta\Gamma t}{2} +({\rm Re\lambda_f+2 Re\delta})\,\sinh\frac{\Delta\Gamma t}{2}\right.\nn\\
            &+& \left.\left(\frac{1-|\lambda_f|^2}{2}-2\,{\rm Re\lambda_f Re\delta}-2\,{\rm Im\lambda_f Im\delta}\right)\,\cos\Delta m t-({\rm Im\lambda_f+ 2 Im\delta})
            \,\sin\Delta m t\right\}\, ,\nn\\
  N_{\Bq f}(t) &=& \frac{|\Aq_f|^2}{|\lambda_f|^2}\,\re^{-\Gamma t}\left\{\left(\frac{|\lambda_f|^2+1}{2}-2\,{\rm Re\lambda_f Re\delta}+2\,{\rm Im\lambda_f Im\delta}\right)
                \,\cosh\frac{\Delta\Gamma t}{2} +({\rm Re\lambda_f-2 Re\delta})\,\sinh\frac{\Delta\Gamma t}{2}\right.\nn\\
            &+& \left.\left(\frac{|\lambda_f|^2-1}{2}+2\,{\rm Re\lambda_f Re\delta}-2\,{\rm Im\lambda_f Im\delta}\right)\,\cos\Delta m t+({\rm Im\lambda_f+ 2 Im\delta})
            \,\sin\Delta m t\right\}\, .\label{Eq-rateBtoF}\eea

We now consider $f$ to be a CP eigenstate. A non-vanishing difference of the two rates $N_{Bf}(t)$ and $N_{\Bq f}(t)$ violates CP symmetry. 
The difference depends on the five parameters 
Re\,$\delta$, Im\,$\delta$, Re\,$\epsilon$, $|\Aq_f/A_f|$, Im\,$\lambda_f$ and also on the sign of Re\,$\lambda_f$.
The modulus of $|\lambda_f|$ is given by $\rm Re\,\epsilon$ and $|\Aq_f/A_f|$, and the sign of Re\,$\lambda_f$ needs to be determined separately.
The parameters Re\,$\delta$ and Im\,$\delta$ describe CPT violation
and Re\,$\epsilon$ T violation in transitions. The fourth parameter $|\Aq_f/A_f|$ describes CPT violation in decays if $f$ is a single state with only one 
final-state-interaction phase, as discussed in Subsection \ref{Sub-CPT}. In the following we assume this property of $f$ and discuss possible deviations 
in Section \ref{Sub-mimiking}.
T violation in decays is not observable by Im\,$(\Aq_f/A_f)$ since the amplitudes and also their ratio have no observable phases owing to arbitrary phases
of the states $|B^0\rangle$ and $|\Bqz\rangle$. The ratio $q/p$ depends on these phases in such a way that the product $\lambda_f = q\Aq_f/p A_f$ is an observable.
T symmetry requires that its phase has to be 0 or $\pi$, i.~e.~Im\,$\lambda_f =0$; this T-symmetry property in the
``interplay of decay and transitions" replaces the unobservable T-symmetry property in decays.

Since $N_{Bf}(0)=|A_f|^2$ and $N_{\Bq f}(0)=|\Aq_f|^2$, the ratio of the two rates at $t=0$ determines the parameter $|\Aq_f/A_f|$. 
As already discussed in Section \ref{Sub-B0}, the parameters $\rm Re\,\epsilon$ and $\rm Im\,\delta$  are determined by rate differences in $B^0\Bqz$ transitions, 
e.~g.~using di-lepton events. The two remaining CP-violating parameters Re\,$\delta$  and Im\,$\lambda_f$ are determined by the time dependences
in Eqs.~\ref{Eq-rateBtoF}. The product $\rm Re\,\lambda_f\,Re\,\delta$ enters with different signs in the cosine term, 
and $\rm Im\,\lambda_f$ with different signs
in the sine term of the two time dependences. A combined fit determines both parameters, and a combined fit to the two rates together with
the two di-lepton rates determines all five CP-violating parameters and $\Delta\Gamma$. Such a combined fit has been performed with the data of BABAR \cite{2004-BabarCPTprd} 
and Belle \cite{2012-BelleCPT} using the two final states $\jpsi K^0_S$ and $\jpsi K^0_L$. For the signs of $\Delta\Gamma$ and Re\,$\delta$, it is important to know the sign of
Re\,$\lambda_f$. This has been determined in the final state $\jpsi K^{*0}$ \cite{2005-BABAR-cos2beta}. The present best values for $\Delta\Gamma$, Re\,$\delta$,
Im\,$\delta$ and Re\,$\epsilon$ have been given in Section \ref{Sub-B0}, see Eqs.~\ref{Eq-7-DeltaGammaB0}, \ref{Eq-7-deltaB} and \ref{Eq-6-epsB}. 
The determination of $|\Aq_f/A_f|$ and Im\,$\lambda_f$ for the two states $f=\jpsi K^0_S$ and $\jpsi K^0_L$ will be discussed in the next Section \ref{Sub-BtoJpsiK}. 

\subsection{The CP Eigenstates $\jpsi K^0_S$ and $\jpsi K^0_L$} \label{Sub-BtoJpsiK}

On the level of $10^{-3}$, the two final states $\jpsi K^0_S$ and $\jpsi K^0_L$ are CP eigenstates. Their eigenvalues are obtained by decomposing the states,
\be |\jpsi K^0_{S,L}\rangle = |\jpsi \rangle \times |K^0_{S,L}\rangle \times Y_\ell^m(\theta,\phi)~,\ee
with $\ell =1$ and $m=0$. Because of CP($\jpsi$) = CP($K^0_S$) $=+1$ and CP($K^0_L$) = CP($ Y_1^0$) $=-1$, the CP eigenvalues of the two states are 
CP($\jpsi K^0_S$) $=-1$ and CP ($\jpsi K^0_L$) $=+1$. 

\begin{figure}[h]
\centering
\begin{minipage}{0.9\textwidth}
\includegraphics[width=\textwidth]{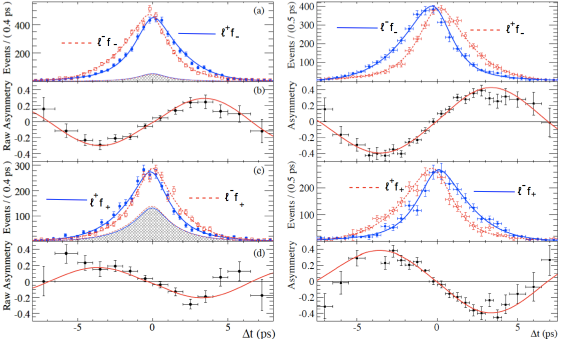}
\caption{Final CP- and T-violation results of BABAR \cite{2009-BABAR-sin2beta} on the left-hand side and Belle \cite{2012-Belle-sin2beta} on the right-hand side.
The four time dependences for each experiment are labeled $\ell^\pm, f_\pm$, where entangled $B^0\Bqz$ pairs decay into one flavor-specific final state
($\ell^+X$ or $\ell^- X$) and one CP eigenstate($f_+=J/\psi K_L$ or $f_-=J/\psi K_S$) and $\Delta t = t(f_\pm {\rm~decay})-t(\ell^\pm~{\rm decay})$.} 
\label{Fig-8-FinalSin2beta} 
\end{minipage}
\end{figure}

Large CP violation in decays of initial $B^0$ and $\Bqz$ states into these two final states has been observed in 2001 by BABAR \cite{2001-BABAR} and Belle \cite{2001-Belle} in the
reaction $e^+e^-\to\Upsilon(4S)\to B^0\Bqz$. Final results in 2009 from BABAR \cite{2009-BABAR-sin2beta} with $465\times 10^6$ $\Upsilon(4S)$ decays and in 2012 from Belle 
\cite{2012-Belle-sin2beta} with $772\times 10^6$ are shown in Fig.~\ref{Fig-8-FinalSin2beta}. The figure shows rates and asymmetries for the sum of final states 
$c{\overline c}\,K^0_S, CP=-1$ and $c{\overline c}\,K^0_L, CP=+1$, where $CP=+1$ contains in addition events with $\jpsi$ and $K^{*0}\to K^0_S\pi_0$. The parameter values of the
time-dependent rates have been determined for all final states separately, their compatibility allows to quote here the values for the sums with $CP=+1$ and $-1$.

As already presented in the previous Section \ref{Sub-BtoF}, combined fits to $c{\overline c}\,K$ CP eigenstates and flavor-specific final states resulted in values
for $\rm Re\,\epsilon$, $\rm Re\,\delta$, $\rm Im\,\delta$ and $\Delta\Gamma$ which are compatible with zero. Without influencing any conclusions on CP, T and CPT violation in decays, 
we can set the four parameters to zero in Eqs.~\ref{Eq-rateBtoF} and discuss the rates in their simplified form 
\bea N_{Bf}(t) &=& |A_f|^2\,\re^{-\Gamma t}\left(\frac{1+|\lambda_f|^2}{2}+\frac{1-|\lambda_f|^2}{2}\cos\Delta m t-{\rm Im\,}\lambda_f\,\sin\Delta m t\right)~,\nn\\
  N_{\Bq f}(t) &=& \left|\frac{\Aq_f}{\lambda_f}\right|^2¸\,\re^{-\Gamma t}\left(\frac{1+|\lambda_f|^2}{2}-\frac{1-|\lambda_f|^2}{2}\cos\Delta m t 
                    +{\rm Im\,}\lambda_f\,\sin\Delta m t\right)~,\label{Eq-rateBtoFsimpl}\eea
with $N_{Bf}(0) = |A_f|^2$ and $N_{\Bq f}(0) = |\Aq_f|^2$. We see that the expected rates depend on only two parameters, $|\lambda_f|$ and ${\rm Im\,}\lambda_f$ for each final
state. Care must be taken for the sign of ${\rm Im\,}\lambda_f$ since it depends on the sign of $q$ which differs
in different textbooks, reviews or analyses. In this review we consistently use the convention of Eq.~\ref{Eq-4-2-1a} for all four neutral-meson systems, i.~e.~with $\delta=0$,
\be M_h = p\,M -q\,{\overline M}~,~~M_\ell = p\,M +q\,{\overline M}~.\ee 
Eqs.~\ref{Eq-rateBtoFsimpl} show two different possibilities for CP violation. The two rates are different, i.~e.~CP-symmetry violating, if $|\lambda_f|^2\ne 1$ or/and if 
Im\,$\lambda_f\ne 0$. Since $|q/p|=1$, we have $|\lambda_f|=|\Aq_f/A_f|$ and
the first case violates CP and CPT if $A_f$ is a single amplitude, as derived in Section \ref{Sub-CPT}. 
The second case is CP and T violation since the condition Im\,$\lambda_f\ne 0$ is insensitive to $|\Aq_f/A_f|$, i.~e.~not sensitive to CPT symmetry; CP violation without
CPT violation means T violation. 

The BABAR and Belle results in Fig.~\ref{Fig-8-FinalSin2beta} are obtained from entangled $B^0\Bqz$ pairs with reconstruction of two decays. One final state is  the CP eigenstate
$f_-=c{\overline c}\,K^0_S$ or $f_+=c{\overline c}\,K^0_L,K^{*0}$. The other decay is flavor-specific, $\ell^+$ ($\ell^-$) like $\ell^+\nu X$ ($\ell^-\nu X$) indicating that the
decaying particle was in the state $B^0$ ($\Bqz$)  
and, if that was the first decay, leaving the other $B$ meson in the state $\Bqz$ ($B^0$). The time
difference $\Delta t$ is defined as $t(f_+,f_-)- t(\ell^+,\ell^-)$. 
For $\Delta t > 0$, the observed time-dependent rates are given by the expressions in 
Eqs.~\ref{Eq-rateBtoFsimpl} with $t=+\Delta t$, $f=f_+$ or $f_-$ and $N_{Bf}(t)$ for the preceding $\ell^-$ decay, $N_{\Bq f}(t)$ for the preceding $\ell^+$ decay. 

For $\Delta t < 0$, the first decay is $f_+$ ($f_-$), leaving the other $B$ meson in an orthogonal state $B_{\perp +}$ ($B_{\perp -}$) which is strictly forbidden to decay 
into $f_+$ ($f_-$) at $\Delta t = 0$. The shown time-dependent
rates for $\Delta t <0$ are the rates $|\langle B^0~{\rm or}~\Bqz|D|B_{\perp +} ~{\rm or}~B_{\perp -}(t)\rangle|^2$ with $t=-\Delta t$. 
A remarkable property of entanglement is the common description of the data with $\Delta t >0$ and $\Delta t <0$ by the so-called two-decay-time formula
\cite{BrancoLavouraSilva,1961-Day,1968-Lipkin}. It extends Eq.~\ref{Eq-rateBtoFsimpl} over the full range of $t$ by the replacement $t\to\Delta t$, and $f=f_+$ of $f_-$:
\bea N(\Bqz~{\rm at}~t_1,f~{\rm at}~t_2) =  |A_f|^2\,\re^{-\Gamma t}\left[\frac{1+|\lambda_f|^2}{2}+\frac{1-|\lambda_f|^2}{2}\cos(\Delta m \Delta t) 
                                          -{\rm Im\,}\lambda_f\,\sin(\Delta m \Delta t) \right],\nn\\
  N(B^0~{\rm at}~t_1,f~{\rm at}~t_2) = \left|\frac{\Aq_f}{\lambda_f}\right|^2¸\,\re^{-\Gamma t}\left[\frac{1+|\lambda_f|^2}{2}-\frac{1-|\lambda_f|^2}{2}\cos(\Delta m \Delta t)  
                    +{\rm Im\,}\lambda_f\,\sin(\Delta m \Delta t) \right].\label{Eq-8-TwoDecayTime}\eea

However, all properties of transitions and of CP, T, CPT symmetry 
in transitions and in decays are single-particle properties and not properties of entangled pairs. Therefore, we first discuss here the results with only $\Delta t >0$. 
They are also obtained from single $B$ mesons with incoherent flavor tagging, e.~g.~by LHCb \cite{2013-LHCb-sin2beta}.
Using the notations
\be 
A_-=\langle c\cq K_S|D|B^0\rangle~,~A_+=\langle c\cq K_L|D|B^0\rangle~,~\Aq_-=\langle c\cq K_S|D|\Bqz\rangle~,~\Aq_+=\langle c\cq K_L|D|\Bqz\rangle,
\label{Eq-8-2-1}\ee 
the main qualitative features of the results in Fig.~\ref{Fig-8-FinalSin2beta} with $t=\Delta t > 0$ are the following: 
\begin{enumerate}\itemsep -2pt
\item At $t=0$, $|A_-|^2= |\Aq_-|^2$ and $|A_+|^2= |\Aq_+|^2$,
\item at $t=0$, $|A_-|^2= |A_+|^2$ and $|\Aq_-|^2= |\Aq_+|^2$,
\item all four rates are not exponentials but well described by $\re^{-\Gamma t}(1+C_f\cos\Delta m t+S_f\sin\Delta mt)$,
\item for all four rates, $|\lambda_f|$ is compatible with 1,
\item $|{\rm Im}~\lambda_f|\ne 0$ is equal for all rates, with
\item ${\rm Im}~\lambda_{f+}=-{\rm Im}~\lambda_{f-}$.
\end{enumerate}
From both (1) and (4) we conclude that there is no visible CPT violation; the case with $A_f$ containing two final-state-interaction phases will be discussed in Section
\ref{Sub-mimiking}. From (5) we conclude that there is large T violation \cite{2013-Gerber}, for both $f_-$ and $f_+$. The observation (6) that the two T 
violations are equal with opposite
signs and the observation (2) are not understandable by Eqs.~\ref{Eq-rateBtoFsimpl}. They require an additional property of $B$ meson decays. Introducing the four amplitudes
and a new $\lambda$ parameter, 
\be
A = \langle c\cq K^0|D|B^0\rangle~,~x\,A =\langle c\cq \Kqz|D|B^0\rangle~,~\Aq =\langle c\cq \Kqz|D|\Bqz\rangle~,~\xq\,\Aq =\langle c\cq K^0|D|\Bqz\rangle~,~
     \lambda=\frac{q\,\Aq}{p\,A}~,\label{Eq-8-2-2}
\ee
we obtain with $|x|\ll 1$, $|\xq|\ll 1$, $|\lambda|=1$, $K_S = (K^0+\Kqz)/\sqrt{2}$ and $K_L = (K^0-\Kqz)/\sqrt{2}$:
\bea
A_- = A(1+x)/\sqrt{2}~,~A_+ = A(1-x)/\sqrt{2}~&,&~\Aq_- = \Aq(1+\xq)/\sqrt{2}~,~\Aq_+ = -\Aq(1-\xq)/\sqrt{2}\,\nn\\
      |A_-|^2 =\frac{|A|^2}{2}\,(1+2\,{\rm Re\,}x)~&,&~|A_+|^2 =\frac{|A|^2}{2}\,(1-2\,{\rm Re\,}x)\,\nn\\
      |\Aq_-|^2 =\frac{|\Aq|^2}{2}\,(1+2\,{\rm Re\,}\xq)~&,&~|\Aq_+|^2 =\frac{|\Aq|^2}{2}\,(1-2\,{\rm Re\,}\xq)\,\nn\\
      |\lambda_{f-}|^2 = 1-2\,{\rm Re\,}(x-\xq)~&,&~|\lambda_{f+}|^2 = 1+2\,{\rm Re\,}(x-\xq)\,\nn\\
      {\rm Im\,}\lambda_{f-} = {\rm Im\,}[\lambda (1-x+\xq)]~&,&
      ~{\rm Im\,}\lambda_{f+} = -{\rm Im\,}[\lambda (1+x-\xq)]~.
\eea
Observation (2) is equivalent to ${\rm Re\,}x = {\rm Re\,}\xq =0$, and observation (6) is equivalent to ${\rm Im\,}(x-\xq)=0$ if ${\rm Re\,}(x-\xq)=0$.

The hypothesis $x=\xq=0$ is called the ``$\Delta S = \Delta b$" rule and has an explanation in the perturbative description of $b\to c\cq s$ decays in the quark model.
Unfortunately, the rule is not well tested experimentally. The best experimental limit exists for ${\rm Im\,}(x-\xq)$ from 
${\rm Im\,}\lambda_{f-}+{\rm Im\,}\lambda_{f+}= 0.002\pm0.047$ \cite{2013-HFAG}, and it seems that no values for ${\rm Re\,}x$  and ${\rm Re\,}\xq$ have been
determined so far. But the $\Delta S = \Delta b$ rule
is only required for the relation between $\lambda_{f-}$ and $\lambda_{f+}$, it is irrelevant for the main conclusion of this Section:
The data of BABAR \cite{2001-BABAR,2009-BABAR-sin2beta} and Belle \cite{2001-Belle,2012-Belle-sin2beta} do not show CPT violation but show large T 
violation in decays $B^0,\Bqz\to\jpsi K^0_S, \jpsi K^0_L$.   

The conclusion of large T-symmetry violation, independent of any assumption on CPT symmetry, is reached quantitatively if the parameters $|\lambda_f|$ and Im\,$\lambda_f$
are obtained by a fit to the data of the B-meson-factory experiments BABAR and Belle for the combination of $\Delta t >0$ and $\Delta t<0$ using Eqs.~\ref{Eq-8-TwoDecayTime}. 
Both experiments cannot measure 
$t_1$ and $t_2$ owing to the wide spread of $\Upsilon(4S)$ production points. However, they can measure well the separation of the two decay points yielding $\Delta t=t_2-t_1$. 
The mean values from the final BABAR and Belle fits over the full range $-\infty<\Delta t<\infty$ are given in Table \ref{Tab-8-FinalResults}, where the last 
line assumes strict validity of $\Delta S=\Delta b$.
\begin{table}[h]
\setlength{\belowcaptionskip}{2mm}
\begin{center}
\caption{Mean values \cite{2013-HFAG} of the final results of BABAR \cite{2009-BABAR-sin2beta} and Belle \cite{2012-Belle-sin2beta}. The last line combines $\jpsi K_S$,
         $\jpsi K_L$ and additional $c\cq K$ and $K^*$ results.} 
\label{Tab-8-FinalResults}
\begin{tabular}{|c|c|c|}
\hline
& Im\,$\lambda_f$ & $(1-|\lambda_f|^2)/2$ \\
\hline
$f=\jpsi K_S$ & $-0.665\pm 0.024$ & $0.024 \pm 0.026$ \\
$f=\jpsi K_L$ & $+0.663\pm 0.041$ & $-0.023 \pm 0.030$ \\
combined & $0.677\pm 0.020$ & $0.006\pm 0.017$\\
\hline\end{tabular}
\end{center}\end{table}

Direct T violation, as discussed in the next Section \ref{Sub-mimiking}, contributes to $|\lambda_f|^2$ if $A_f$ is not a single amplitude. Therefore, the measured value of $|\lambda|$ does 
not prove the absence of CPT violation on the $o(10^{-2})$ level.  BABAR and Belle can only conclude that there is no visible CPT violation and that the observed
CP violation is compatible with T violation alone. If $A_f$ is not a single amplitude, direct CPT violation could contribute to Im\,$\lambda$, as also discussed in the next Section.

\subsection{More than one Final-State-Interaction Phase} \label{Sub-mimiking}

In this Section, we discuss decays where the amplitude does not obey $\Aq = \re^{2\ri\,\delta}\,A^*$ with a single final-state-interaction (FSI) phase $\delta$. 
The presence of two or more FSI phases can lead to apparent CPT violation owing to T violation, apparent T violation owing to CPT violation, and to
$|\Aq/A|= 1$ in spite of CPT violation.

The first case, $|\Aq/A|\ne 1$ in spite of CPT symmetry, is known as ``direct CP violation", we may also call it ``direct T violation". It requires at least
two contributions to $A$ with different weak and different FSI phases:
\be A=A_1\,\re^{\ri\,\phi_1}\,\re^{\ri\,\delta_1}+A_2\,\re^{\ri\,\phi_2}\,\re^{\ri\,\delta_2},\ee
with $A_1$, $A_2$ real, weak-interaction phases $\phi_i$ and strong-interaction phases $\delta_i$. CPT symmetry requires 
\be \Aq=A_1\,\re^{-\ri\,\phi_1}\,\re^{\ri\,\delta_1}+A_2\,\re^{-\ri\,\phi_2}\,\re^{\ri\,\delta_2}.\ee
With $a=A_2/A_1,~\Delta\phi=\phi_2-\phi_1,~\Delta\delta=\delta_2-\delta_1$ and $a\ll 1$ we obtain the T-violating asymmetry 
\be \frac{|\Aq|^2-|A|^2}{|\Aq|^2+|A|^2}\approx 2\,a\,\sin\Delta\phi\,\sin\Delta\delta~,\ee
which does not vanish if $\Delta\phi\ne 0$ and $\Delta\delta\ne 0$, in spite of CPT symmetry. 

The third case, $|\Aq/A|= 1$ in spite of CPT violation, can be reached by the addition of a third amplitude $A_3\,\re^{\ri(\phi_3+\delta_3)}$  
violating CPT symmetry, i.~e.~$\Aq_3\ne A_3$.
If $a\,\sin\Delta\phi\,\sin\Delta\delta$ is small, a simple ansatz is
\be A= A_1\,\re^{\ri\phi_1}\,\re^{\ri\delta_1}\,[1+a\,\re^{\ri(\Delta\phi+\Delta\delta)}+b]~,~~\Aq= A_1\,\re^{-\ri\phi_1}\,\re^{\ri\delta_1}\,
       [1+a\,\re^{\ri(-\Delta\phi+\Delta\delta)}+\bq]~,\ee
with complex amplitude ratios $b$ and $\bq$ and $\delta_3=\delta_1$. CPT violation requires $\bq\ne b^*$.
With $|b|$ and $|\bq| \ll 1$, this leads to an asymmetry
\be \frac{|\Aq|^2-|A|^2}{|\Aq|^2+|A|^2}\approx 2\,a\,\sin\Delta\phi\,\sin\Delta\delta+{\rm Re}\,(\bq-b)~.\ee
Since $b$ and $\bq$ are arbitrary, they can cancel the contribution of direct T violation, leading to $|\Aq/A|= 1$
in spite of CPT violation in the decay amplitudes.

The second case can also be reached by an amplitude with at least two FSI phases. Apparent T violation
with Im\,$\lambda_f\ne 0$ in spite of T symmetry is obtained by an amplitude with a CPT-symmetric and a CPT-violating contribution \cite{2014-Gerber}:
\bea p\,A = A_1\,\re^{\ri\phi}\,\re^{\ri\delta_1}[1+b\,\re^{\ri\,(\delta_2-\delta_1)}]/\sqrt{2}~,
       ~~q\,\Aq = A_1\,\re^{\ri\phi}\,\re^{\ri\delta_1}[1+\bq\,\re^{\ri\,(\delta_2-\delta_1)}]/\sqrt{2}~,\nn\\
    {\rm Im\,}\lambda = \frac{(\bq-b)\,\sin(\delta_2-\delta_1)}{1+b^2+2\,b\,\cos(\delta_2-\delta_1)}~,\label{Eq-CptMimicsT}\eea
where $b=A_2/A_1$ and $\bq=\Aq_2/A_1$ have to be real if we impose T symmetry of the weak Hamiltonian.
With $\bq\ne b$ and $\sin(\delta_2-\delta_1)\ne 0$, CPT violation mimics T violation owing to final-state interactions.

The largest value of Im\,$\lambda$ is reached with $\sin(\delta_2-\delta_1)=1$, leading to
\be \lambda =\frac{1+\ri\,\bq}{1+\ri\,b}~,~~|\lambda|=\sqrt{\frac{1+\bq^2}{1+b^2}}~,~~{\rm Im\,}\lambda = \frac{\bq-b}{1+b^2}
          ~,~~{\rm Re\,}\lambda = \frac{1+b\bq}{1+b^2}~.\ee
A very special choice $\bq = -b =0.2$ fulfills, together with  $\sin(\delta_2-\delta_1)=1$, all three observations, 
$|\lambda|=1$, ${\rm Im\,}\lambda =0.7$, and also \cite{2005-BABAR-cos2beta} ${\rm Re\,}\lambda/{\rm Im\,}\lambda > 0$.
 
CPT violation could fake the observed large T violation in $B^0\to J/\psi K^0$ decays if $A$ consists of two amplitudes, 
one CPT-conserving, one CPT-violating on the $20\%$ level,
and with a very narrow window for their final-state-interaction phase difference. In all rigour, this construction could only be falsified with
``now impossible" experiments like measuring the cross sections for inverse decays
$c\cq K^0_S,K^0_L \to B^0,\Bqz$ \cite{2014-Nir}, or measuring the time dependences of the rates for $\Upsilon(4S)\to (c\cq K^0_S,K^0_L)(c\cq K^0_S,K^0_L)$
events \cite{2014-Gerber}. 

\subsection{Motion-Reversal Violation in Decays $B\to\jpsi K^0_{S,L}$} \label{Sub-TinBtoJpsiK}

Following the proposal of J.~Bernabeu et al.~in 1999 \cite{1999-BanulsBernabeu,2004-AlvarezBernabeu}, T violation in $B^0\to\jpsi K$ decays has been demonstrated two times by
``direct observation" of motion-reversal violation. The first demonstration was performed in 2008 by E.~Alvarez and A.~Szynkman \cite{2008-AlvarezSzynkman}
with a re-interpretation of BABAR and Belle results, the second by BABAR in 2012  \cite{2012-BABAR-T} with a re-analysis of the final BABAR data \cite{2009-BABAR-sin2beta}.

The four curves in Fig.~\ref{Fig-8-FinalSin2beta} are labeled ($\ell^+,f_-$), ($\ell^-,f_-$), ($\ell^+,f_+$) and ($\ell^-,f_+$) according to the two decays of the $B^0\Bqz$ pair.
One meson decays into a flavor-specific final state $\ell^+$ or $\ell^-$, the other one into a CP eigenstate 
$f_- = c\cq K^0_S$ or $f_+ = c\cq K^0_L$. For $\Delta t>0$ the $\ell^\pm$ decay occurs before and for $\Delta t<0$ after the $f_\pm$ decay. 
If the first decay is a $f_-$ decay, the other $B$ at that time is in a state which is not allowed to decay into $f_-$ owing to the 
antisymmetry of the entangled two-particle state; this state is given by $B_+ = (\Aq\,B^0-A\,\Bqz)/\sqrt{|A|^2+|\Aq|^2}$ using the notations in
Eq.~\ref{Eq-8-2-2}  with $x ={\overline x} = 0$, i.~e.~assuming $\Delta S = \Delta b$. 
With $f_+$ as first decay, the surviving state is $B_-= (\Aq\,B^0+A\,\Bqz)/\sqrt{|A|^2+|\Aq|^2}$. If $|\Aq/A|=1$, which
is well fulfilled experimentally ($\pm 1.7\%$) independent of assuming CPT symmetry, we have $\langle B_+|B_-\rangle = 0$ and the entangled pair is  
$(B_+ B_- - B_- B_+)/\sqrt{2}$. A first decay into $f_-$ prepares the state $B_+$, a decay into $f_+$ prepares the state $B_-$.

\begin{table}[h]
\setlength{\belowcaptionskip}{2mm}
\begin{center}
\caption{Correspondences between observed decay pairs and transitions for demonstrating the violation of motion-reversal
symmetry \cite{2008-AlvarezSzynkman,2012-BABAR-T}. The last column gives the fitted values for the coefficients $S_i$ in ref.~\cite{2012-BABAR-T}.} 
\label{Tab-8-4}
\begin{tabular}{|c|c|c|c|}
\hline
$i$ & Decay pairs & Transitions & $S_i$\\
\hline
1 & $\ell^-,f_-,\Delta t>0$ & $B^0\to B_-$ & $-0.76\pm 0.06\pm 0.06$\\
2 & $\ell^-,f_-,\Delta t<0$ & $B_+\to\Bqz$ & $+0.67\pm 0.10\pm 0.08$\\
3 & $\ell^+,f_+,\Delta t>0$ & $\Bqz\to B_+$ & $-0.69\pm 0.11\pm 0.04$\\
4 & $\ell^+,f_+,\Delta t<0$ & $B_-\to B^0$ & $+0.70\pm 0.19\pm 0.12$\\
5 & $\ell^+,f_-,\Delta t>0$ & $\Bqz\to B_-$ & $+0.55\pm 0.09\pm 0.06$\\
6 & $\ell^+,f_-,\Delta t<0$ & $B_+\to B^0$ & $-0.66\pm 0.06\pm 0.04$\\
7 & $\ell^-,f_+,\Delta t>0$ & $B^0\to B_+$ & $+0.51\pm 0.17\pm 0.11$\\
8 & $\ell^-,f_+,\Delta t<0$ & $B_-\to \Bqz$ & $-0.83\pm 0.11\pm 0.06$\\
\hline\end{tabular}
\end{center}\end{table}

Motion-reversal symmetry requires that the rates for the transitions $B^0\to B_-$ and $B_-\to B^0$ are equal at any time, also the rates for $\Bqz\leftrightarrow B_+$,  
$B^0\leftrightarrow B_+$ and $\Bqz\leftrightarrow B_-$. The correspondences between transition rates and the observed decay-pair rates are given in
Table \ref{Tab-8-4}. 
The rates for the observed decay-pairs are equal to the rates of the corresponding transitions if $|\Aq/A|=1$ \cite{2013-MITPsummary} and if
$\Gamma(B^0\to\ell^+)=\Gamma(\Bqz\to\ell^-)$. 

In their motion-reversal analysis, BABAR \cite{2012-BABAR-T} determines the four time-dependent asymmetries
\be A_T(t|i,j) =\frac{N_i(t)-N_j(t)}{N_i(t)+N_j(t)}~,\ee
using the eight measured rates in Table \ref{Tab-8-4} with the pairings $(i,j)=(1,4),(3,2),(5,8)$ and $(7,6)$. For separating the rates into their contributions 
with $\Delta t > 0$ and $\Delta t <0$, BABAR has unfolded the resolution function ${\cal R}(\Delta t - \Delta t_{\rm true})$ in the measured rates $N(\Delta t)$ 
with $-\infty<\Delta t<+\infty$. The unfolding requires an ansatz for the true rates $N_i(t)$. BABAR has chosen
\be N_i(t) \propto \re^{-\Gamma t}\,(1+C_i\cos\Delta m t +S_i\sin \Delta m t)~,\ee
where $t=-\Delta t_{\rm true}$ or $+\Delta t_{\rm true}$ depending on $i$. The eight obtained coefficients $C_i$ are compatible with zero, the eight $S_i$ 
are given in Table \ref{Tab-8-4}. The large difference $S_1-S_4$ violates motion-reversal symmetry, as well as the three other asymmetries. 
Their combined significance for being different from zero is 14 $\sigma$, a clear violation of motion-reversal symmetry.   

The 2008 analysis of Alvarez and Szynkman \cite{2008-AlvarezSzynkman} follows the same strategy \cite{1999-BanulsBernabeu,2004-AlvarezBernabeu} but does not separate 
the rates $N(\ell^\pm, f_\pm)$ into their parts with $\Delta t >0$ and $\Delta t <0$. 
As demonstrated in the following, this leads to the same conclusion as reached by BABAR in 2012.

Because of entanglement, the two rates $N_7$ and $N_8$ for the transitions $B^0\to B_+$ and $B_-\to \Bqz$ respectively, using the notations of Table 4, are described by the
same two-decay-time expression in Eq.~\ref{Eq-8-TwoDecayTime} with two parameters $|\lambda_{f+}|$ and Im\,$\lambda_{f+}$.
Since $|\lambda_{f+}|=1$ within errors and taking the 2007 average of BABAR and Belle,
\be N_7 = N_8 = const.\times \re^{-\Gamma|\Delta t|}\,[1+(0.666\pm 0.046)\sin\Delta m\Delta t]~.\label{Eq-F-1}\ee
For $N_7(B^0\to B_+)$ with $t=+\Delta t$, this leads to $S_7=+0,666\pm 0.046$. The time-reversed transition $B_+\to B^0$ is described by
\be N_6 (B_+\to B^0)=N_5(\Bqz\to B_-)= const.\times \re^{-\Gamma|\Delta t|}\,[1+(0.666\pm 0.028)\sin\Delta m\Delta t]~.\label{Eq-F-2}\ee
For $N_6$ with $t=-\Delta t$, $S_6=-0.666 \pm 0.028$, and the motion-reversal asymmetry is found to be
\be \frac{N_7(B^0\to B_+)-N_6(B_+\to B^0)}{N_7(B^0\to B_+)+N_6(B_+\to B^0)}=\frac{(S_7-S_6)\sin\Delta m t}{2+(S_7+S_6)\sin\Delta m t}
      =(0.666\pm 0.027)\sin\Delta m t~.\label{Eq-F-3}\ee
The values and errors presented in Eqs.~\ref{Eq-F-1}, \ref{Eq-F-2} and \ref{Eq-F-3} include the results from the other three motion-reversal
asymmetries in the data.

The conclusion of Alvarez and Szynkman in 2008 is the same as that of BABAR in 2012: The large asymmetry of $0.7\times\sin\Delta m t$ in Eq.~\ref{Eq-F-3} is an
unambiguous observation of motion-reversal violation. As discussed in Section \ref{Sub-mimiking} of this review: unambiguous if $A(B^0\to c\cq K^0)$ is a single amplitude, and not
rigorously excluded to be faked by CPT violation in a very narrow and very unlikely range of CPT-violating and final-state-interaction parameters.

\section{Outlook and Conclusions} \label{Sec-Outlook}

The lack of observing any violation of CPT symmetry is linked to all other successes of quantum field theory in which the CPT theorem 
\cite{1954-Lueders,1955-Pauli} follows from very basic principles. CPT-violating effects could occur if quarks or leptons are not point-like
or if Lorentz invariance is broken, e.~g.~owing to the appearance of Lorentz tensors \cite{2008-Kostelecky}. Motivated by the arguments of
V.~A.~Kosteleck\'y \cite{2008-Kostelecky}, BABAR \cite{2008-BABAR-Kostelecky} has searched for a special breakdown of Lorentz invariance by allowing a time-dependent
variation of the CPT-violating parameter $\delta$ in $B^0\Bqz$ transitions according to
\be z=-2\delta = z_0+z_1\times \cos(\Omega t_{\rm sid}+\Phi)~,\ee
where $\Omega$ describes the earth's rotation in the galactic frame and $t_{\rm sid}$ is the sidereal time. Fits to the di-lepton asymmetry in Eq.~\ref{Eq-ACPTB0}
give results for $z_0$ and $z_1$ which deviate slightly from zero. The significance of the deviation is, however, smaller than $3~\sigma$.

No T violation has been observed in strong and electromagnetic interactions. In weak interactions, T violation is well established. In 1970, it has 
been observed in $K^0\Kqz$ transitions by using unitarity and the existing CP-violation results and limits in all $K^0$ decay modes; Re~$\epsilon(K^0) = (1.68 \pm 0.30)~10^{-3}$, 
Eq.~\ref{Eq-KRSresult}. This Bell-Steinbergerv unitarity analysis has been repeated many times until recently. Today's best result is Re~$\epsilon(K^0) = (1.611 \pm 0.005)~10^{-3}$,
Eq.~\ref{Eq-137}. The result has been confirmed in 1998 by a direct motion-reversal observation, comparing the rates of $K^0\to\Kqz$ and $\Kqz\to K^0$, independent of unitarity
but assuming CPT symmetry in the semileptonic $K^0$ decay amplitudes, Re~$\epsilon(K^0) = (1.55 \pm 0.42)~10^{-3}$, Eq.~\ref{Eq-here-45}. 
No T violation has been observed in
$D^0\Dqz$, $B^0\Bqz$ and $B_s\Bqs$ transitions, but there are good prospects that the experiments LHCb and Belle-II will reach a sensitivity on Re~$\epsilon(B^0)$ on
the level of the expected Standard-Model T violation, Eq.~\ref{Eq-LenzNierste}. T violation has also been observed in weak decay amplitudes: 
T is violated in $K^0\to\pi\pi,\,I=2$ decays,
where the phase of $\epsilon^\prime/\epsilon_0$ is incompatible with T symmetry as shown in Fig.~\ref{Fig-6-KTeV}. 
And T is violated in $B^0\to J/\psi~K^0$ decays, observed in 
2001 since the CP-violating parameter $\lambda_f$ in these decays is also T-violating, see Section \ref{Sub-BtoJpsiK}.  
Entanglement of $B^0$ and $\Bqz$ states in $\Upsilon(4S)$ decays
allowed a direct demonstration of motion-reversal violation with these decays one decade later, see Section \ref{Sub-TinBtoJpsiK}.

No CPT violation has been observed, but 
future experiments with neutral-meson systems have the duty to test CPT symmetry with ever increasing precision. 
Nature may neither have point-like quarks and leptons nor fully Lorentz-invariant couplings of them to vector and scalar fields. \\[8mm]

{\Large{\bf Acknowledgements}}\\[4mm]
I would like to thank E.~Alvarez, J.~Bernabeu, A.~Denig, A.~Di Domenico, F.~Martinez, U.~Nierste, H.~Quinn, P.~Villanueva, R.~Waldi and especially
H.-J.~Gerber and T.~Ruf for many helpful discussions. I gratefully remember earlier discussions with J.~Bell, M.-K.~Gaillard, O.~Nachtmann and
B.~Stech.  I thank Frieder Niebergall
for carefully reading the manuscript and Philipp Schubert for his valuable help with applying Mathematica. I also thank M.~Neubert and T.~Hurth for
arranging the workshop ``T Violation and CPT Tests in Neutral-Meson Systems" in April 2013 at MITP Mainz where many items of this review have been
discussed.

\end{document}